\documentclass[aps,onecolumn,10pt]{revtex4}
\usepackage{amsmath}
\usepackage{amssymb}
\usepackage{hyperref}
\hypersetup{colorlinks=true} 
\numberwithin{equation}{section}
\numberwithin{equation}{section}

\begin{document}
\allowdisplaybreaks
\setcounter{equation}{0}

\title{Exact solution to perturbative conformal cosmology in the recombination era}

\author{Philip D. Mannheim}
\affiliation{Department of Physics, University of Connecticut, Storrs, CT 06269, USA \\
 philip.mannheim@uconn.edu\\ }

\date{November 15 2020}

\begin{abstract}

We study cosmological perturbation theory in the cosmology associated with conformal gravity, establish the validity of the decomposition theorem for it, and then use the theorem to provide an exact solution to the theory in the recombination era. Central to our approach is the use of a fully gauge invariant formulation of the cosmological fluctuations equations. In the recombination era not only is perturbation theory applicable, because of its specific structure in the conformal case the fluctuation equations are found to greatly simplify. Using a master equation for scalar, vector and tensor fluctuation modes we show that the radial equations for the three-dimensional vector and tensor modes are respectively the same as those of scalar modes in five and seven spatial dimensions. This enables us to construct normalization conditions for the three-dimensional modes.

\end{abstract}

\maketitle

\section{Introduction}
\label{S1}

\subsection{Motivation}
\label{S1az}

Since the discovery of the cosmic microwave background (CMB) a central focus of cosmological research has been  the study of fluctuations around that background (see e.g.  \cite{Dodelson2003,Mukhanov2005,Weinberg2008,Lyth2009,Ellis2012,Maggiore2018}). With the background itself being homogeneous and isotropic these fluctuations are associated with anisotropies and inhomogeneities in the background. While such fluctuations eventually grow non-perturbatively into the galaxies  that we see  today, at the time of recombination of electrons and baryons into atoms (typical temperatures of order one $eV$ or  $10^{4}$$^{\circ}K$),  these fluctuations were very small and could thus be explored perturbatively. The actual behavior of the perturbations depends on the dynamics of the particular gravitational theory under consideration, with the standard treatment of  these fluctuations being based on the Newton-Einstein gravitational theory, viz. the standard cosmological model. However, use of this model requires the inclusion of unobserved dark matter particles, of a not as of yet understood, highly fine-tuned  dark energy or cosmological constant component, and a presumption that the classical treatment of the model that is made would not be destroyed by quantum radiative corrections even though these corrections are known to lead to uncontrollable infinities \cite{footnoteY}.  In response to these concerns some candidate alternative proposals have been advanced in the literature and in this paper we consider one specific alternative, namely conformal gravity. 

As shown in  \cite{Mannheim1989,Mannheim1994,Mannheim2006,Mannheim2012b,Mannheim2017} and references therein, conformal gravity eliminates the need for galactic dark matter by providing fits to a wide class of galactic rotation curves without any need for dark matter. Moreover, conformal gravity has a local conformal symmetry (invariance under $g_{\mu\nu}(x)\rightarrow e^{2\alpha(x)}g_{\mu\nu}(x)$) that controls the cosmological constant without fine tuning. And with the conformal symmetry requiring the gravitational coupling constant, $\alpha_g$, to be dimensionless, the conformal theory is renormalizable. With conformal gravity also being quantum-mechanically ghost free and unitary \cite{Bender2008a,Bender2008b,Mannheim2011a,Mannheim2018}, conformal gravity  provides a consistent quantum gravity theory in  four spacetime dimensions. (The gravitational coupling constant associated with the conformal gravity action $I_{\rm W}$ that is given below is only dimensionless in four spacetime dimensions.)  The view of conformal gravity is that the dark matter, dark energy and quantum gravity problems are not three separate problems, but that since they all have the same common origin, namely the extrapolation of Newton-Einstein gravity beyond its solar system origins, they can have a common solution, with conformal gravity endeavoring to provide such a solution through a different extrapolation of solar system wisdom. However, in order for the conformal theory to be viable it needs to address the other regime in which the standard model needs dark matter, namely cosmology. And even though the conformal theory  has successfully done this for the homogeneous and isotropic background by providing \cite{Mannheim1992,Mannheim2006,Mannheim2012b,Mannheim2017} a horizon-free background cosmology with no flatness problem \cite{footnoteYY}, while providing a very good, non-fine-tuned, dark matter free  fit to the accelerating universe supernovae data of \cite{Riess1998,Perlmutter1999}, it still needs to do so for the fluctuations around that background. The development of the cosmological perturbation theory that is required for this has been presented in general in \cite{Mannheim2012a,Amarasinghe2019,Phelps2019,Amarasinghe2020}, and in this paper we take a further step by  providing a new exact solution to the conformal cosmological fluctuation equations in the recombination era. In regard to conformal gravity we note also that various other studies of conformal gravity and of higher derivative gravity theories in general can be found in \cite{Hoyle1964,Stelle1977,Stelle1978,Adler1982,Lee1982,Zee1983,Riegert1984a,Riegert1984b,Teyssandier1989,'tHooft2010a,'tHooft2010b,'tHooft2011,'tHooft2015a,Maldacena2011}.

In order to be able to derive solutions to the conformal gravity cosmological fluctuation equations, we first need to derive the equations themselves, and in so doing we actually obtain conformal gravity fluctuation equations that hold in any cosmological epoch, and not just at recombination. Moreover, even though negative spatial three-curvature is preferred for conformal gravity itself \cite{Mannheim2006,Mannheim2012b,Mannheim2017}, something we elaborate on Secs. \ref{S1} and \ref{S2}, the equations that we obtain [viz. (\ref{2.31y}), (\ref{2.32y}) and (\ref{2.33y})] are generic to conformal gravity in the sense that they hold for general background matter sources and for any general Robertson-Walker background with arbitrary expansion radius $a(t)$ and arbitrary spatial three-curvature $k$. To actually study the gravitational fluctuation equations we shall use the scalar, vector, tensor expansion of the fluctuation metric that was first introduced in  \cite{Lifshitz1946}  and \cite{Bardeen1980} and then widely applied in  perturbative cosmological studies  (see e.g. \cite{Kodama1984,Mukhanov1992,Stewart1990,Ma1995,Bertschinger1996,Zaldarriaga1998,Straumann2008,Szapudi2012} and \cite{Dodelson2003,Mukhanov2005,Weinberg2008,Lyth2009,Ellis2012,Maggiore2018}). This expansion is based on quantities that transform as three-dimensional scalars, vectors and tensors, and as such it is particularly well suited to Robertson-Walker geometries because such geometries have a spatial sector that is maximally three-symmetric. While not manifestly covariant, the scalar, vector, tensor expansion is covariant as it leads to equations that involve appropriate combinations of the scalars, vectors and tensors that are fully four-dimensionally gauge invariant, this being all that one needs  for covariance \cite{footnoteZ}. To be as general as possible we shall both derive and solve the fluctuation equations in a procedure in which this full gauge invariance is maintained at each stage of the process,  and shall make no restriction to any particularly convenient gauge that might facilitate finding a solution.

Since the appropriate gauge invariant combinations of the scalar, vector and tensor components of the fluctuation metric are coupled in the fluctuation equations, our strategy is to first manipulate these equations so that we  obtain equations in Sec. \ref{S3} in which these various components are decoupled. Since this decoupling is only achievable at a higher-derivative level, it is these higher-derivative equations that we will need to integrate. And to be able to do so  we will need to introduce spatial boundary conditions, with it turning out that we will need  boundary conditions not just at $r=\infty$ but also at  $r=0$. Asymptotic boundedness is not actually a new dynamical  assumption, since, as noted in \cite{Amarasinghe2019,Phelps2019},  it is actually needed in order to be able to make the scalar, vector, tensor expansion in the first place. For the typical case for instance of the decomposition of a three-dimensional Cartesian vector $A_i$ into its transverse and longitudinal components one wants to be able to set  $A_i=\partial_iV+V_i$ where $\partial^iV_i=0$.  On applying $\partial^i$ to $A_i$ we obtain 

\begin{eqnarray}
\partial^iA_i=\partial^i\partial_iV. 
\label{1.1y}
\end{eqnarray}
On introducing the Green's function $D^{(3)}(\mathbf{x}-\mathbf{y})$ that obeys 
\begin{eqnarray}
\partial_i\partial^iD^{(3)}(\mathbf{x}-\mathbf{y})=\delta^3(\mathbf{x}-\mathbf{y}),
\label{1.2y}
\end{eqnarray}
$V$ is given by 
\begin{eqnarray}
V({\bf x})=\int d^3yD^{(3)}(\mathbf{x}-\mathbf{y})\partial^iA_i({\bf y}),
\label{1.3y}
\end{eqnarray}
with $V_i$ then being given by
\begin{eqnarray}
V_i({\bf x})=A_i(x)-\partial_i\int d^3yD^{(3)}(\mathbf{x}-\mathbf{y})\partial^jA_j({\bf y}).
\label{1.4y}
\end{eqnarray}
Thus in order to be able to decompose a vector into its transverse and longitudinal components in the first place  one requires $A_i$ to be well enough behaved at spatial infinity so that the integral in (\ref{1.3y}) actually exists. In \cite{Amarasinghe2019,Phelps2019} we have carried out an analogous analysis for the full  scalar, vector, tensor expansion, and discuss it further in Sec. \ref{S7b}.

In the literature it is standard practice not to decouple the fluctuation equations at some higher-derivative level  but to treat the various components as evolving independently at the level of the fluctuation equations themselves, the so-called decomposition theorem. The basic idea behind the decomposition theorem is that in an equation such as 
\begin{eqnarray}
B_i+\partial_iB=C_i+\partial_iC,
\label{1.5y}
\end{eqnarray}
where $B$ and $C$ are three-scalars and $B_i$ and $C_i$ are transverse three-vectors that obey $\partial^iB_i=0$, $\partial^iC_i=0$ the solutions are taken to obey 
\begin{eqnarray}
B=C, \quad B_i=C_i,
\label{1.6y}
\end{eqnarray}
with the scalar and vector sectors of  (\ref{1.5y}) solving (\ref{1.5y}) separately. However, if we apply $\partial^i$  we obtain 
\begin{eqnarray}
\partial^i\partial_i(B-C)=0,
\label{1.7y}
\end{eqnarray}
an equation that  can admit of solutions other than $B=C$. In fact, in general we can obtain $B-C=c+c^ix_i$, where   $c$ and  $c^i$ are constants. To exclude such an outcome we impose boundary conditions that the solutions vanish at spatial infinity. This then sets $B-C=0$, and then from the initial $B_i+\partial_iB=C_i+\partial_iC$ we infer that $B_i-C_i=0$ too. With asymptotic  boundary conditions we thus establish the validity of the decomposition theorem [that the only allowed  solution to (\ref{1.5y}) is (\ref{1.6y})] in this particular case. In this simple example we see that the same asymptotic boundary condition needed to decompose a vector into its separate longitudinal and transverse components in the first place is the same as the one needed to establish the validity of the decomposition theorem.  It is thus only through the use of boundary conditions that we can obligate the separate scalar, vector and tensor sectors to propagate independently. For fluctuations in the standard cosmological theory it was shown  \cite{Phelps2019} that, as augmented by conditions at the origin of coordinates, this analysis generalizes to the full scalar, vector, tensor cosmological expansion, to thus establish the validity of the decomposition theorem in the standard case. In the present paper we show in Secs. \ref{S4} and \ref{S5} that the decomposition theorem also generalizes to the conformal case, again in any cosmological epoch. Armed with this  theorem we can then proceed in Secs. \ref{S6}, \ref{S7}, \ref{S8} and an Appendix to solve the conformal gravity fluctuation equations. And we find that in the conformal gravity case the fluctuation equations simplify so much in the recombination era that one is able to find exact analytic solutions.

As well as impose boundary conditions on the fluctuation modes, in Sec. \ref{S4d} we present a master equation for scalar, vector and tensor fluctuation modes and show that the radial equations for the three-dimensional vector and tensor modes are respectively the same as those of scalar modes in five and seven spatial dimensions. This enables us to construct normalization conditions for the three-dimensional vector and tensor modes (construction of normalization conditions for the three-dimensional scalar modes can be achieved with three-dimensional information alone). To ensure that these modes are normalizable we will again require spatially asymptotic boundary conditions, conditions that will turn out to be more stringent than just having the modes vanish at infinity as minimally as possible. Specifically, for the negative spatial curvature  Robertson-Walker cosmology that we study in detail in this paper, this being the one of relevance to conformal gravity, we find that in terms of the radial coordinate $r=\sinh \chi$ normalizability requires that the scalar, vector and tensor modes respectively behave as $e^{-\chi}$, $e^{-2\chi}$, $e^{-3\chi}$ as $\chi \rightarrow \infty$.

\subsection{The Background Conformal Gravity Cosmology -- Gravity Sector}
\label{S1bz}

Conformal gravity is a pure metric theory of gravity that possesses all of the general coordinate invariance and equivalence principle structure of standard gravity while augmenting it with an additional symmetry, local conformal invariance, in which  the action is left invariant under local conformal transformations on the metric of the form $g_{\mu\nu}(x)\rightarrow e^{2\alpha(x)}g_{\mu\nu}(x)$ with arbitrary local phase $\alpha(x)$. Under such a symmetry a gravitational action that is to be a polynomial function of the Riemann tensor is uniquely prescribed, and with use of the Gauss-Bonnet theorem is given by (see e.g. \cite{Mannheim2006}) 
\begin{eqnarray}
I_{\rm W}=-\alpha_g\int d^4x\, (-g)^{1/2}C_{\lambda\mu\nu\kappa}
C^{\lambda\mu\nu\kappa}
\equiv -2\alpha_g\int d^4x\, (-g)^{1/2}\left[R_{\mu\kappa}R^{\mu\kappa}-\frac{1}{3} (R^{\alpha}_{\phantom{\alpha}\alpha})^2\right].
\label{1.8y}
\end{eqnarray}
Here $\alpha_g$ is a dimensionless  gravitational coupling constant, and
\begin{eqnarray}
C_{\lambda\mu\nu\kappa}= R_{\lambda\mu\nu\kappa}
-\frac{1}{2}\left(g_{\lambda\nu}R_{\mu\kappa}-
g_{\lambda\kappa}R_{\mu\nu}-
g_{\mu\nu}R_{\lambda\kappa}+
g_{\mu\kappa}R_{\lambda\nu}\right)
+\frac{1}{6}R^{\alpha}_{\phantom{\alpha}\alpha}\left(
g_{\lambda\nu}g_{\mu\kappa}-
g_{\lambda\kappa}g_{\mu\nu}\right)
\label{1.9y}
\end{eqnarray}
is the conformal Weyl tensor.  

The conformal Weyl tensor has two features that are not possessed by the Einstein tensor $G_{\mu\nu}=R_{\mu\nu}-\tfrac{1}{2}g_{\mu\nu} R^{\alpha}_{\phantom{\alpha}\alpha}$, namely that $C^{\lambda}_{\phantom{\lambda}\mu\nu\kappa}$  vanishes in geometries that are conformal to flat (this precisely being the case for  the Robertson-Walker and de Sitter geometries that are of relevance to cosmology),  and that for any arbitrary metric  $C^{\lambda}_{\phantom{\lambda}\mu\nu\kappa}$ transforms as  $C^{\lambda}_{\phantom{\lambda}\mu\nu\kappa} \rightarrow  C^{\lambda}_{\phantom{\lambda}\mu\nu\kappa}$ under $g_{\mu\nu}(x)\rightarrow e^{2\alpha(x)}g_{\mu\nu}(x)$, with all derivatives of $\alpha(x)$ dropping out. With all of these derivatives dropping out $I_{\rm W}$ is locally conformal invariant \cite{footnoteA}.

With the Weyl action $I_{\rm W}$ given in  (\ref{1.8y}) being a fourth-order derivative function of the metric, functional variation with respect to the metric $g_{\mu\nu}(x)$ generates fourth-order derivative gravitational equations of motion of the form \cite{Mannheim2006} 
\begin{eqnarray}
-\frac{2}{(-g)^{1/2}}\frac{\delta I_{\rm W}}{\delta g_{\mu\nu}}=4\alpha_g W^{\mu\nu}=4\alpha_g\left[2\nabla_{\kappa}\nabla_{\lambda}C^{\mu\lambda\nu\kappa}-
R_{\kappa\lambda}C^{\mu\lambda\nu\kappa}\right]=4\alpha_g\left[W^{\mu
\nu}_{(2)}-\frac{1}{3}W^{\mu\nu}_{(1)}\right]=T^{\mu\nu},
\label{1.10y}
\end{eqnarray}
where the functions $W^{\mu \nu}_{(1)}$ and $W^{\mu \nu}_{(2)}$ (respectively associated with the $(R^{\alpha}_{\phantom{\alpha}\alpha})^2$ and $R_{\mu\kappa}R^{\mu\kappa}$ terms in (\ref{1.8y})) are given by
\begin{eqnarray}
W^{\mu \nu}_{(1)}&=&
2g^{\mu\nu}\nabla_{\beta}\nabla^{\beta}R^{\alpha}_{\phantom{\alpha}\alpha}                                             
-2\nabla^{\nu}\nabla^{\mu}R^{\alpha}_{\phantom{\alpha}\alpha}                          
-2 R^{\alpha}_{\phantom{\alpha}\alpha}R^{\mu\nu}                              
+\frac{1}{2}g^{\mu\nu}(R^{\alpha}_{\phantom{\alpha}\alpha})^2,
\nonumber\\
W^{\mu \nu}_{(2)}&=&
\frac{1}{2}g^{\mu\nu}\nabla_{\beta}\nabla^{\beta}R^{\alpha}_{\phantom{\alpha}\alpha}
+\nabla_{\beta}\nabla^{\beta}R^{\mu\nu}                    
 -\nabla_{\beta}\nabla^{\nu}R^{\mu\beta}                       
-\nabla_{\beta}\nabla^{\mu}R^{\nu \beta}                          
 - 2R^{\mu\beta}R^{\nu}_{\phantom{\nu}\beta}                                    
+\frac{1}{2}g^{\mu\nu}R_{\alpha\beta}R^{\alpha\beta},
\label{1.11y}
\end{eqnarray}                                 
and where $T^{\mu\nu}$ is the conformal invariant energy-momentum tensor associated with a conformal matter source.  Since $W^{\mu\nu}=W^{\mu\nu}_{(2)}-(1/3)W^{\mu\nu}_{(1)}$, known as the Bach tensor \cite{Bach1921},  is obtained from an action that is both general coordinate invariant and conformal invariant, in consequence, and without needing to impose any equation of motion or stationarity condition, $W^{\mu\nu}$ is automatically covariantly conserved and covariantly traceless and obeys $\nabla_{\nu}W^{\mu\nu}=0$, $g_{\mu\nu}W^{\mu\nu}=0$ on every variational path used for the functional variation of $I_{\rm W}$. While this is not necessarily the case for the arbitrary matter field, it is the case for massless gauge fields since, as noted in \cite{footnoteA}, they have a conformal structure analogous to that of  the gravitational field of a conformal theory. However, for non-gauge fields we note that  in a conformal invariant theory the relevant $T_{\mu\nu}$ is still conformal invariant. Then, with $W_{\mu\nu}$ being both conserved and traceless, in solutions to the gravitational equations of motion the conformal invariant  $T^{\mu\nu}$ is covariantly conserved and covariantly traceless too. Without the imposition of the gravitational equations of motion a non-gauge matter field $T_{\mu\nu}$ would still be covariantly conserved and covariantly traceless in solutions to matter field equations of motion as long as they are conformal invariant.

\subsection{The Background Conformal Gravity Cosmology -- Matter Sector}
\label{S1cz}

As well as being covariantly conserved and covariantly traceless in any geometry, because a general Robertson-Walker geometry is conformal to flat, in such a geometry the Weyl tensor vanishes identically, and thus from (\ref{1.10y}) it follows that $W_{\mu\nu}$ vanishes identically too.  Thus given the conformal gravity field equation $4\alpha_gW_{\mu\nu}=T_{\mu\nu}$, it follows that in a background Robertson-Walker conformal cosmology the matter sector $T_{\mu\nu}$ also vanishes. While this would seem to imply that the matter source is trivial,  this is not in fact necessarily the case. Specifically, in the literature two ways in which a background $T_{\mu\nu}$  could vanish non-trivially have been identified, one involving a conformally coupled elementary scalar field \cite{Mannheim1990}, and the other involving a conformal perfect fluid \cite{Mannheim2000}. We describe both of the cases now since even though they both were developed for the background,  we shall have occasion to discuss aspects of both of them when we study fluctuations below.

For a conformally coupled scalar field $S(x)$ the matter action is
\begin{eqnarray}
I_S&=&-\int d^4x(-g)^{1/2}\left[\frac{1}{2}\nabla_{\mu}S
\nabla^{\mu}S-\frac{1}{12}S^2R^\mu_{\phantom         
{\mu}\mu}+\lambda S^4\right]
\nonumber\\
&=&\int d^4x(-g)^{1/2}\left[\frac{1}{2c^2}\dot{S}^2-\frac{1}{2}(\vec{\nabla} S)^2
+\frac{1}{12}S^2R^\mu_{\phantom         
{\mu}\mu}-\lambda S^4\right],
\label{1.12y}
\end{eqnarray}                                 
where  $\lambda$ is a dimensionless coupling constant. (Since we use the convention given in  \cite{Weinberg1972} where $g_{00}$ is taken to have negative signature, and where the proper time is written as $ds^2=-g_{\mu\nu}dx^{\mu}dx^{\nu}$, (\ref{1.12y}) thus corresponds to a scalar field with a normal positive signatured kinetic energy.) As such, the $I_{\rm S}$ action is the most general curved space polynomial matter action for the $S(x)$ field that is invariant under both general coordinate transformations and local conformal transformations of the form
$S(x)\rightarrow e^{-\alpha(x)}S(x)$,  $g_{\mu\nu}(x)\rightarrow e^{2\alpha(x)}g_{\mu\nu}(x)$. Variation of the $I_S$ action with respect to  $S(x)$ yields the scalar field equation of motion
\begin{eqnarray}
\nabla_{\mu}\nabla^{\mu}S+\frac{1}{6}SR^\mu_{\phantom{\mu}\mu}
-4\lambda S^3=0,
\label{1.13y}
\end{eqnarray}                                 
while variation with respect to the metric yields a matter field  energy-momentum tensor 
\begin{eqnarray}
T_{\rm S}^{\mu \nu}&=&\frac{2}{3}\nabla^{\mu}S \nabla^{\nu} S
-\frac{1}{6}g^{\mu\nu}\nabla_{\alpha}S\nabla^{\alpha}S
-\frac{1}{3}S\nabla^{\mu}\nabla^{\nu}S
\nonumber \\             
&+&\frac{1}{3}g^{\mu\nu}S\nabla_{\alpha}\nabla^{\alpha}S                           
-\frac{1}{6}S^2\left(R^{\mu\nu}
-\frac{1}{2}g^{\mu\nu}R^\alpha_{\phantom{\alpha}\alpha}\right)-g^{\mu\nu}\lambda S^4. 
\label{1.14y}
\end{eqnarray}                                 
Use of the matter field equation of motion then confirms that this energy-momentum tensor obeys the tracelessness condition $g_{\mu\nu}T_{\rm S}^{\mu\nu}=0$, just as it should do in  a conformal invariant theory.

In the presence of a spontaneously broken, scale-setting,  non-zero constant vacuum expectation
value $S_0$ for the scalar field, the scalar field wave equation and the energy-momentum tensor are then found to simplify to  
\begin{eqnarray}
R^\alpha_{\phantom{\alpha}\alpha}&=&24\lambda S_0^2,
\nonumber\\
T_{\rm S}^{\mu \nu}&=& 
-\frac{1}{6} S_0^2\left(R^{\mu\nu}-\frac{1}{2}g^{\mu\nu}
R^\alpha_{\phantom{\alpha}\alpha}\right)-g^{\mu\nu}\lambda S_0^4.
\label{1.15y}
\end{eqnarray}                                 
With the Ricci scalar being non-zero in this solution, we see immediately that once $S_0$ is non-zero  the geometry is necessarily non-trivial \cite{footnoteA0}. Moreover, if we take the geometry to be a de Sitter geometry in which $R^{\lambda\mu\sigma\nu}=K[g^{\mu \sigma}g^{\lambda \nu}-g^{\mu \nu}g^{\lambda \sigma}]$, $R^{\mu\nu}=-3Kg^{\mu\nu}$, $R^\alpha_{\phantom{\alpha}\alpha}=-12K$, $G_{\mu\nu}=R_{\mu\nu}-\tfrac{1}{2}g_{\mu\nu} R^{\alpha}_{\phantom{\alpha}\alpha}=3Kg_{\mu\nu}$,
then since $W^{\mu \nu}$ will vanish identically in such a de Sitter geometry, it follows that 
 $T_{\rm S}^{\mu \nu}$ must vanish identically too. And with $K=-2\lambda S_0^2$ it can readily be checked that it in fact does. Thus even though $W^{\mu\nu}$ and $T^{\mu\nu}$ both vanish identically, as noted in \cite{Mannheim1990} the conformal cosmology governed by $4\alpha_gW^{\mu\nu}=T^{\mu\nu}$ admits of a non-trivial de Sitter geometry solution, with a non-vanishing four-curvature $K=-2\lambda S_0^2$.

A second way in which $T^{\mu\nu}$ can vanish non-trivially was given in \cite{Mannheim2000}. If we drop the $\lambda$-dependent term in $I_S$, then in a generic Robertson-Walker geometry with metric
\begin{eqnarray}
ds^2=c^2dt^2-a^2(t)\left[\frac{dr^2}{1-kr^2}+r^2d\theta^2+r^2\sin^2\theta d\phi^2\right]
=c^2dt^2-a^2(t)\tilde{\gamma}_{ij}dx^idx^j,
\label{1.16y}
\end{eqnarray}
solutions to the scalar field wave equation (\ref{1.13y})  obey \cite{Mannheim1988}
\begin{eqnarray}
\frac{1}{f(\tau)}\left[\frac{1}{c^2}\frac{d^2f}{d\tau^2}+kf(\tau)\right]=\frac{1}{g(r,\theta,\phi)}\tilde{\gamma}^{-1/2}\partial_i[\tilde{\gamma}^{1/2}\tilde{\gamma}^{ij}\partial_jg(r,\theta,\phi)]=-\zeta^2,
\label{1.17y}
\end{eqnarray}
where $\tau=\int dt/a(t)$, $S=f(\tau)g(r,\theta,\phi)/a(\tau)$, $\tilde{\gamma}^{ij}$ is the metric of the spatial part of the Robertson-Walker metric, and $\zeta^2$ is a separation constant. From (\ref{1.17y}) we see that $f(\tau)$ is harmonic with frequencies that obey $\omega^2/c^2=\zeta^2+k$, while we can set $g(r,\theta,\phi)=g^{\ell}_{\zeta}(r)Y^m_{\ell}(\theta,\phi)$, where $g^{\ell}_{\zeta}(r)$ obeys
\begin{eqnarray}
\left[(1-kr^2)\frac{\partial^2}{\partial r^2}+\frac{(2-3kr^2)}{r}\frac{\partial}{\partial r}-\frac{\ell(\ell+1)}{r^2}+\zeta^2\right]g^{\ell}_{\zeta}(r)=0.
\label{1.18y}
\end{eqnarray}

To form a perfect fluid energy-momentum tensor, in $T^{\mu\nu}_S$ we make an incoherent averaging over all allowed spatial modes associated with a given $\omega$  (this is equivalent to calculating statistical averages using a density matrix that is proportional to the unit matrix and normalized to one). And on doing the sum over all modes,  for each $\omega$ we obtain  \cite{Mannheim1988} the automatically traceless
\begin{eqnarray}
T_S^{\mu\nu}=\frac{\omega^4(g^{\mu\nu}+4U^{\mu}U^{\nu})}{6c^4\pi^2a^4(t)}=
\frac{(\zeta^2+k)^2(g^{\mu\nu}+4U^{\mu}U^{\nu})}{6\pi^2a^4(t)},
\label{1.19y}
\end{eqnarray}
where $U^{\mu}$ is a unit timelike vector. This $T^{\mu\nu}_S$ vanishes if $\omega^2=0$, and with $\omega^2/c^2=\zeta^2+k$, we can thus satisfy $T^{\mu\nu}_S=0$ non-trivially  if and only if $k$ is negative. In doing the incoherent averaging when $\omega=0$, for $T^{00}_S$ for instance we obtain
\begin{eqnarray}
T_S^{00}=\frac{1}{6}\sum_{\ell,m}\left[\sum _{i=1}^3\tilde{\gamma}^{ii}|\partial_i(g^{\ell}_{(-k)^{1/2}}Y^{m}_{\ell}(\theta,\phi))|^2+k|g^{\ell}_{(-k)^{1/2}}Y^{m}_{\ell}(\theta,\phi)|^2\right]
\label{1.20y}
\end{eqnarray}
when $k$ is negative, with it being shown in \cite{Mannheim2000} that the sum in (\ref{1.20y}) vanishes identically. Essentially what happens is that a positive contribution to $T^{\mu\nu}_S$ by the scalar field modes is cancelled by a negative contribution from the gravitational field due to its negative spatial three-curvature. With negative $k$, solutions to (\ref{1.18y}) are associated Legendre functions, and even though we have now fixed $\zeta^2$ to $-k$, (\ref{1.18y}) still possesses an infinite number of solutions labelled by $\ell$ and $m$. An incoherent averaging over all of these solutions then causes $T^{\mu\nu}_S$ to vanish non-trivially. Thus as we see, it is negative $k$ that is selected. 

\subsection{Phenomenological Justification for Negative Spatial Three-Curvature}
\label{S1dz}

In applications of conformal gravity to astrophysical and cosmological data it has been found that phenomenologically $k$ actually should be negative. In conformal cosmology very good non-fine-tuned, negative $k$  fits to the accelerating universe Hubble plot data have been presented in \cite{Mannheim2006,Mannheim2012b,Mannheim2017} and will be described below. Similarly, very good negative $k$ conformal gravity fits to the rotation curves of 138 galaxies have been presented in \cite{Mannheim2011b,Mannheim2012c,O'Brien2012}.  That galactic rotation curves would even be sensitive to cosmology is initially somewhat puzzling since this is not the case in standard  Newton-Einstein gravity, and so we clarify the point. Without needing to impose any gravitational equation of motion, in conformal gravity the metric associated with a static, spherically symmetric system can be brought to the form $ds^2=B(r)c^2dt^2-dr^2/B(r)-r^2d\theta^2-r^2\sin^2\theta d\phi^2$ via a sequence of general coordinate and conformal transformations \cite{Mannheim1989}, with the relation $4\alpha_gW_{\mu\nu}=T_{\mu\nu}$ taking the exact form \cite{Mannheim1994}
\begin{equation}                                                                               
\nabla^4 B(r) = \frac{3}{4\alpha_g B(r)}\left(T^0_{{\phantom 0} 0} -
T^r_{{\phantom r} r}\right) =f(r)
\label{1.21y}
\end{equation}      
in such a geometry without approximation, with (\ref{1.21y}) serving to define $f(r)$. The solution to (\ref{1.21y}) can be written as \cite{Mannheim1994} 
\begin{eqnarray}
B(r)&=&-\frac{r}{2}\int_0^r
dr^{\prime}r^{\prime 2}f(r^{\prime})
-\frac{1}{6r}\int_0^r
dr^{\prime}r^{\prime 4}f(r^{\prime})
-\frac{1}{2}\int_r^{\infty}
dr^{\prime}r^{\prime 3}f(r^{\prime})
-\frac{r^2}{6}\int_r^{\infty}
dr^{\prime}r^{\prime }f(r^{\prime}),
\nonumber\\
B^{\prime}(r)&=&-\frac{1}{2}\int_0^r
dr^{\prime}r^{\prime 2}f(r^{\prime})
+\frac{1}{12r^2}\int_0^r
-\frac{r}{3}\int_r^{\infty}
dr^{\prime}r^{\prime }f(r^{\prime}).
\label{1.22y}
\end{eqnarray}                                 
In (\ref{1.22y}) we recognize two potential terms coming from matter localized to a finite region and one potential term coming from global matter that is distributed all the way to $r=\infty$. For localized matter the  potential of a star of radius $R^*$ is given by  $V^*(r>R^*)=-\beta^* c^2/r+\gamma^* c^2 r/2$ \cite{Mannheim1989}, where \cite{Mannheim1994}
\begin{equation}
\gamma^*= -\frac{1}{2}\int_0^{R^*}
dr^{\prime}r^{\prime 2}f(r^{\prime}),\quad 2\beta^*=\frac{1}{6}\int_0^{R^*}
dr^{\prime}r^{\prime 4}f(r^{\prime}).
\label{1.23y}
\end{equation}                                 

While the global curvature of the universe plays no role in dark matter fits to galactic rotation curves (for a $1/r$ potential one only needs to consider sources within individual galaxies), in the conformal gravity (\ref{1.22y}) there are contributions coming from material not just outside of a given star of interest but from the global sources in the entire rest of the universe. (If the potential of a given source is growing with distance then the potentials of sources very distant from the given source are also growing with distance, to thus impact the given source.) These global sources provide two forms of contributions that are associated with conformal cosmology, namely the contribution of the Hubble flow and the contribution of inhomogeneities in it. Since galactic motions are determined in the rest frames of galaxies one has to transform the comoving Hubble flow to each local galactic rest frame. On doing this one finds \cite{Mannheim1989} that a negative three-curvature (and only a negative three-curvature) background Robertson-Walker cosmology generates a universal linear potential $\gamma_0c^2/r$ where $\gamma_0=(-4k)^{1/2}$. Similarly, inhomogeneities in the Hubble flow are found \cite{Mannheim2011b} to generate a universal quadratic potential $-\kappa c^2 r^2$ (the last term in $B(r)$ in (\ref{1.22y})). 

For a spiral disk galaxy with surface brightness $\Sigma (R)=\Sigma_0e^{-R/R_0}$, where $R$ is the radial distance in the plane of the disk and $R_0$ is the disk scale length, the rotation velocity is given by \cite{Mannheim2006} 
\begin{eqnarray}
v^2(R)&=&
\frac{N^*\beta^*c^2 R^2}{2R_0^3}\bigg{[}I_0\left(\frac{R}{2R_0}
\right)K_0\left(\frac{R}{2R_0}\right)
-I_1\left(\frac{R}{2R_0}\right)
K_1\left(\frac{R}{2R_0}\right)\bigg{]}
\nonumber\\
&+&\frac{N^*\gamma^* c^2R^2}{2R_0}I_1\left(\frac{R}{2R_0}\right)
K_1\left(\frac{R}{2R_0}\right)+\frac{\gamma_0c^2 R}{2}-\kappa c^2 R^2,
\label{1.24y}
\end{eqnarray} 
where $N^*$ is the number of stars in the galaxy in solar mass units, $\beta^*$ is the Schwarzschild radius of the sun, and $I_0$, $I_1$, $K_0$ and $K_1$ are modified Bessel functions.   Very good fitting to the rotation curves of the 138 galaxies is obtained in \cite{Mannheim2011b,Mannheim2012c,O'Brien2012} with fixed, universal (i.e., galaxy-independent) parameters
\begin{eqnarray}
\beta^*&=&1.48\times 10^5 {\rm cm},\quad \gamma^*=5.42\times 10^{-41} {\rm cm}^{-1},
\nonumber\\
\gamma_0&=&3.06\times
10^{-30} {\rm cm}^{-1},\quad \kappa = 9.54\times 10^{-54} {\rm cm}^{-2},
\label{1.25y}
\end{eqnarray} 
and with there being no need to introduce any dark matter. Since current dark matter fits require two free parameters per galactic halo, the galaxy-dependent 276 free dark matter halo parameters that are needed for the 138 galaxy sample are replaced by just the three universal parameters: $\gamma^*$, $\gamma_0$ and $\kappa$. (The luminous Newtonian $N^*\beta^*$-dependent  contribution in (\ref{1.24y}) is common to both dark matter and conformal gravity fits and is included in both cases.) With $\gamma_0$ being of order the inverse of the Hubble radius and with $\kappa$ being of order a typical cluster of galaxies scale, the values for $\gamma^0$ and $\kappa$ that are obtained show that they are indeed of the cosmological scales associated with the homogeneous Hubble flow and the inhomogeneities in it. We can thus use stars in galaxies to serve as test particles that measure the global geometry of the universe. From the perspective of a local $1/r$ Newtonian potential the fact that the measured velocities exceed the luminous Newtonian expectation is described as the missing mass problem, with undetected or dark matter within the galaxies  themselves being needed in order to be able to account for the shortfall \cite{footnoteA1}. From the perspective of conformal gravity the shortfall is explained by the rest of the visible mass in the universe. The missing mass is thus not missing at all, it is the rest of the visible universe and it has been hiding in plain sight all along.

Now in the standard gravity inflationary universe model \cite{Guth1981} fits to accelerating universe data, to properties of clusters of galaxies and to the anisotropy of the CMB lead \cite{Bahcall2000,deBernardis2000,Tegmark2004}
to a spatially flat three-geometry. It is thus paramount to determine the conformal gravity expectations for the anisotropy to see if the data could support a $k<0$ universe, and the objective of this paper is to prepare some of the needed groundwork by studying an exact solution to fluctuations around a $k<0$ conformal cosmology. While beyond the scope of the present paper this groundwork will also enable us to analyze baryon acoustic oscillations in the CMB and analyze the galaxy correlation function, and for the moment we note only that both are associated with a 150 Mpc scale, viz.  an inhomogeneity  scale that is of the same order as the scale associated with the  inhomogeneity-generated $\kappa$ ($\kappa^{-1/2} \sim 100$ Mpc) that is measured in conformal gravity fits to galactic rotation curves. 

\subsection{The Background Cosmological Model}
\label{S1ez}

To construct a background cosmological model we combine the scalar field and perfect fluid models described above, but now look for a vanishing of the total $T_{\mu\nu}$ by an interplay between them. We thus take the total background matter sector energy-momentum tensor to be of the form
\begin{equation}
T^{\mu \nu} = \frac{1}{c}\left[(\rho_m+p_m)U^{\mu}U^{\nu}+p_mg^{\mu\nu}\right] 
-\frac{1}{6}S_0^2\left(R^{\mu\nu}
-\frac{1}{2}g^{\mu\nu}R^\alpha_{\phantom{\alpha}\alpha}\right)         
-g^{\mu\nu}\lambda S_0^4,
\label{1.26y}
\end{equation}                                 
where the suffix $m$ denotes matter. 
On taking the background geometry to be the comoving Robertson-Walker metric given in (\ref{1.16y}), the background 
$W_{\mu\nu}$ thus vanishes, so that the background $T_{\mu\nu}=W_{\mu\nu}/4\alpha_g$ then vanishes too. We can  rewrite the equation $T_{\mu\nu}=0$ in the instructive form 
\begin{equation}
\frac{1}{6}S_0^2\left(R^{\mu\nu}
-\frac{1}{2}g^{\mu\nu}R^\alpha_{\phantom{\alpha}\alpha}\right) = 
\frac{1}{c}\left[(\rho_m+p_m)U^{\mu}U^{\nu}+p_mg^{\mu\nu}\right]  -g^{\mu\nu}\lambda S_0^4.
\label{1.27y}
\end{equation}                                 
We thus recognize the conformal cosmological evolution equation given in  (\ref{1.27y}) as being of the form
of none other than the cosmological evolution equation of the standard theory, viz. (on setting $\Lambda =\lambda S_0^4$) 
\begin{eqnarray}
-\frac{c^3}{8\pi G}\left(R^{\mu\nu}
-\frac{1}{2}g^{\mu\nu}R^\alpha_{\phantom{\alpha}\alpha}\right)=\frac{1}{c}\left[(\rho_m+p_m)U^{\mu}U^{\nu}+p_mg^{\mu\nu}\right]  -g^{\mu\nu}\Lambda,
\label{1.28y}
\end{eqnarray}
save only for the fact that the standard $G$ has been replaced by
an effective, dynamically induced one given by
\begin{equation}
G_{{\rm eff}}=-\frac{3c^3}{4\pi S_0^2},
\label{1.29y}
\end{equation}                                 
viz.  by an effective gravitational constant that,  as noted in \cite{Mannheim1992}, is expressly negative. 
Conformal cosmology is thus 
controlled by an effective gravitational coupling constant that is
repulsive rather than attractive, and which becomes smaller the larger
$S_0$ might be. 

In the conformal theory local non-relativistic solar system gravity is controlled by the parameter $\beta^*$ that appears in (\ref{1.23y}). With the conformal coupling constant $\alpha_g$ not participating in homogeneous
geometries such as the cosmological one in which the Weyl tensor is zero, while participating in the inhomogeneous  (\ref{1.21y}) where the Weyl tensor  is non-zero (static, spherically symmetric geometries not being conformal to flat), $G_{{\rm eff}}$ is completely decoupled from $\alpha_g$, and thus completely decoupled from the local Newtonian $G=\beta^* c^2/M^*$ associated with a source of mass $M^*$. Thus in the conformal gravity theory the sign of the local $\beta^*$ is  related to the sign of $\alpha_g$ while the sign of $G_{{\rm eff}}$ is not. Consequently, a negative effective global cosmological $G_{{\rm eff}}$ is not in conflict with the existence of a positive local $G$. The fact
that the dynamically induced $G_{{\rm eff}}$ is negative in the
conformal theory had been thought of as being a disadvantage since it
seemed to imply that the local $G$ would be given by the same negative
$G_{{\rm eff}}$, to then be repulsive too. (This even prompted many authors to flip the overall sign of $I_S$ even though that would then make the kinetic energy ghostlike.) However, as we see, a
repulsive global cosmological $G_{{\rm eff}}$ and an attractive local
$G$ can coexist in one and the same theory, an aspect of the theory
which can now actually be regarded as a plus 
since  a repulsive component to gravity causes cosmic acceleration rather than deceleration. 

To see how central the negative sign of $G_{{\rm eff}}$ is to cosmic acceleration 
we define 
\begin{equation}
\bar{\Omega}_{M}(t)=\frac{8\pi G_{{\rm eff}}\rho_{m}(t)}{3c^2H^2(t)}, \quad
\bar{\Omega}_{\Lambda}(t)=\frac{8\pi G_{{\rm
eff}}\Lambda}{3cH^2(t)},\quad \bar{\Omega}_k(t)=-\frac{kc^2}{\dot{a}^2(t)},
\label{1.30y}
\end{equation}                                 
where $H=\dot{a}/a$. And on introducing the deceleration parameter $q=-a\ddot{a}/\dot{a}^2$, from (\ref{1.27y}) we obtain
\begin{eqnarray}
\dot{a}^2(t) +kc^2
&=&\dot{a}^2(t)\left(\bar{\Omega}_{M}(t)+
\bar{\Omega}_{\Lambda}(t)\right),\quad \bar{\Omega}_M(t)+
\bar{\Omega}_{\Lambda}(t)+\bar{\Omega}_k(t)=1,
\nonumber \\
q(t)&=&\frac{1}{2}\left(1+\frac{3p_m}{\rho_m}\right)\bar{\Omega}_M(t)
-\bar{\Omega}_{\Lambda}(t)
\label{1.31y}
\end{eqnarray}
as the evolution equations of conformal cosmology. To solve (\ref{1.31y}) we need to specify an equation of state for the matter field, and since we will momentarily find that it will not matter whether we use a massless or a massive field equation of state, we set $\rho_m=3p_m=A/a^4(t)=\sigma T^4$, and with $k<0$ obtain \cite{Mannheim2006}
\begin{equation}
a^2(t)= -\frac{k(\beta-1)}{2\alpha}
-\frac{k\beta{\rm sinh}^2 (\alpha^{1/2} ct)}{\alpha},
\label{1.32y}
\end{equation}
where 
\begin{equation}
\alpha =-2\lambda S_0^2=\frac{8\pi G_{\rm eff}\Lambda}{3c},\quad
\beta=\left(1- \frac{16A\lambda}{k^2
c}\right)^{1/2}.
\label{1.33y}
\end{equation}

Since $\Lambda$ represents the free energy that is released in the phase transition that generated $S_0$ in the first place, $\Lambda $ is necessarily negative. Then with $G_{{\rm eff}}$ also being negative the quantity $\bar{\Omega}_{\Lambda}(t)$ is positive, i.e., the conformal theory needs a negative $G_{{\rm eff}}$ in order to obtain a positive $\bar{\Omega}_{\Lambda}(t)$. (In contrast the standard model rationale for positive $\Omega_{\Lambda}=8\pi G\Lambda/3c^2H^2$ is that since the Newtonian $G$ is positive $\Lambda$ has to be taken to be  positive too.) While the standard model cannot accommodate a large $\Lambda$ the conformal theory can since $G_{{\rm eff}}$ can be much smaller than $G$. In fact as $S_0$ gets bigger $\Lambda$ gets bigger too but $G_{{\rm eff}}$ gets smaller, with $\bar{\Omega}_{\Lambda}(t)$ self quenching. To see by how much we note that if we set $\Lambda=-\sigma T_V^4/c$ ($V$ denotes vacuum) where $T_V$ is the large temperature at which the $S_0$ generating phase transition occurs,  then with $\bar{\Omega}_M(t)$ being of order $\sigma T^4/c$, in the current era the ratio $\bar{\Omega}_M(t)/\bar{\Omega}_{\Lambda}(t)=T^4/T_V^4$ is completely negligible. Moreover, since the temperature at recombination is only of order 1 $eV$, at recombination $\bar{\Omega}_M(t)/\bar{\Omega}_{\Lambda}(t)$ is negligible too. Thus we have to go into the very early universe to obtain a temperature at which $\bar{\Omega}_M(t)/\bar{\Omega}_{\Lambda}(t)=T^4/T_V^4$ would not be negligible. In the very early universe we can use $\rho_m=3p_m$ as the equation of state, and while massive matter would be non-relativistic at recombination, it would be irrelevant as to what equation of state we were to use for it since $\bar{\Omega}_M(t)/\bar{\Omega}_{\Lambda}(t)$  is negligible at recombination. With the matter contribution being negligible at recombination, for all temperatures from recombination until the current era (\ref{1.31y}) reduces to
\begin{eqnarray}
\bar{\Omega}_{\Lambda}(t)+\bar{\Omega}_k(t)=1,\quad q(t)=-\bar{\Omega}_{\Lambda}(t).
\label{1.34y}
\end{eqnarray}
With $k$ being negative the quantity $\bar{\Omega}_k(t)$ must be positive. Thus with $\bar{\Omega}_{\Lambda}(t)$ also being positive, other than in the early universe it must lie in the interval $0\leq \bar{\Omega}_{\Lambda}(t) \leq 1$, and thus it is indeed self-quenched sufficiently. Similarly, $\bar{\Omega}_k(t)$ must lie in the range $0\leq \bar{\Omega}_k(t)\leq 1$. Moreover, from recombination onward the deceleration parameter must lie in the interval $-1\leq q(t)\leq 0$, to not only be accelerating but to be so without any need for fine tuning.

For evolution in the region from recombination until the current time $t_0$ the matter density plays no role and so we we can approximate $\beta=1$ in this region, with $a(t)$ then being given by 
\begin{equation}
a(t)= \frac{(-k)^{1/2}\sinh (\alpha^{1/2} ct)}{\alpha^{1/2}}.
\label{1.35y}
\end{equation}
With such an $a(t)$ we obtain 
\begin{equation}
\bar{\Omega}_{\Lambda}(t ) = {\rm tanh}^2 (\alpha^{1/2}ct),\quad  \bar{\Omega}_k(t)= {\rm
sech}^2 (\alpha^{1/2}ct), \quad q(t) =-\tanh^2 (\alpha^{1/2}ct),
\label{1.36y}
\end{equation}
and a luminosity distance redshift relation of the form  \cite{Mannheim2006}

\begin{equation} 
d_L=-\frac{c}{H(t_0)}\frac{(1+z)^2}{q_0}\left[1-\left(1+q_0-
\frac{q_0}{(1+z)^2}\right)^{1/2}\right],
\label{1.37y}
\end{equation}
where $q_0=q(t_0)$ is the current era value of the deceleration parameter and $H(t_0)$ is the current era value of the Hubble parameter.

Fitting the  type 1A supernovae accelerating universe data with (\ref{1.37y}) gives a fit  \cite{Mannheim2006,Mannheim2012b,Mannheim2017} that is comparable in quality with that of the standard model $\Omega_M=0.3$, $\Omega_{\Lambda}=0.7$ dark matter dark energy paradigm. In the conformal gravity fit $q_0$ is fitted to the value $-0.37$, i.e., quite non-trivially found to be right in the allowed $-1\leq q_0\leq 0$ range. Since $\bar{\Omega}_M$ is negligible no dark matter is needed, and since $q_0$ and $\bar{\Omega}_{\Lambda}=-q_0$ fall right in the allowed region, no fine tuning is needed either.

With ${\rm tanh}^2(\alpha^{1/2}ct_0)=0.37$ we determine ${\rm tanh}(\alpha^{1/2}ct_0)=0.61$, $\sinh(\alpha^{1/2}ct_0)=0.77$, $\alpha^{1/2}ct_0=0.71$, $H(t_0)=\alpha^{1/2}c/{\rm tanh} (a^{1/2}ct_0)=1.16/t_0$. With $H_0=72$ km/sec/Mpc we obtain $t_0=4.83\times 10^{17}$ sec, a perfectly acceptable value for the age of the universe. Similarly, we obtain $\alpha^{1/2}c=0.15\times 10^{-17}$ ${\rm sec}^{-1}$, $\alpha^{1/2}=0.50\times 10^{-28}$ ${\rm cm}^{-1}$. Recalling that $(-k)^{1/2}=\gamma_0/2=1.53\times 10^{-30}  {\rm cm}^{-1}$, we obtain $a(t_0)=2.36\times10^{-2}$, so the current era expansion radius itself is also small.

If we extrapolate back to the recombination time $t_R$ we obtain $a(t_R)/a(t_0)=T_0/T_R=O(10^{-4})$. Consequently,  with $\sinh(\alpha^{1/2}ct_0)=0.77$ we obtain  $\sinh(\alpha^{1/2}ct_R)=0.77\times 10^{-4}$. Thus we can approximate $\sinh(\alpha^{1/2}ct_R)$ by $\alpha^{1/2}ct_R$ itself at recombination. Finally then,  to one part in $10^4$ for both $\bar{\Omega}_{\Lambda}(t_R )$ and $ \bar{\Omega}_{k}(t_R )$ we have
\begin{equation} 
a(t_R)=(-k)^{1/2}ct_R,\quad \bar{\Omega}_{\Lambda}(t_R )\approx 0,\quad \bar{\Omega}_{k}(t_R )\approx 1
\label{1.38y}
\end{equation}
at recombination, with $\alpha$ dropping out of $a(t_R)$, and with the numerical value of $a(t_R)$ being $2.36\times 10^{-6}$. As we see, at recombination the conformal universe is curvature dominated. We thus recognize three epochs for conformal cosmology:  radiation dominated early universe, curvature dominated recombination universe, cosmological constant dominated late universe. While there will always be a trace of $\bar{\Omega}_{M}(t)$  in any non-early universe epoch, and while non-early universe propagating matter fields will respond to a geometry that they are not affecting in any substantial way, at recombination we see that $a(t_R)$ as given in (\ref{1.38y}) is  independent not just of $\bar{\Omega}_{M}(t_R)$ but even of $\bar{\Omega}_{\Lambda}(t_R)$ as well. 

Now a geometry in which $k$ is negative and  $a(t)$ is linear in $t$  can formally actually be brought to a locally four-flat form (though not globally four-flat since the  negative three-curvature global topology does not change under a coordinate transformation). Specifically,  under $t^{\prime}=(1+r^2)^{1/2}t$, $r^{\prime}=rt$ we obtain
\begin{eqnarray}
dt^2-t^2\left[\frac{dr^2}{1+r^2}+r^2d\theta^2+r^2\sin^2\theta d\phi^2\right]\rightarrow dt^{\prime 2}-dr^{\prime 2}-r^{\prime 2}d\theta^2-r^{\prime 2}\sin^2\theta d\phi^2.
\label{1.39y}
\end{eqnarray}
However, since $t^{\prime 2}-r^{\prime 2}=t^2$, only the region with $t^{\prime 2}-r^{\prime 2}\geq 0$ is mapped this way, with the region that would map into the unobservable spacelike $t^{\prime 2}-r^{\prime 2}<0$ region being associated with a Euclidean $t$ region \cite{footnoteA2}. We described this transformation as being formal since the $t\rightarrow 0$ limit of (\ref{1.35y}) is quite delicate. As noted in \cite{footnoteA0}, the Ricci scalar is given by $R^{\alpha}_{\phantom{\alpha}\alpha}=-6(a\ddot{a}+\dot{a}^2+k)/a^2$, and evaluating it for $a=\sinh t$, $k=-1$ yields $R^{\alpha}_{\phantom{\alpha}\alpha}=-12$. Now under a coordinate transformation $R^{\alpha}_{\phantom{\alpha}\alpha}$ cannot change, and thus it cannot be brought to a flat form in which $R^{\alpha}_{\phantom{\alpha}\alpha}=0$. To see what is happening we work to order $t^2$ and set $a\dot{a}=\sinh^2t\rightarrow t^2$, $\dot{a}^2=\cosh^2t\rightarrow 1+t^2$, with the Ricci scalar thus limiting to 
\begin{eqnarray}
R^{\alpha}_{\phantom{\alpha}\alpha}\rightarrow -6\frac{t^2+1+t^2-1}{t^2}=-6\frac{1-1}{t^2}-6\frac{t^2+t^2}{t^2}=\frac{0}{t^2}-12.
\label{1.40y}
\end{eqnarray}
Thus because of the $t^2$ factor in the denominator of $R^{\alpha}_{\phantom{\alpha}\alpha}$ we must work to order $t^2$ in the numerator. Without including this term we would be lead to the erroneous conclusion that $R^{\alpha}_{\phantom{\alpha}\alpha}$ is zero. In other words, if we simply set $a(t)=t$, $k=-1$ in $R^{\alpha}_{\phantom{\alpha}\alpha}=-6(a\ddot{a}+\dot{a}^2+k)/a^2$ we would indeed get zero. However we are working in a geometry in which $a(t)$ limits to $t$, not in one in which it is identically equal to $t$, and in the limit we need to carry the order $t^2$ term. Moreover, regardless of this concern, we note that in the  $t^{\prime 2}-r^{\prime 2}\geq 0$ region, the only region that is observable, the cosmological recombination era geometry is anyway not exactly locally four-flat, but we can approximate it as such to one part in $10^4$. And even so, current era observers are looking at anisotropies in the CMB  through a geometry in which $a(t)$ is given by the non-flat (\ref{1.35y}). Given the negative signs for $G_{{\rm eff}}$ and $k$ and given the form for $a(t_R)$, we can now proceed to study conformal cosmological fluctuations around the Robertson-Walker metric given in (\ref{1.16y}) at recombination. 

However, before doing so we note that conformal models in which the scalar field is not an elementary field but actually a vacuum expectation value $\langle \Omega|\bar{\psi}\psi|\Omega\rangle$ of a fermion bilinear have also been considered \cite{Mannheim2012b,Mannheim2017}. In these models it is possible for the matter sources to make a more substantial contribution to cosmic expansion at recombination than in the elementary scalar field case.  These dynamical models are not as straightforward to handle as the elementary scalar field model and will be considered elsewhere. Nonetheless, in these models the gauge invariant  evolution equations given in Sec. \ref{S2} and the decomposition theorem structure that we derive in Sec. \ref{S5} also hold just as exactly (these equations being generic to any conformal cosmology). It is just that the form of the background $a(t_R)$ at recombination is not as straightforward to deal with as in the elementary scalar field model. And indeed, it is the simplicity of $a(t_R)=(-k)^{1/2}ct_R$ in the elementary scalar field model at recombination that enables us to solve the model  completely analytically  at recombination, just as we now do.

\section{The Fluctuations}
\label{S2}

\subsection{Converting the Background to Conformal Time}
\label{S2az}

While the above phenomenological discussion was developed for a specific background conformal cosmology with $k<0$, we now discuss the fluctuation equations for arbitrary $a(t)$, arbitrary $k$ and arbitrary background matter sources. Rather than work in comoving time we have found it more convenient to work in conformal time. Thus on defining
\begin{eqnarray}
\tau =\int \frac{dt}{a(t)},\quad \Omega(\tau)=a(t),
\label{2.1z}
\end{eqnarray}
we replace the background (\ref{1.16y}) by
\begin{eqnarray}
ds^2=\Omega^2(\tau)\left[c^2d\tau^2-\frac{dr^2}{1-kr^2}-r^2d\theta^2-r^2\sin^2\theta d\phi^2\right]
=\Omega^2(\tau)[c^2d\tau^2-\tilde{\gamma}_{ij}dx^idx^j],
\label{2.2z}
\end{eqnarray}
with $\tilde{\gamma}_{ij}$ being the metric of the spatial sector, and with $(i,j,k)=(r,\theta,\phi)$.   In conformal time the background Einstein tensor is given by
\begin{align}
G_{00}&= -3k- \frac{3}{c^2} \dot{\Omega}^2\Omega^{-2},\quad G_{0i} =0,
\quad G_{ij} = \tilde{\gamma}_{ij}\left[k - \frac{1}{c^2}\dot\Omega^2\Omega^{-2}+ \frac{2}{c^2}\ddot\Omega \Omega^{-1}\right],\quad R^{\alpha}_{\phantom{\alpha}\alpha}=-\frac{6}{\Omega^2}\left[k+ \frac{1}{c^2}\ddot\Omega \Omega^{-1}\right],
\label{2.3z}
\end{align}
where the dot now denotes the derivative with respect to $\tau$. In conformal time  a generic background perfect matter fluid is described by
\begin{align}
T^m_{\mu\nu}&=\frac{1}{c}\left[(\rho_m+p_m)U_{\mu}U_{\nu}+p_mg_{\mu\nu}\right],\quad g^{\mu\nu}U_{\mu}U_{\nu}=-1, \quad U^{0}=\Omega^{-1}(\tau), \quad U_0=-\Omega(\tau), \quad U^{i}=0, \quad U_i=0,
\label{2.4z}
\end{align}
with covariant conservation condition
\begin{align}
 \dot{\rho}_m+3\frac{\dot{\Omega}}{\Omega}(\rho_m+p_m)=0.
\label{2.5z}
\end{align}
For a conformal time radiation fluid with $3p_m=\rho_m$ we obtain $\rho_m=A/\Omega^4$, and for a non-relativistic  fluid with $p_m=0$ we obtain $\rho_m=B/\Omega^3$, viz. the same relations as obtained in comoving time. The background evolution equations are of the form
\begin{align}
4\alpha_g W_{\mu\nu}&=\frac{1}{c}\left[(\rho_m+p_m)U_{\mu}U_{\nu}+p_mg_{\mu\nu}\right] 
-\frac{1}{6}S_0^2G_{\mu\nu}         
-g_{\mu\nu}\lambda S_0^4. 
\label{2.6z}
\end{align}
In a conformal to flat background geometry in which $W_{\mu\nu}=0$ the background evolution equations take the form
\begin{align}
\frac{1}{2c^2}S_0^2(kc^2 +\dot{\Omega}^2\Omega^{-2})+\frac{\rho_m}{c}\Omega^2+\Omega^2\Lambda=0,\quad
-\frac{1}{6c^2}S_0^2(kc^2 -\dot{\Omega}^2\Omega^{-2}+2\ddot{\Omega}\Omega^{-1})+\frac{p_m}{c}\Omega^2-\Omega^2\Lambda=0.
\label{2.7z}
\end{align}
For $\rho_m=A/\Omega^{4}$ we obtain
\begin{align}
-\frac{S_0^2}{2c^2\Lambda}\dot{\Omega}^2=\left[\Omega^2+\frac{kS_0^2}{4\Lambda}+
\left(\frac{k^2S_0^4}{16\Lambda^2}-\frac{A}{\Lambda c}\right)^{1/2}\right]
\left[\Omega^2+\frac{kS_0^2}{4\Lambda}-
\left(\frac{k^2S_0^4}{16\Lambda^2}-\frac{A}{\Lambda c}\right)^{1/2}\right].
\label{2.8z}
\end{align}
While integrating (\ref{2.8z}) gives a somewhat intractable elliptic integral, in the non-early conformal gravity universe we can ignore radiation and set $A=0$, and with $k$ and $\Lambda$ both being negative then obtain 
\begin{align}
\Omega(\tau)=\frac{S_0(k/2\Lambda)^{1/2}}{\sinh(-(-k)^{1/2}c\tau)}.
\label{2.9z}
\end{align}
To relate the conformal $\tau$ and the comoving $t$, from $a(t)=(-k/\alpha)^{1/2}\sinh(\alpha^{1/2}ct)$ as given in (\ref{1.35y}) we set
\begin{align}
\tau=\int \frac{dt}{(-k/\alpha)^{1/2}\sinh\alpha^{1/2}ct}=\frac{1}{(-kc^2)^{1/2}}\log\tanh(\alpha^{1/2}ct/2),\quad
e^{(-kc^2)^{1/2}\tau}=\tanh(\alpha^{1/2}ct/2),
\label{2.10z}
\end{align}
as normalized so that $\tau=-\infty$ when $t=0$ and $\tau=0$ when $t=\infty$. (With the range of $\tau$ being negative, as given in (\ref{2.9z}) $\Omega(\tau)$ is positive everywhere within the range.) With $\Omega(\tau)=a(t)$,  from (\ref{2.10z}) and $a(t)=(-k/\alpha)^{1/2}\sinh(\alpha^{1/2}ct)$ (\ref{2.9z}) then follows since $\alpha=-2\Lambda/S_0^2$ \cite{footnoteB}. Finally, since at small comoving $t$ the conformal time $\tau_R$  goes to minus infinity, at recombination we can set $\Omega(\tau_R)=2S_0(k/2\Lambda)^{1/2}\exp[(-k)^{1/2}c\tau_R]$.

\subsection{The Scalar, Vector, Tensor Basis for Fluctuations}
\label{S2bz}

In analyzing cosmological perturbations it is very convenient to use the scalar, vector, tensor (SVT) basis for the fluctuations as developed in  \cite{Lifshitz1946}  and \cite{Bardeen1980}. In this basis the fluctuations are characterized according to how they transform under three-dimensional rotations, and in this form the basis  has been applied extensively in cosmological perturbation theory (see e.g. \cite{Kodama1984,Mukhanov1992,Stewart1990,Ma1995,Bertschinger1996,Zaldarriaga1998,Straumann2008,Szapudi2012} and \cite{Dodelson2003,Mukhanov2005,Weinberg2008,Lyth2009,Ellis2012,Maggiore2018}) since the background Robertson-Walker geometry itself has an underlying maximal spatial symmetry. With the background metric being written with an overall conformal factor $\Omega^2(\tau)$ in (\ref{2.2z}) we shall take the fluctuation metric to also have an overall conformal factor, with the full metric  thus being of the form \cite{footnoteC}
\begin{align}
ds^2&=-(g_{\mu\nu}+h_{\mu\nu})dx^{\mu}dx^{\nu}=\Omega^2(\tau)\left[d\tau^2-\frac{dr^2}{1-kr^2}-r^2d\theta^2-r^2\sin^2\theta d\phi^2\right]
\nonumber\\
&+\Omega^2(\tau)\left[2\phi d\tau^2 -2(\tilde{\nabla}_i B +B_i)d\tau dx^i - [-2\psi\tilde{\gamma}_{ij} +2\tilde{\nabla}_i\tilde{\nabla}_j E + \tilde{\nabla}_i E_j + \tilde{\nabla}_j E_i + 2E_{ij}]dx^i dx^j\right].
\label{2.11z}
\end{align}
In (\ref{2.11z})  $\tilde{\nabla}_i=\partial/\partial x^i$ and  $\tilde{\nabla}^i=\tilde{\gamma}^{ij}\tilde{\nabla}_j$  (with Latin indices) are defined with respect to the background three-space metric $\tilde{\gamma}_{ij}$, and $(1,2,3)=(r,\theta,\phi)$. And with
\begin{eqnarray}
\tilde{\gamma}^{ij}\tilde{\nabla}_j V_i=\tilde{\gamma}^{ij}[\partial_j V_i-\tilde{\Gamma}^{k}_{ij}V_k]
\label{2.12z}
\end{eqnarray}
for any three-vector $V_i$ in a three-space with three-space connection $\tilde{\Gamma}^{k}_{ij}$, the elements of (\ref{2.11z}) are required to obey
\begin{eqnarray}
\tilde{\gamma}^{ij}\tilde{\nabla}_j B_i = 0,\quad \tilde{\gamma}^{ij}\tilde{\nabla}_j E_i = 0, \quad E_{ij}=E_{ji},\quad \tilde{\gamma}^{jk}\tilde{\nabla}_kE_{ij} = 0, \quad\tilde{\gamma}^{ij}E_{ij} = 0.
\label{2.13z}
\end{eqnarray}
With the  three-space sector of the background geometry being maximally three-symmetric, it is described by a Riemann tensor of the form
\begin{eqnarray}
\tilde{R}_{ijk\ell}=k[\tilde{\gamma}_{jk}\tilde{\gamma}_{i\ell}-\tilde{\gamma}_{ik}\tilde{\gamma}_{j\ell}].
\label{2.14z}
\end{eqnarray}
As written, (\ref{2.11z}) contains ten elements, whose transformations are defined with respect to the background spatial sector as four three-dimensional scalars ($\phi$, $B$, $\psi$, $E$) each with one degree of freedom, two transverse three-dimensional vectors ($B_i$, $E_i$) each with two independent degrees of freedom, and one symmetric three-dimensional transverse-traceless tensor ($E_{ij}$) with two degrees of freedom. The great utility of this basis is that since the cosmological fluctuation equations are gauge invariant, only gauge-invariant scalar, vector, or tensor combinations of the components of the scalar, vector, tensor basis can appear in the fluctuation equations. 
In \cite{Amarasinghe2019} it was shown that for the fluctuations associated with the metric given in (\ref{2.11z}) and with $\Omega(\tau)$ being an arbitrary function of $\tau$, the gauge-invariant metric combinations are 
\begin{align}
\alpha=\phi + \psi + \dot B - \ddot E,\quad  \gamma= - \dot\Omega^{-1}\Omega \psi + B - \dot E,\quad  B_i-\dot{E}_i,  \quad E_{ij},
\label{2.15z}
\end{align}
for a total of six degrees of freedom, just as required since one can make four coordinate transformations  on the initial ten fluctuation components. As we shall see below, the fluctuation equations will explicitly depend on these specific combinations.

Given the fluctuation basis we evaluate the fluctuation Einstein tensor, and obtain \cite{Phelps2019} 
\begin{eqnarray}
\delta G_{00}&=& -6 k \phi - 6 k \psi + 6 \dot{\psi} \dot{\Omega} \Omega^{-1} + 2 \dot{\Omega} \Omega^{-1} \tilde{\nabla}_{a}\tilde{\nabla}^{a}B - 2 \dot{\Omega} \Omega^{-1} \tilde{\nabla}_{a}\tilde{\nabla}^{a}\dot{E} - 2 \tilde{\nabla}_{a}\tilde{\nabla}^{a}\psi, 
 \nonumber\\ 
\delta G_{0i}&=& 3 k \tilde{\nabla}_{i}B -  \dot{\Omega}^2 \Omega^{-2} \tilde{\nabla}_{i}B + 2 \overset{..}{\Omega} \Omega^{-1} \tilde{\nabla}_{i}B - 2 k \tilde{\nabla}_{i}\dot{E} - 2 \tilde{\nabla}_{i}\dot{\psi} - 2 \dot{\Omega} \Omega^{-1} \tilde{\nabla}_{i}\phi +2 k B_{i} -  k \dot{E}_{i} \nonumber \\ 
&& -  B_{i} \dot{\Omega}^2 \Omega^{-2} + 2 B_{i} \overset{..}{\Omega} \Omega^{-1} + \frac{1}{2} \tilde{\nabla}_{a}\tilde{\nabla}^{a}B_{i} -  \frac{1}{2} \tilde{\nabla}_{a}\tilde{\nabla}^{a}\dot{E}_{i},
 \nonumber\\ 
\delta G_{ij}&=& -2 \overset{..}{\psi}\tilde{\gamma}_{ij} + 2 \dot{\Omega}^2\tilde{\gamma}_{ij} \phi \Omega^{-2} + 2 \dot{\Omega}^2\tilde{\gamma}_{ij} \psi \Omega^{-2} - 2 \dot{\phi} \dot{\Omega}\tilde{\gamma}_{ij} \Omega^{-1} - 4 \dot{\psi} \dot{\Omega}\tilde{\gamma}_{ij} \Omega^{-1} - 4 \overset{..}{\Omega}\tilde{\gamma}_{ij} \phi \Omega^{-1} \nonumber \\ 
&& - 4 \overset{..}{\Omega}\tilde{\gamma}_{ij} \psi \Omega^{-1} - 2 \dot{\Omega}\tilde{\gamma}_{ij} \Omega^{-1} \tilde{\nabla}_{a}\tilde{\nabla}^{a}B - \tilde{\gamma}_{ij} \tilde{\nabla}_{a}\tilde{\nabla}^{a}\dot{B} +\tilde{\gamma}_{ij} \tilde{\nabla}_{a}\tilde{\nabla}^{a}\overset{..}{E} + 2 \dot{\Omega}\tilde{\gamma}_{ij} \Omega^{-1} \tilde{\nabla}_{a}\tilde{\nabla}^{a}\dot{E} 
\nonumber \\ 
&& - \tilde{\gamma}_{ij} \tilde{\nabla}_{a}\tilde{\nabla}^{a}\phi +\tilde{\gamma}_{ij} \tilde{\nabla}_{a}\tilde{\nabla}^{a}\psi + 2 \dot{\Omega} \Omega^{-1} \tilde{\nabla}_{j}\tilde{\nabla}_{i}B + \tilde{\nabla}_{j}\tilde{\nabla}_{i}\dot{B} -  \tilde{\nabla}_{j}\tilde{\nabla}_{i}\overset{..}{E} - 2 \dot{\Omega} \Omega^{-1} \tilde{\nabla}_{j}\tilde{\nabla}_{i}\dot{E} \nonumber \\ 
&& + 2 k \tilde{\nabla}_{j}\tilde{\nabla}_{i}E - 2 \dot{\Omega}^2 \Omega^{-2} \tilde{\nabla}_{j}\tilde{\nabla}_{i}E + 4 \overset{..}{\Omega} \Omega^{-1} \tilde{\nabla}_{j}\tilde{\nabla}_{i}E + \tilde{\nabla}_{j}\tilde{\nabla}_{i}\phi -  \tilde{\nabla}_{j}\tilde{\nabla}_{i}\psi +\dot{\Omega} \Omega^{-1} \tilde{\nabla}_{i}B_{j} + \frac{1}{2} \tilde{\nabla}_{i}\dot{B}_{j}
\nonumber \\ 
&& -  \frac{1}{2} \tilde{\nabla}_{i}\overset{..}{E}_{j} -  \dot{\Omega} \Omega^{-1} \tilde{\nabla}_{i}\dot{E}_{j} + k \tilde{\nabla}_{i}E_{j} -  \dot{\Omega}^2 \Omega^{-2} \tilde{\nabla}_{i}E_{j} + 2 \overset{..}{\Omega} \Omega^{-1} \tilde{\nabla}_{i}E_{j} + \dot{\Omega} \Omega^{-1} \tilde{\nabla}_{j}B_{i} + \frac{1}{2} \tilde{\nabla}_{j}\dot{B}_{i} \nonumber \\ 
&& -  \frac{1}{2} \tilde{\nabla}_{j}\overset{..}{E}_{i} -  \dot{\Omega} \Omega^{-1} \tilde{\nabla}_{j}\dot{E}_{i} + k \tilde{\nabla}_{j}E_{i} -  \dot{\Omega}^2 \Omega^{-2} \tilde{\nabla}_{j}E_{i} + 2 \overset{..}{\Omega} \Omega^{-1} \tilde{\nabla}_{j}E_{i}- \overset{..}{E}_{ij} - 2 \dot{\Omega}^2 E_{ij} \Omega^{-2} \nonumber \\ 
&& - 2 \dot{E}_{ij} \dot{\Omega} \Omega^{-1} + 4 \overset{..}{\Omega} E_{ij} \Omega^{-1} + \tilde{\nabla}_{a}\tilde{\nabla}^{a}E_{ij},
 \nonumber\\
g^{\mu\nu}\delta G_{\mu\nu} &=& 6 \dot{\Omega}^2 \phi \Omega^{-4} + 6 \dot{\Omega}^2 \psi \Omega^{-4} - 6 \dot{\phi} \dot{\Omega} \Omega^{-3} - 18 \dot{\psi} \dot{\Omega} \Omega^{-3} - 12 \overset{..}{\Omega} \phi \Omega^{-3} - 12 \overset{..}{\Omega} \psi \Omega^{-3} - 6 \overset{..}{\psi} \Omega^{-2} + 6 k \phi \Omega^{-2} \nonumber \\ 
&& + 6 k \psi \Omega^{-2} - 6 \dot{\Omega} \Omega^{-3} \tilde{\nabla}_{a}\tilde{\nabla}^{a}B - 2 \Omega^{-2} \tilde{\nabla}_{a}\tilde{\nabla}^{a}\dot{B} + 2 \Omega^{-2} \tilde{\nabla}_{a}\tilde{\nabla}^{a}\overset{..}{E} + 6 \dot{\Omega} \Omega^{-3} \tilde{\nabla}_{a}\tilde{\nabla}^{a}\dot{E} \nonumber \\ 
&& - 2 \dot{\Omega}^2 \Omega^{-4} \tilde{\nabla}_{a}\tilde{\nabla}^{a}E + 4 \overset{..}{\Omega} \Omega^{-3} \tilde{\nabla}_{a}\tilde{\nabla}^{a}E + 2 k \Omega^{-2} \tilde{\nabla}_{a}\tilde{\nabla}^{a}E - 2 \Omega^{-2} \tilde{\nabla}_{a}\tilde{\nabla}^{a}\phi + 4 \Omega^{-2} \tilde{\nabla}_{a}\tilde{\nabla}^{a}\psi. 
\label{2.16z}
\end{eqnarray}

For fluctuations in the matter field $T^m_{\mu\nu}$ we obtain
\begin{eqnarray}
\delta T^m_{\mu\nu}=\frac{1}{c}\left[(\delta\rho_m+\delta p_m)U_{\mu}U_{\nu}+\delta p_mg_{\mu\nu}+(\rho_m+p_m)(\delta U_{\mu}U_{\nu}+U_{\mu}\delta U_{\nu})+p_mh_{\mu\nu}\right].
\label{2.17z}
\end{eqnarray}
With $g^{\mu\nu}U_{\mu}U_{\nu}=-1$ we obtain 
\begin{eqnarray}
 \delta g^{00}U_{0}U_{0}+2g^{00}U_{0}\delta U_{0}=0,
\label{2.18z}
\end{eqnarray}
which entails that 
\begin{eqnarray}
\delta U_{0}=-\frac{1}{2}(g^{00})^{-1}(-g^{00}g^{00}\delta g_{00})U_{0}=-\Omega(\tau)\phi,
\label{2.19z}
\end{eqnarray}
with $\delta U_0$ thus not being an independent degree of freedom. With $\delta U_i$ being a three-vector we shall decompose it into its transverse and longitudinal parts as $\delta U_i=V_i+\tilde{\nabla}_iV$, where now $\tilde{\gamma}^{ij}\tilde{\nabla}_j V_i=\tilde{\gamma}^{ij}[\partial_j V_i-\tilde{\Gamma}^{k}_{ij}V_k]=0$. As constructed, in general we have 11 fluctuation variables, the six from the metric together with $\delta\rho_m$,  $\delta p_m$ and  the three $\delta U_i$. But we only have ten fluctuation equations. Thus to solve the theory when there is both a $\delta \rho_m$ and a $\delta p_m$ we will need some constraint between $\delta p_m$ and $\delta \rho_m$. However, while this would be required if we want to obtain the general solution, as we had noted above, at recombination both $\delta p_m$ and $\delta \rho_m$ are suppressed in the conformal case, so no constraint between $\rho_m$ and $p_m$ is needed for our purposes here. Finally, we note that the fluctuation in the cosmological constant term is just  $-\lambda S_0^4 h_{\mu\nu}$.

The fluctuation $\delta W_{\mu\nu}$ in the Bach tensor $W_{\mu\nu}$ is of the form \cite{Phelps2019}
\begin{eqnarray}
\delta W_{00}&=& - \frac{2}{3\Omega^2} (\tilde\nabla_a\tilde\nabla^a + 3k)\tilde\nabla_b\tilde\nabla^b \alpha,
 \nonumber\\ 
\delta W_{0i}&=& -\frac{2}{3\Omega^2}  \tilde\nabla_i (\tilde\nabla_a\tilde\nabla^a + 3k)\dot\alpha
+\frac{1}{2\Omega^2}(\tilde\nabla_b \tilde\nabla^b-\partial_{\tau}^2-2k)(\tilde\nabla_c \tilde\nabla^c+2k)(B_i-\dot{E}_i),
  \nonumber\\ 
\delta W_{ij}&=& -\frac{1}{3 \Omega^2} \left[ \tilde{\gamma}_{ij} \tilde\nabla_a\tilde\nabla^a (\tilde\nabla_b \tilde\nabla^b +2k-\partial_{\tau}^2)\alpha - \tilde\nabla_i\tilde\nabla_j(\tilde\nabla_a\tilde\nabla^a - 3\partial_{\tau}^2)\alpha \right]
\nonumber\\
&& +\frac{1}{2 \Omega^2} \left[ \tilde\nabla_i (\tilde\nabla_a\tilde\nabla^a -2k-\partial_{\tau}^2) (\dot{B}_j-\ddot{E}_j) 
+  \tilde\nabla_j ( \tilde\nabla_a\tilde\nabla^a -2k-\partial_{\tau}^2) (\dot{B}_i-\ddot{E}_i)\right]
\nonumber\\
&&+ \frac{1}{\Omega^2}\left[ (\tilde\nabla_b \tilde\nabla^b-\partial_{\tau}^2-2k)^2+4k\partial_{\tau}^2 \right] E_{ij}.
\label{2.20z}
\end{eqnarray}
The structure of $\delta W_{\mu\nu}$  is noteworthy in two regards: first $\delta W_{\mu\nu}$ is built out of gauge invariant quantities alone, even though this is not the case for $\delta G_{\mu\nu}$, and second it obeys the tracelessness condition $g^{\mu\nu}\delta W_{\mu\nu}=0$. Neither of these two features is generic for any $\delta W_{\mu\nu}$, but they do  hold for fluctuations around a background in which $W_{\mu\nu}$ is zero. Specifically, when $W_{\mu\nu}$ is zero any general background $T_{\mu\nu}$ must be zero too. Now while we had noted above that $T_{\mu\nu}$ could vanish non-trivially, it could of course also vanish trivially if there are no matter sources. The structure of $\delta W_{\mu\nu}$ is not sensitive to how the background $T_{\mu\nu}$ vanishes, and it would  have the same form in either case. However, $4\alpha_g \delta W_{\mu\nu}-\delta T_{\mu\nu}$ is always gauge invariant, and thus it would be gauge invariant if $T_{\mu\nu}$ vanishes trivially and there is no $\delta T_{\mu\nu}$  at all. Thus for fluctuations around any background in which $W_{\mu\nu}$ vanishes $\delta W_{\mu\nu}$ will always be gauge invariant on its own.

Now regardless of whether or not the background $W_{\mu\nu}$ vanishes,  because of the underlying conformal invariance of the theory it will still obey the tracelessness condition $g^{\mu\nu} W_{\mu\nu}=0$. Thus one will always have $\delta[g^{\mu\nu}W_{\mu\nu}]=0$, i.e., $g^{\mu\nu} \delta W_{\mu\nu}-h^{\mu\nu}W_{\mu\nu}=0$. Then if the background is such that $W_{\mu\nu}=0$ one would have $g^{\mu\nu}\delta W_{\mu\nu}=0$. This then is the case for fluctuations around any Robertson-Walker background, and thus  $g^{\mu\nu}\delta W_{\mu\nu}=0$ does hold for the $\delta W_{\mu\nu}$ given in (\ref{2.20z}). Since $g^{\mu\nu}\delta W_{\mu\nu}$ does vanish, the tensor $\delta W_{\mu\nu}$ can only have nine independent components. With four coordinate invariances  $\delta W_{\mu\nu}$ can only depend on five gauge invariant degrees of freedom, and as we see, they are $\alpha$, $B_i-\dot{E}_i$ and $E_{ij}$.

From (\ref{2.6z}) we obtain  background and fluctuation equations of the form
\begin{align}
4\alpha_g W_{\mu\nu}&=\frac{1}{c}\left[(\rho_m+p_m)U_{\mu}U_{\nu}+p_mg_{\mu\nu}\right] 
-\frac{1}{6}S_0^2G_{\mu\nu}         
-g_{\mu\nu}\Lambda,
\nonumber\\
4\alpha_g \delta W_{\mu\nu}&=\frac{1}{c}\left[(\delta\rho_m+\delta p_m)U_{\mu}U_{\nu}+\delta p_mg_{\mu\nu}+(\rho_m+p_m)(\delta U_{\mu}U_{\nu}+U_{\mu}\delta U_{\nu})+p_mh_{\mu\nu}\right]
-\frac{1}{6}S_0^2\delta G_{\mu\nu} -h_{\mu\nu}\Lambda.
\label{2.21z}
\end{align}
It is convenient to define 
\begin{eqnarray}
\eta=-\frac{24\alpha_g}{S_0^2}, \quad \rho=-\frac{6(\rho_m+c\Lambda)}{S_0^2}, \quad p=-\frac{6(p_m-c\Lambda)}{S_0^2}, \quad \delta \rho=-\frac{6\delta \rho_m}{S_0^2}, \quad \delta p=-\frac{6\delta p_m}{S_0^2}. 
\label{2.22y}
\end{eqnarray}
The background and fluctuation equations then take the form
\begin{align}
\eta W_{\mu\nu}&= G_{\mu\nu}+ \frac{1}{c}\left[(\rho+ p)U_{\mu}U_{\nu}+ pg_{\mu\nu}\right]=\Delta^{(0)}_{\mu\nu},
\label{2.23y}
\end{align}
\begin{align}
\eta \delta W_{\mu\nu}=\delta G_{\mu\nu}+ \frac{1}{c}\left[(\delta \rho+\delta p)U_{\mu}U_{\nu}+\delta pg_{\mu\nu}+(\rho+p)(\delta U_{\mu}U_{\nu}+U_{\mu}\delta U_{\nu})+ph_{\mu\nu}\right]=\Delta_{\mu\nu},
\label{2.24y}
\end{align}
with (\ref{2.23y}) and (\ref{2.24y}) serving to define $\Delta^{(0)}_{\mu\nu}$ and $\Delta_{\mu\nu}$.
With use of $\Delta^{(0)}_{\mu\nu}=0$ (which follows here since  $W_{\mu\nu}=0$),  and with $\delta G_{\mu\nu}$ being given in (\ref{2.16z}), the components of $\Delta_{\mu\nu}$ have been given in \cite{Phelps2019} and are of the form:
\begin{eqnarray}
\Delta_{00}&=& 6 \dot{\Omega}^2 \Omega^{-2}(\alpha-\dot\gamma) + \delta \hat{\rho} \Omega^2 + 2 \dot{\Omega} \Omega^{-1} \tilde{\nabla}_{a}\tilde{\nabla}^{a}\gamma, 
\label{2.25y}
\end{eqnarray}
\begin{eqnarray}
\Delta_{0i}&=& -2 \dot{\Omega} \Omega^{-1} \tilde{\nabla}_{i}(\alpha - \dot\gamma) + 2 k \tilde{\nabla}_{i}\gamma 
+(-4 \dot{\Omega}^2 \Omega^{-3}  + 2 \overset{..}{\Omega} \Omega^{-2}  - 2 k \Omega^{-1}) \tilde{\nabla}_{i}\hat{V}
\nonumber\\
&& +k(B_i-\dot E_i)+ \frac{1}{2} \tilde{\nabla}_{a}\tilde{\nabla}^{a}(B_{i} - \dot{E}_{i})
+ (-4 \dot{\Omega}^2 \Omega^{-3} + 2 \overset{..}{\Omega} \Omega^{-2} - 2 k \Omega^{-1})V_{i},
\label{2.26y}
\end{eqnarray}
\begin{eqnarray}
\Delta_{ij}&=& \tilde{\gamma}_{ij}\big[ 2 \dot{\Omega}^2 \Omega^{-2}(\alpha-\dot\gamma)
-2  \dot{\Omega} \Omega^{-1}(\dot\alpha -\ddot\gamma)-4\ddot\Omega\Omega^{-1}(\alpha-\dot\gamma)+ \Omega^2 \delta \hat{p}-\tilde\nabla_a\tilde\nabla^a( \alpha + 2\dot\Omega \Omega^{-1}\gamma) \big] 
\nonumber\\
&&+\tilde\nabla_i\tilde\nabla_j( \alpha + 2\dot\Omega \Omega^{-1}\gamma)
+\dot{\Omega} \Omega^{-1} \tilde{\nabla}_{i}(B_{j}-\dot E_j)+\frac{1}{2} \tilde{\nabla}_{i}(\dot{B}_{j}-\ddot{E}_j)
+\dot{\Omega} \Omega^{-1} \tilde{\nabla}_{j}(B_{i}-\dot E_i)+\frac{1}{2} \tilde{\nabla}_{j}(\dot{B}_{i}-\ddot{E}_i)
\nonumber\\
&&- \overset{..}{E}_{ij} - 2 k E_{ij} - 2 \dot{E}_{ij} \dot{\Omega} \Omega^{-1} + \tilde{\nabla}_{a}\tilde{\nabla}^{a}E_{ij},
\label{2.27y}
\end{eqnarray}
\begin{eqnarray}
\tilde{\gamma}^{ij}\Delta_{ij} &=&  6 \dot{\Omega}^2 \Omega^{-2}(\alpha-\dot\gamma)
-6  \dot{\Omega} \Omega^{-1}(\dot\alpha -\ddot\gamma)-12\ddot\Omega\Omega^{-1}(\alpha-\dot\gamma)+ 3\Omega^2 \delta \hat{p}-2\tilde\nabla_a\tilde\nabla^a( \alpha + 2\dot\Omega \Omega^{-1}\gamma),
\label{2.28y}
\end{eqnarray}
\begin{eqnarray}
g^{\mu\nu}\Delta_{\mu\nu}&=& 3 \delta \hat{p} -  \delta \hat{\rho}
-12 \overset{..}{\Omega}  \Omega^{-3}(\alpha - \dot\gamma) -6 \dot{\Omega} \Omega^{-3}(\dot{\alpha} -\ddot\gamma)
-2 \Omega^{-2} \tilde{\nabla}_{a}\tilde{\nabla}^{a}(\alpha +3\dot\Omega\Omega^{-1}\gamma),
\label{2.29y}
\end{eqnarray}
where
\begin{eqnarray}
\Omega^2\rho&=&3k+3\dot{\Omega}^2\Omega^{-2},\quad \Omega^2 p=-k+\dot{\Omega}^2\Omega^{-2}
-2\ddot{\Omega}\Omega^{-1},\quad \dot{\rho}+3\dot{\Omega}(\rho+p)\Omega^{-1}=0,
\nonumber\\
\alpha  &=& \phi + \psi + \dot B - \ddot E,\quad \gamma = - \dot\Omega^{-1}\Omega \psi + B - \dot E,\quad \hat{V} = V-\Omega^2 \dot \Omega^{-1}\psi,
 \nonumber\\
\delta \hat{\rho}&=&\delta \rho - 12 \dot{\Omega}^2 \psi \Omega^{-4} + 6 \overset{..}{\Omega} \psi \Omega^{-3} - 6 k \psi \Omega^{-2}=\delta\rho +\dot{\Omega}^{-1}\dot{\rho}\psi\Omega=\delta \rho-3(\rho+p)\psi,
\nonumber\\
\delta \hat{p}&=&\delta p - 4 \dot{\Omega}^2 \psi \Omega^{-4} + 8 \overset{..}{\Omega} \psi \Omega^{-3} + 2 k \psi \Omega^{-2} - 2 \overset{...}{\Omega} \dot{\Omega}^{-1} \psi \Omega^{-2}=\delta p +\dot{\Omega}^{-1}\dot{p}\psi \Omega.
\label{2.30y}
\end{eqnarray}
(The first three expressions in (\ref{2.30y}) hold for the background and follow from $\Delta^{(0)}_{\mu\nu}=0$.) 
With $\eta\delta W_{\mu\nu}-\Delta_{\mu\nu}$ being gauge invariant and with $\delta W_{\mu\nu}$ being gauge invariant on its own, it follows that $\Delta_{\mu\nu}$ is gauge invariant too, and thus its dependence on the metric sector fluctuations must be solely on the metric combinations $\alpha$, $\gamma$, $B_i-\dot{E}_i$ and $E_{ij}$, just as we see. Then since the metric sector $\alpha$, $\gamma$, $B_i-\dot{E}_i$ and $E_{ij}$ are gauge invariant, from the gauge invariance of $\Delta_{\mu\nu}$  it follows that $\delta \hat{\rho}$, $\delta \hat{p}$, $\hat{V}$ and $V_i$ are gauge invariant too \cite{footnoteC2}. We thus have expressed the fluctuation equations entirely in terms of gauge invariant combinations without needing to specify any particular gauge \cite{footnoteD}. Given (\ref{2.20z}) and (\ref{2.25y}) - (\ref{2.27y}) the fluctuation equations take the form 
\begin{eqnarray}
\eta \delta W_{00}&=& - \frac{2\eta}{3\Omega^2} (\tilde\nabla_a\tilde\nabla^a + 3k)\tilde\nabla_b\tilde\nabla^b \alpha
 \nonumber\\ 
 &=&\Delta_{00}=6 \dot{\Omega}^2 \Omega^{-2}(\alpha-\dot\gamma) + \delta \hat{\rho} \Omega^2 + 2 \dot{\Omega} \Omega^{-1} \tilde{\nabla}_{a}\tilde{\nabla}^{a}\gamma, 
\label{2.31y}
\end{eqnarray}
\begin{eqnarray}
\eta\delta W_{0i}&=& -\frac{2\eta}{3\Omega^2}  \tilde\nabla_i (\tilde\nabla_a\tilde\nabla^a + 3k)\dot\alpha
+\frac{\eta}{2\Omega^2}(\tilde\nabla_b \tilde\nabla^b-\partial_{\tau}^2-2k)(\tilde\nabla_c \tilde\nabla^c+2k)(B_i-\dot{E}_i)
  \nonumber\\ 
  &=&\Delta_{0i}= -2 \dot{\Omega} \Omega^{-1} \tilde{\nabla}_{i}(\alpha - \dot\gamma) + 2 k \tilde{\nabla}_{i}\gamma 
+(-4 \dot{\Omega}^2 \Omega^{-3}  + 2 \overset{..}{\Omega} \Omega^{-2}  - 2 k \Omega^{-1}) \tilde{\nabla}_{i}\hat{V}
\nonumber\\
&& +k(B_i-\dot E_i)+ \frac{1}{2} \tilde{\nabla}_{a}\tilde{\nabla}^{a}(B_{i} - \dot{E}_{i})
+ (-4 \dot{\Omega}^2 \Omega^{-3} + 2 \overset{..}{\Omega} \Omega^{-2} - 2 k \Omega^{-1})V_{i},
\label{2.32y}
\end{eqnarray}
\begin{eqnarray}
\eta \delta W_{ij}&=& -\frac{\eta}{3 \Omega^2} \left[ \tilde{\gamma}_{ij} \tilde\nabla_a\tilde\nabla^a (\tilde\nabla_b \tilde\nabla^b +2k-\partial_{\tau}^2)\alpha - \tilde\nabla_i\tilde\nabla_j(\tilde\nabla_a\tilde\nabla^a - 3\partial_{\tau}^2)\alpha \right]
\nonumber\\
&& +\frac{\eta}{2 \Omega^2} \left[ \tilde\nabla_i (\tilde\nabla_a\tilde\nabla^a -2k-\partial_{\tau}^2) (\dot{B}_j-\ddot{E}_j) 
+  \tilde\nabla_j ( \tilde\nabla_a\tilde\nabla^a -2k-\partial_{\tau}^2) (\dot{B}_i-\ddot{E}_i)\right]
\nonumber\\
&&+ \frac{\eta}{\Omega^2}\left[ (\tilde\nabla_b \tilde\nabla^b-\partial_{\tau}^2-2k)^2+4k\partial_{\tau}^2 \right] E_{ij}
\nonumber\\
&=&\Delta_{ij}=\tilde{\gamma}_{ij}\big[ 2 \dot{\Omega}^2 \Omega^{-2}(\alpha-\dot\gamma)
-2  \dot{\Omega} \Omega^{-1}(\dot\alpha -\ddot\gamma)-4\ddot\Omega\Omega^{-1}(\alpha-\dot\gamma)+ \Omega^2 \delta \hat{p}-\tilde\nabla_a\tilde\nabla^a( \alpha + 2\dot\Omega \Omega^{-1}\gamma) \big] 
\nonumber\\
&&+\tilde\nabla_i\tilde\nabla_j( \alpha + 2\dot\Omega \Omega^{-1}\gamma)
+\dot{\Omega} \Omega^{-1} \tilde{\nabla}_{i}(B_{j}-\dot E_j)+\frac{1}{2} \tilde{\nabla}_{i}(\dot{B}_{j}-\ddot{E}_j)
+\dot{\Omega} \Omega^{-1} \tilde{\nabla}_{j}(B_{i}-\dot E_i)+\frac{1}{2} \tilde{\nabla}_{j}(\dot{B}_{i}-\ddot{E}_i)
\nonumber\\
&&- \overset{..}{E}_{ij} - 2 k E_{ij} - 2 \dot{E}_{ij} \dot{\Omega} \Omega^{-1} + \tilde{\nabla}_{a}\tilde{\nabla}^{a}E_{ij}.
\label{2.33y}
\end{eqnarray}
In the conformal  gravity theory these cosmological fluctuation equations are completely general, and hold for any possible matter source and any possible $a(t)$ and $k$.

We had noted above that the only difference between the conformal gravity (\ref{1.27y}) and the standard Einstein gravity 
(\ref{1.28y}) was in the replacement of the Newtonian $G$ by the conformal gravity $G_{\rm eff}$ given in (\ref{1.29y}).  We can thus treat both $\Delta^{\mu\nu}_{(0)}$ and $\Delta^{\mu\nu}$ as being generic to both theories. Consequently, Einstein gravity fluctuation theory can be recognized as the $\eta=0$ limit of the conformal gravity $\eta W^{\mu\nu}-\Delta_{(0)}^{\mu\nu}=0$, $\eta \delta W^{\mu\nu}-\Delta^{\mu\nu}=0$  in which $\Delta^{\mu\nu}_{(0)}=0$ and $\Delta^{\mu\nu}=0$. As we will see in Secs. \ref{S3} and \ref{S4}, some of the analysis  obtained in the Einstein gravity study given in \cite{Phelps2019} will thus carry over to our present conformal gravity study. 

We should also add that the parameter $\alpha_g$ is actually known to be negative \cite{Mannheim2011a,Mannheim2016}. The parameter $\eta=-24\alpha_g/S_0^2$ is thus positive. This will prove to be a key feature of the development below as it will lead us to solutions to the fluctuation equations that oscillate in time rather than grow or decay exponentially. To actually find the solutions we need first to manipulate the fluctuation equations so as to find equations for the individual gauge invariant combinations (this will be done by taking the same judicious choice of derivatives of the fluctuation equations as was done in \cite{Phelps2019}), and to then solve the equations that are obtained by such a technique.

\section{Separating the Fluctuation Equations}
\label{S3}

\subsection{The Scalar Sector}
\label{S3a}

With the $\tilde{\nabla}_i$ derivatives acting in a maximally symmetric three-space with three-curvature equal to $k$ the following relations hold for any three-scalar $S$ \cite{Phelps2019}
\begin{eqnarray}
\tilde{\nabla}_a\tilde{\nabla}^a\tilde{\nabla}_iS&=&\tilde{\nabla}_i\tilde{\nabla}_a\tilde{\nabla}^aS+2k\tilde{\nabla}_iS,\quad \tilde{\nabla}_a\tilde{\nabla}^a\tilde{\nabla}_i\tilde{\nabla}_{j}S=\tilde{\nabla}_i\tilde{\nabla}_{j}\tilde{\nabla}_a\tilde{\nabla}^aS
+6k\tilde{\nabla}_i\tilde{\nabla}_{j}S-2k\tilde{\gamma}_{ij}\tilde{\nabla}_a\tilde{\nabla}^aS.
\label{3.1z}
\end{eqnarray}
Similarly for any three-vector $A_i$ we have \cite{Phelps2019}
\begin{eqnarray}
&&\tilde\nabla_i\tilde\nabla_a\tilde\nabla^aA_j-\tilde\nabla_a\tilde\nabla^a\tilde\nabla_iA_j 
= 2k\tilde{\gamma}_{ij}\tilde{\nabla}_aA^a-2k(\tilde\nabla_i A_j + \tilde\nabla_j A_i),
\nonumber\\
&&\tilde\nabla^j\tilde\nabla_a\tilde\nabla^aA_j=
(\tilde\nabla_a\tilde\nabla^a+2k)\tilde\nabla^j A_j,\quad \tilde{\nabla}^j\tilde{\nabla}_iA_j=\tilde{\nabla}_i\tilde{\nabla}^jA_j+2kA_i,
\label{3.2z}
\end{eqnarray}
with (\ref{3.2z}) stating that if $A_j$ is transverse then so is $\tilde\nabla_a\tilde\nabla^aA_j$. 
For a general symmetric rank two tensor in a (more general) maximally symmetric $D$-dimensional space with curvature $K$ we have \cite{Mannheim2012a}
\begin{eqnarray}
\nabla_{P}\nabla_{N}\nabla^{N}A^{P}_{\phantom{P}M}
=[\nabla_{N}\nabla^{N}+K(D+1)]\nabla_{P}A^{P}_{\phantom{P}M}
-2K\nabla_{M}A^{P}_{\phantom{P}P}.
\label{3.3z}
\end{eqnarray}
Thus for $D=3$, we see that if  $A_{ij}$ is transverse and traceless then so is $\tilde\nabla_a\tilde\nabla^aA_{ij}$.

We shall now use this information to obtain equations that do not mix scalars, vectors and tensors. For the scalar sector we already have two such relations already, $\eta \delta W_{00}=\Delta_{00}$ and also $g^{\mu\nu}\Delta_{\mu\nu}=0$ since $g^{\mu\nu}\delta W_{\mu\nu}=0$, viz.
\begin{eqnarray}
\eta \delta W_{00}-\Delta_{00}=- \frac{2\eta}{3 \Omega^2} (\tilde\nabla_a\tilde\nabla^a + 3k)\tilde\nabla_b\tilde\nabla^b \alpha
-6 \dot{\Omega}^2 \Omega^{-2}(\alpha-\dot\gamma) - \delta \hat{\rho} \Omega^2 - 2 \dot{\Omega} \Omega^{-1} \tilde{\nabla}_{a}\tilde{\nabla}^{a}\gamma=0,
\label{3.4z}
\end{eqnarray}
\begin{eqnarray}
g^{\mu\nu}(\eta\delta W_{\mu\nu}-\Delta_{\mu\nu}) =-3 \delta \hat{p} +\delta \hat{\rho}
+12 \overset{..}{\Omega}  \Omega^{-3}(\alpha - \dot\gamma) +6 \dot{\Omega} \Omega^{-3}(\dot{\alpha} -\ddot\gamma)
+2 \Omega^{-2} \tilde{\nabla}_{a}\tilde{\nabla}^{a}(\alpha +3\dot\Omega\Omega^{-1}\gamma)=0.
\label{3.5z}
\end{eqnarray}
From the $(0,i)$ sector we have  $\tilde{\nabla}^i(\delta W_{0i}-\Delta_{0i})=0$, and with $\tilde\nabla^i\tilde\nabla_a\tilde\nabla^a(B_i-\dot{E}_i)=(\tilde\nabla_a\tilde\nabla^a+2k)\tilde\nabla^i(B_i-\dot{E}_i)=0$ thus obtain 
\begin{align}
&\tilde{\nabla}^i(\eta\delta W_{0i}-\Delta_{0i})=-\frac{2\eta}{3 \Omega^2} \tilde\nabla^i \tilde\nabla_i (\tilde\nabla_a\tilde\nabla^a + 3k)\dot\alpha
\nonumber\\
&+2 \dot{\Omega} \Omega^{-1} \tilde\nabla^i \tilde{\nabla}_{i}(\alpha - \dot\gamma) - 2 k \tilde\nabla^i \tilde{\nabla}_{i}\gamma 
-(-4 \dot{\Omega}^2 \Omega^{-3}  + 2 \overset{..}{\Omega} \Omega^{-2}  - 2 k \Omega^{-1})\tilde\nabla^i \tilde{\nabla}_{i}\hat{V}=0.
\label{3.6z}
\end{align}

With $\tilde{\nabla}^i\tilde{\nabla}^j(\tilde{\nabla}_iA_j+\tilde{\nabla}_jA_i)=2(\tilde{\nabla}_i\tilde{\nabla}_i+2k)\tilde{\nabla}^jA_j$ and $\tilde{\nabla}^i\tilde{\nabla}^j\tilde{\nabla}_i\tilde{\nabla}_jS=(\tilde{\nabla}^i\tilde{\nabla}_i+2k)\tilde{\nabla}^j\tilde{\nabla}_jS$,  in the $(i,j)$ sector we obtain
\begin{align}
&\tilde{\nabla}^i\tilde{\nabla}^j(\eta\delta W_{ij}-\Delta_{ij})
\nonumber\\
&=-\frac{2\eta}{3 \Omega^2}\tilde{\nabla}_i\tilde{\nabla}^i (\tilde{\nabla}_a \tilde{\nabla}^a +3k)\partial_{\tau}^2\alpha
\nonumber\\
&-\tilde{\nabla}_{i}\tilde{\nabla}^i\big[ 2 \dot{\Omega}^2 \Omega^{-2}(\alpha-\dot\gamma)
-2  \dot{\Omega} \Omega^{-1}(\dot\alpha -\ddot\gamma)-4\ddot\Omega\Omega^{-1}(\alpha-\dot\gamma)+ \Omega^2 \delta \hat{p}+2k( \alpha + 2\dot\Omega \Omega^{-1}\gamma) \big] =0,
\label{3.7z}
\end{align}
\begin{align}
&\tilde{\gamma}^{ij}(\eta\delta W_{ij}-\Delta_{ij})
\nonumber\\
&=-\frac{2\eta}{3 \Omega^2}\tilde{\nabla}_i\tilde{\nabla}^i (\tilde{\nabla}_a \tilde{\nabla}^a +3k)\alpha 
\nonumber\\
&- \big[ 6\dot{\Omega}^2 \Omega^{-2}(\alpha-\dot\gamma)
-6  \dot{\Omega} \Omega^{-1}(\dot\alpha -\ddot\gamma)-12\ddot\Omega\Omega^{-1}(\alpha-\dot\gamma)+ 3\Omega^2 \delta \hat{p}-2\tilde\nabla_a\tilde\nabla^a( \alpha + 2\dot\Omega \Omega^{-1}\gamma)\big]=0.
\label{3.8z}
\end{align}
To separate out the various combinations we evaluate
\begin{align}
&3\tilde{\nabla}^i\tilde{\nabla}^j(\eta\delta W_{ij}-\Delta_{ij})-\tilde{\nabla}_a\tilde{\nabla}^a\tilde{\gamma}^{ij}(\eta\delta W_{ij}-\Delta_{ij})
\nonumber\\
&=-\frac{2\eta}{3 \Omega^2}\tilde{\nabla}_i\tilde{\nabla}^i(\tilde{\nabla}_j\tilde{\nabla}^j+3k)( 3\partial_{\tau}^2 -\tilde\nabla_a\tilde\nabla^a)\alpha-2\tilde{\nabla}_i\tilde{\nabla}^i(\tilde{\nabla}_j\tilde{\nabla}^j+3k)(\alpha + 2\dot\Omega \Omega^{-1}\gamma)=0,
\label{3.9z}
\end{align}
\begin{align}
&\tilde{\nabla}^i\tilde{\nabla}^j(\eta\delta W_{ij}-\Delta_{ij})+k\tilde{\gamma}^{ij}(\eta \delta W_{ij}-\Delta_{ij})
\nonumber\\
&=-\frac{2\eta}{3\Omega^2}\tilde{\nabla}_i\tilde{\nabla}^i(\tilde\nabla_a\tilde\nabla^a+3k)(k+\partial_{\tau}^2)\alpha 
\nonumber\\
&-[\tilde\nabla_a\tilde\nabla^a+3k][2 \dot{\Omega}^2 \Omega^{-2}(\alpha-\dot\gamma)
-2  \dot{\Omega} \Omega^{-1}(\dot\alpha -\ddot\gamma)-4\ddot\Omega\Omega^{-1}(\alpha-\dot\gamma)+ \Omega^2 \delta \hat{p}]=0.
\label{3.10z}
\end{align}

Thus for the five scalar functions $\alpha$, $\gamma$, $\delta\hat{\rho}$, $\delta \hat{p}$ and $\hat{V}$ we initially appear to have obtained what would be a requisite five equations for them, viz. (\ref{3.4z}), (\ref{3.5z}), (\ref{3.6z}), (\ref{3.9z})  and (\ref{3.10z}), with one of them, viz. (\ref{3.9z}), not depending on any of the matter sources. However the trace condition given in (\ref{3.5z}) is not independent of the other conditions. Specifically,  we already have $g^{\mu\nu}\Delta^{(0)}_{\mu\nu}=0$ since the background matter sector is conformal. Now in general in any background that does obey $g^{\mu\nu}\Delta^{(0)}_{\mu\nu}=0$ we can set
\begin{eqnarray}
0=\delta[g^{\mu\nu}\Delta^{(0)}_{\mu\nu}]=g^{\mu\nu}\Delta_{\mu\nu}-h^{\mu\nu}\Delta^{(0)}_{\mu\nu},
\label{3.11z}
\end{eqnarray}
and so in general $g^{\mu\nu}\Delta_{\mu\nu}$ will not  be zero. However, if we now impose the background equations of motion $\eta W_{\mu\nu}=\Delta^{(0)}_{\mu\nu}$ given in (\ref{2.23y}), then since the background $W_{\mu\nu}$ is zero (the background being conformal to flat), it follows that the background $\Delta^{(0)}_{\mu\nu}
$ is zero too, and thus the fluctuation trace $g^{\mu\nu}\Delta_{\mu\nu}$ is automatically zero in solutions to the background equations of motion. With the form of the fluctuation $\Delta_{\mu\nu}$ given in (\ref{2.25y}) to (\ref{2.29y}) having been derived under the imposition of $\Delta^{(0)}_{\mu\nu}=0$, (\ref{3.5z}) is automatically obeyed for the $\Omega(\tau)$ that obeys the background (\ref{2.7z}). Thus in the scalar sector we only have four independent fluctuation equations [(\ref{3.4z}), (\ref{3.6z}), (\ref{3.9z})  and (\ref{3.10z})], but we have five dynamical variables in the sector: $\alpha$, $\gamma$, $\delta \hat{\rho}$, $\delta \hat{p}$ and $\hat{V}$. Thus without further information we cannot solve completely. This is a common feature of all fluctuation studies, and the additional information that is ordinarily assumed in fluctuation theory is a relation between $\delta \rho$ and $\delta p$ of the form $\delta p/\delta \rho=v^2/c^2$ where $v$ is a matter fluid  fluctuation velocity. (While this relation is ordinarily imposed in $k=0$ backgrounds, as noted in \cite{Phelps2019} in the $k<0$ case of interest to us here this relation would need to be generalized using fluids built out of incoherent averages of modes that obey (\ref{1.18y}) and the analogs of it that we encounter below.)  

While we do need more information to solve the scalar sector completely, this is not the case for either the vector or the tensor sectors as there we have just the right number of degrees of freedom [four in the vector sector ($B_i-\dot{E}_i$ and $V_i$) and two in the tensor sector ($E_{ij}$)]. In fact since we have the ten fluctuation equations given in (\ref{2.31y}) - (\ref{2.33y}) and 11 dynamical degrees of freedom we only have a shortfall of one degree of freedom, and thus no more than one of the scalar, vector or tensor sectors can be affected by this concern. However, since we could not have just a single extraneous degree of freedom in either the vector or the tensor sectors (we would need an even number since $B_i-\dot{E}_i$, $V_i$ and $E_{ij}$ each have two components), the shortfall would have to be in the scalar sector, just as we have found. Thus whatever goes on in the vector and tensor sectors could not be sensitive to this shortfall,  and thus these two sectors can always be solved without needing to specify any relation between $\delta p$ and $\delta \rho$.

\subsection{The Vector Sector}
\label{S3b}

For the vector sector the gauge invariant combination $B_i-\dot{E}_i$ appears in the $(0,i)$ and $(i,j)$ sectors.  For the $(0,i)$ sector first, we note that, given (\ref{2.32y}), we see that $\eta\delta W_{0i}-\Delta_{0i}$ can be written symbolically as the derivative of a scalar plus a transverse vector, viz.  as $\eta\delta W_{0i}-\Delta_{0i}=\tilde{\nabla}_iX+X_i$, where $\tilde{\nabla}^iX_i=0$. Thus $\tilde{\nabla}^i(\eta\delta W_{0i}-\Delta_{0i})=\tilde{\nabla}_i\tilde{\nabla}^iX=0$. Then given the relation  $\tilde{\nabla}_i\tilde{\nabla}_a\tilde{\nabla}^aS=(\tilde{\nabla}_a\tilde{\nabla}^a-2k)\tilde{\nabla}_iS$ that holds for any three-scalar in the background associated with (\ref{2.14z}), we can set 
\begin{align}
(\tilde{\nabla}_k\tilde\nabla^k -2k)(\eta\delta W_{0i}-\Delta_{0i})=(\tilde{\nabla}_i\tilde{\nabla}^i-2k)(\tilde{\nabla}_iX+X_i)=\tilde{\nabla}_i\tilde{\nabla}_a\tilde{\nabla}^aX+(\tilde{\nabla}_i\tilde{\nabla}^i-2k)X_i=(\tilde{\nabla}_i\tilde{\nabla}^i-2k)X_i=0.
\label{3.12z}
\end{align}
Thus  we directly obtain 
\begin{align}
&(\tilde{\nabla}_k\tilde\nabla^k -2k)(\eta\delta W_{0i}-\Delta_{0i})
\nonumber\\
&= (\tilde{\nabla}_k\tilde\nabla^k -2k)\frac{\eta}{2\Omega^2} (\tilde\nabla_b \tilde\nabla^b-\partial_{\tau}^2-2k)(\tilde\nabla_c \tilde\nabla^c+2k) (B_i-\dot{E}_i)
\nonumber\\
&-(\tilde{\nabla}_k\tilde\nabla^k-2k)\left[k(B_i-\dot E_i)+ \frac{1}{2} \tilde{\nabla}_{a}\tilde{\nabla}^{a}(B_{i} - \dot{E}_{i})
+ (-4 \dot{\Omega}^2 \Omega^{-3} + 2 \overset{..}{\Omega} \Omega^{-2} - 2 k \Omega^{-1})V_{i}\right]
=0,
\label{3.13z}
\end{align}
a relation that only involves vectors, with no scalars being present.

A second relation that we can obtain is given by noting that $\epsilon^{k ji}\tilde{\nabla}_{j}(\tilde{\nabla}_iX+X_i)=\epsilon^{k ji}\tilde{\nabla}_{j}X_i$. Thus we obtain 
\begin{eqnarray}
&&\epsilon^{kji}\tilde{\nabla}_j\frac{\eta}{2\Omega^2} \left[ (\tilde\nabla_b \tilde\nabla^b-\partial_{\tau}^2-2k)(\tilde\nabla_c \tilde\nabla^c+2k)\right] (B_i-\dot{E}_i)
\nonumber\\
&&-\epsilon^{kji}\tilde{\nabla}_j\left[k(B_i-\dot E_i)+ \frac{1}{2} \tilde{\nabla}_{a}\tilde{\nabla}^{a}(B_{i} - \dot{E}_{i})
+ (-4 \dot{\Omega}^2 \Omega^{-3} + 2 \overset{..}{\Omega} \Omega^{-2} - 2 k \Omega^{-1})V_{i}\right]=0.
\label{3.14z}
\end{eqnarray}
We thus have $(\tilde{\nabla}_k\tilde\nabla^k -2k)X_i=0$ and  $\epsilon^{k ji}\tilde{\nabla}_{j}X_i=0$, relations that can lead to $X_i=0$ or to $X_i=\tilde{\nabla}_i\chi$ where $\chi$ is a scalar that obeys $\tilde{\nabla}^iX_i=\tilde{\nabla}^i\tilde{\nabla}_i\chi=0$ since $X_i$ is transverse. (A transverse vector can be equal to the gradient of a scalar $\chi$ without being longitudinal if $\tilde{\nabla}^i\tilde{\nabla}_i\chi=0$.)
The $\tilde{\nabla}^i\tilde{\nabla}_i\chi=0$ condition  is consistent with  $(\tilde{\nabla}_k\tilde\nabla^k -2k)X_i=0$ when $X_i=\tilde{\nabla}_i\chi$ since $(\tilde{\nabla}_k\tilde\nabla^k -2k)X_i=(\tilde{\nabla}_k\tilde\nabla^k -2k)\tilde{\nabla}_i\chi= \tilde{\nabla}_i\tilde{\nabla}_k\tilde\nabla^k \chi=0$. Consequently, (\ref{3.13z}) and (\ref{3.14z}) are not independent, and we thus need more information in order to be able to solve for the vector sector.

This extra information comes from the $(i,j)$ sector. To discuss the $(i,j)$  sector  it is convenient to define 
\begin{eqnarray}
&&A=2 \dot{\Omega}^2 \Omega^{-2}(\alpha-\dot\gamma)-2  \dot{\Omega} \Omega^{-1}(\dot\alpha -\ddot\gamma)-4\ddot\Omega\Omega^{-1}(\alpha-\dot\gamma)+ \Omega^2 \delta \hat{p},\quad C=\alpha+2\dot{\Omega}\Omega^{-1}\gamma,
\nonumber\\
&&P= -\frac{1}{3 \Omega^2}  \tilde\nabla_a\tilde\nabla^a (\tilde\nabla_b \tilde\nabla^b +2k-\partial_{\tau}^2)\alpha,\quad
Q= \frac{1}{3 \Omega^2} (\tilde\nabla_a\tilde\nabla^a - 3\partial_{\tau}^2)\alpha,
\nonumber\\
&&\delta W_{ij}=\tilde{\gamma}_{ij}P+\tilde{\nabla}_i\tilde{\nabla}_jQ+X_{ij},\quad
\Delta_{ij}=\tilde{\gamma}_{ij}(A-\tilde{\nabla}_a\tilde{\nabla}^aC)+\tilde{\nabla}_i\tilde{\nabla}_jC+F_{ij},
\label{3.15z}
\end{eqnarray}
where $X_{ij}$ and $F_{ij}$ are everything other than the scalar part. With $A$, $C$, $P$ and $Q$ all being scalars we can set 
\begin{eqnarray}
&&\tilde{\nabla}^i[\tilde{\gamma}_{ij}(A-\tilde{\nabla}_a\tilde{\nabla}^aC)+\tilde{\nabla}_i\tilde{\nabla}_j C]
=\tilde{\nabla}_{j}( A+ 2 kC),
\nonumber\\
&&\tilde{\nabla}^i(\tilde{\gamma}_{ij}P+\tilde{\nabla}_i\tilde{\nabla}_jQ)=-\frac{2}{3\Omega^2}\tilde{\nabla}_{j}(\tilde{\nabla}_b\tilde{\nabla}^b+3k)\ddot{\alpha},
\label{3.16z}
\end{eqnarray}
and thus obtain
\begin{eqnarray}
(\tilde{\nabla}_a\tilde{\nabla}^a-2k)(\tilde{\nabla}_a\tilde{\nabla}^a+k)\tilde{\nabla}^i(\tilde{\gamma}_{ij}P+\tilde{\nabla}_i\tilde{\nabla}_jQ)&=&
-\frac{2}{3\Omega^2}\tilde{\nabla}_i\tilde{\nabla}_a\tilde{\nabla}^a(\tilde{\nabla}_b\tilde{\nabla}^b+3k)(\tilde{\nabla}_b\tilde{\nabla}^b+3k)\ddot{\alpha},
\nonumber\\
(\tilde{\nabla}_a\tilde{\nabla}^a-2k)(\tilde{\nabla}_a\tilde{\nabla}^a+k)\tilde{\nabla}^i[\tilde{\gamma}_{ij}(A-\tilde{\nabla}_a\tilde{\nabla}^aC)+\tilde{\nabla}_i\tilde{\nabla}_j C]&=&
\tilde{\nabla}_i\tilde{\nabla}_a\tilde{\nabla}^a(\tilde{\nabla}_b\tilde{\nabla}^b+3k)(A+2kC).
\label{3.17z}
\end{eqnarray}
Then with (\ref{3.7z}) and (\ref{3.16z}) giving
\begin{align}
&\tilde{\nabla}^i\tilde{\nabla}^j(\eta\delta W_{ij}-\Delta_{ij})
=-\frac{2\eta}{3 \Omega^{2}}\tilde{\nabla}_i\tilde{\nabla}^i (\tilde{\nabla}_a \tilde{\nabla}^a +3k)\ddot{\alpha}
-\tilde{\nabla}_{i}\tilde{\nabla}^i( A+2kC) =0,
\nonumber\\
&(\tilde{\nabla}_a\tilde{\nabla}^a-2k)\tilde\nabla^j\left[\eta(\tilde{\gamma}_{ij}P+\tilde{\nabla}_i\tilde{\nabla}_jQ)-\tilde{\gamma}_{ij}(A-\tilde{\nabla}_a\tilde{\nabla}^aC)-\tilde{\nabla}_i\tilde{\nabla}_jC\right]
\nonumber\\
&=(\tilde{\nabla}_a\tilde{\nabla}^a-2k)\tilde\nabla^j\left[-\frac{2\eta}{3\Omega^2}(\tilde{\nabla}_b\tilde{\nabla}^b+3k)\ddot{\alpha}-( A+ 2 kC)\right]=\tilde\nabla^j\tilde{\nabla}_a\tilde{\nabla}^a\left[-\frac{2\eta}{3\Omega^2}(\tilde{\nabla}_b\tilde{\nabla}^b+3k)\ddot{\alpha}-( A+ 2 kC)\right]=0,
\label{3.18z}
\end{align}
through use of (\ref{3.2z}) we obtain 
\begin{eqnarray}
&&(\tilde{\nabla}_a\tilde{\nabla}^a-2k)(\tilde{\nabla}_b\tilde{\nabla}^b+k)\tilde\nabla^j(\eta\delta W_{ij}-\Delta_{ij})
=(\tilde{\nabla}_a\tilde{\nabla}^a-2k)(\tilde{\nabla}_b\tilde{\nabla}^b+k)\tilde\nabla^j(\eta X_{ij}-F_{ij})
\nonumber\\
&&=\frac{\eta}{2\Omega^2}(\tilde{\nabla}_a\tilde{\nabla}^a-2k)(\tilde{\nabla}_b\tilde{\nabla}^b+k) (\tilde{\nabla}_{c}\tilde{\nabla}^{c}+2k)(\tilde\nabla_a\tilde\nabla^a -2k-\partial_{\tau}^2) (\dot{B}_i-\ddot{E}_i)
\nonumber\\
&&-(\tilde{\nabla}_a\tilde{\nabla}^a-2k)(\tilde{\nabla}_b\tilde{\nabla}^b+k)(\tilde{\nabla}_{c}\tilde{\nabla}^{c}+2k)\left[\frac{1}{2}(\dot{B}_i-\ddot{E}_i)+\dot{\Omega}\Omega^{-1}(B_i-\dot{E}_i)\right]=0,
\label{3.19z}
\end{eqnarray}
a relation that only involves $B_i-\dot{E}_{i}$. From (\ref{3.19z}) we can determine the two components of the transverse $B_i-\dot{E}_i$, and then use (\ref{3.13z}) to determine the two components of the transverse $V_i$. In the vector sector then we have just the right number of equations needed to fix all of the vector sector degrees of freedom.

\subsection{The Tensor Sector}
\label{S3c}

For the tensor sector the gauge invariant  $E_{ij}$ appears in the $(i,j)$ sector. In the tensor sector it is convenient to define 
\begin{eqnarray}
R_{ij}=\tilde{\gamma}_{ij}P+\tilde{\nabla}_i\tilde{\nabla}_jQ,\quad
D_{ij}=\tilde{\gamma}_{ij}(A-\tilde{\nabla}_a\tilde{\nabla}^aC)+\tilde{\nabla}_i\tilde{\nabla}_jC.
\label{3.20z}
\end{eqnarray}
We next introduce
\begin{eqnarray}
S_{ij}&=&R_{ij}-\frac{1}{3}\tilde{\gamma}_{ij} \tilde{\gamma}^{ab}R_{ab}=(\tilde{\nabla}_i\tilde{\nabla}_j-\frac{1}{3} \tilde{\gamma}_{ij}\tilde{\nabla}_a\tilde{\nabla}^a)Q,
\nonumber\\
A_{ij}&=&D_{ij}-\frac{1}{3}\tilde{\gamma}_{ij} \tilde{\gamma}^{ab}D_{ab}=(\tilde{\nabla}_i\tilde{\nabla}_j-\frac{1}{3} \tilde{\gamma}_{ij}\tilde{\nabla}_a\tilde{\nabla}^a)C.
\label{3.21z}
\end{eqnarray}
Then with 
\begin{eqnarray}
&&\delta W_{ij}=\tilde{\gamma}_{ij}P+\tilde{\nabla}_i\tilde{\nabla}_jQ+X_{ij},\quad
\Delta_{ij}=\tilde{\gamma}_{ij}(A-\tilde{\nabla}_a\tilde{\nabla}^aC)+\tilde{\nabla}_i\tilde{\nabla}_jC+F_{ij},\quad \tilde{\gamma}^{ij}X_{ij}=0, \quad \tilde{\gamma}^{ij}F_{ij}=0,
\label{3.22y}
\end{eqnarray}
we can set
\begin{eqnarray}
\delta W_{ij}-\frac{1}{3}\tilde{\gamma}_{ij} \tilde{\gamma}^{ab}\delta W_{ab}&=&(\tilde{\nabla}_i\tilde{\nabla}_j-\frac{1}{3} \tilde{\gamma}_{ij}\tilde{\nabla}_a\tilde{\nabla}^a)Q+X_{ij}=S_{ij}+X_{ij},
\nonumber\\
\Delta_{ij}-\frac{1}{3}\tilde{\gamma}_{ij} \tilde{\gamma}^{ab}\Delta_{ab}&=&(\tilde{\nabla}_i\tilde{\nabla}_j-\frac{1}{3} \tilde{\gamma}_{ij}\tilde{\nabla}_a\tilde{\nabla}^a)C +F_{ij}=A_{ij}+F_{ij},
\nonumber\\
\eta \delta W_{ij}-\Delta_{ij}-\frac{1}{3}\tilde{\gamma}_{ij} \tilde{\gamma}^{ab}(\eta\delta W_{ab}-\Delta_{ab})&=&
(\tilde{\nabla}_i\tilde{\nabla}_j-\frac{1}{3} \tilde{\gamma}_{ij}\tilde{\nabla}_a\tilde{\nabla}^a)(\eta Q-C) +\eta X_{ij}-F_{ij}.
\label{3.23y}
\end{eqnarray}
With repeated use of the second relation in (\ref{3.1z}) we obtain
\begin{eqnarray}
(\tilde{\nabla}_b\tilde{\nabla}^b-3k)S_{ij}&=&
(\tilde{\nabla}_i\tilde{\nabla}_j-\frac{1}{3} \tilde{\gamma}_{ij}\tilde{\nabla}_a\tilde{\nabla}^a)(\tilde{\nabla}_c\tilde{\nabla}^c+3k)Q,
\nonumber\\
(\tilde{\nabla}_b\tilde{\nabla}^b-3k)A_{ij}&=&
(\tilde{\nabla}_i\tilde{\nabla}_j-\frac{1}{3} \tilde{\gamma}_{ij}\tilde{\nabla}_a\tilde{\nabla}^a)(\tilde{\nabla}_c\tilde{\nabla}^c+3k)C,
\nonumber\\
(\tilde{\nabla}_a\tilde{\nabla}^a-6k)(\tilde{\nabla}_b\tilde{\nabla}^b-3k)S_{ij}&=&
(\tilde{\nabla}_i\tilde{\nabla}_j-\frac{1}{3} \tilde{\gamma}_{ij}\tilde{\nabla}_a\tilde{\nabla}^a)\tilde{\nabla}_b\tilde{\nabla}^b(\tilde{\nabla}_c\tilde{\nabla}^c+3k)Q,
\nonumber\\
(\tilde{\nabla}_a\tilde{\nabla}^a-6k)(\tilde{\nabla}_b\tilde{\nabla}^b-3k)A_{ij}&=&
(\tilde{\nabla}_i\tilde{\nabla}_j-\frac{1}{3} \tilde{\gamma}_{ij}\tilde{\nabla}_a\tilde{\nabla}^a)\tilde{\nabla}_b\tilde{\nabla}^b(\tilde{\nabla}_c\tilde{\nabla}^c+3k)C.
\label{3.24y}
\end{eqnarray}
Through use of (\ref{3.24y}) we obtain
\begin{align}
&(\tilde{\nabla}_a\tilde{\nabla}^a-6k)(\tilde{\nabla}_b\tilde{\nabla}^b-3k)\left(\eta\delta W_{ij}-\Delta_{ij}-\frac{1}{3}\tilde{\gamma}_{ij}\tilde{\gamma}^{ab}(\eta\delta W_{ab}-\Delta_{ab})\right)
\nonumber\\
&=\left(\tilde{\nabla}_i\tilde{\nabla}_j-\frac{1}{3} \tilde{\gamma}_{ij}\tilde{\nabla}_a\tilde{\nabla}^a\right)\tilde{\nabla}_i\tilde{\nabla}^i(\tilde{\nabla}_j\tilde{\nabla}^j+3k)(\eta Q-C)+(\tilde{\nabla}_a\tilde{\nabla}^a-6k)(\tilde{\nabla}_b\tilde{\nabla}^b-3k)(\eta X_{ij}-F_{ij}).
\label{3.25y}
\end{align}
On rewriting (\ref{3.9z}) as
\begin{align}
&3\tilde{\nabla}^i\tilde{\nabla}^j(\eta\delta W_{ij}-\Delta_{ij})-\tilde{\nabla}_a\tilde{\nabla}^a\tilde{\gamma}^{ij}(\eta\delta W_{ij}-\Delta_{ij})=2\tilde{\nabla}_b\tilde{\nabla}^b(\tilde{\nabla}_c\tilde{\nabla}^c+3k)(\eta Q-C)=0,
\label{3.26y}
\end{align}
it then follows that the scalar sector drops out of (\ref{3.25y}), to leave us with
\begin{align}
&(\tilde{\nabla}_a\tilde{\nabla}^a-6k)(\tilde{\nabla}_b\tilde{\nabla}^b-3k)\left(\eta\delta W_{ij}-\Delta_{ij}-\frac{1}{3}\tilde{\gamma}^{ij}\tilde{\gamma}^{ab}(\eta\delta W_{ab}-\Delta_{ab})\right)
=(\tilde{\nabla}_a\tilde{\nabla}^a-6k)(\tilde{\nabla}_b\tilde{\nabla}^b-3k)(\eta X_{ij}-F_{ij})=0,
\label{3.27y}
\end{align}
with the right-hand side of (\ref{3.27y}) only involving the vector and tensor sectors.

To eliminate the vector sector we now note that for any vector $A_i$ that obeys $\tilde{\nabla}^iA_i=0$, through repeated use of the first relation in (\ref{3.2z}) we obtain 
\begin{align}
(\tilde{\nabla}_b\tilde{\nabla}^b-3k)(\tilde{\nabla}_iA_j+\tilde{\nabla}_jA_i)&=
\tilde{\nabla}_i(\tilde{\nabla}_b\tilde{\nabla}^b+k)A_j+
\tilde{\nabla}_j(\tilde{\nabla}_b\tilde{\nabla}^b+k)A_i,
\nonumber\\
(\tilde{\nabla}_a\tilde{\nabla}^a-6k)(\tilde{\nabla}_b\tilde{\nabla}^b-3k)(\tilde{\nabla}_iA_j+\tilde{\nabla}_jA_i)
&=\tilde{\nabla}_i(\tilde{\nabla}_a\tilde{\nabla}^a-2k)(\tilde{\nabla}_b\tilde{\nabla}^b+k)A_j+
\tilde{\nabla}_j(\tilde{\nabla}_a\tilde{\nabla}^a-2k)(\tilde{\nabla}_b\tilde{\nabla}^b+k)A_i.
\label{3.28y}
\end{align}
On using the first relation in (\ref{3.2z}) again,  it follows that 
\begin{eqnarray}
&&(\tilde{\nabla}_c\tilde{\nabla}^c-2k)(\tilde{\nabla}_a\tilde{\nabla}^a-6k)(\tilde{\nabla}_b\tilde{\nabla}^b-3k)(\tilde{\nabla}_iA_j+\tilde{\nabla}_jA_i)
\nonumber\\
&=&\tilde{\nabla}_i(\tilde{\nabla}_c\tilde{\nabla}^c+2k)(\tilde{\nabla}_a\tilde{\nabla}^a-2k)(\tilde{\nabla}_b\tilde{\nabla}^b+k)A_j+
\tilde{\nabla}_j(\tilde{\nabla}_c\tilde{\nabla}^c+2k)(\tilde{\nabla}_a\tilde{\nabla}^a-2k)(\tilde{\nabla}_b\tilde{\nabla}^b+k)A_i.
\label{3.29y}
\end{eqnarray}
On recognizing that the vector sectors of $X_{ij}$ and $F_{ij}$ are precisely of the form $\tilde{\nabla}_iA_j+\tilde{\nabla}_jA_i$, using  (\ref{3.19z}) we can eliminate the dependence on the vector sector by applying $\tilde{\nabla}_c\tilde{\nabla}^c-2k$ to (\ref{3.27y}). Then finally from (\ref{3.27y}) we obtain
\begin{eqnarray}
&&(\tilde{\nabla}_c\tilde{\nabla}^c-2k)(\tilde{\nabla}_a\tilde{\nabla}^a-6k)(\tilde{\nabla}_b\tilde{\nabla}^b-3k)
\left(\eta\delta W_{ij}-\Delta_{ij}-\frac{1}{3}\tilde{\gamma}^{ij}\tilde{\gamma}^{ab}(\eta\delta W_{ab}-\Delta_{ab})\right)
\nonumber\\
&&=(\tilde{\nabla}_c\tilde{\nabla}^c-2k)(\tilde{\nabla}_a\tilde{\nabla}^a-6k)(\tilde{\nabla}_b\tilde{\nabla}^b-3k)
\nonumber\\
&&\times
\bigg{[}\frac{\eta}{\Omega^2}\left[ (\tilde\nabla_b \tilde\nabla^b-\partial_{\tau}^2-2k)^2+4k\partial_{\tau}^2 \right] E_{ij}
+ \overset{..}{E}_{ij} +2 k E_{ij} +2  \dot{\Omega} \Omega^{-1}\dot{E}_{ij} - \tilde{\nabla}_{d}\tilde{\nabla}^{d}E_{ij}\bigg{]}=0.
\label{3.30y}
\end{eqnarray}
We thus obtain a relation that only involves $E_{ij}$. Since there is no tensor part to the matter fluctuation,  (\ref{3.30y}) is all we need for the two degree of freedom tensor sector. Having now decoupled the scalar, vector and tensor sectors from each other in a set of gauge invariant equations that are  completely exact, that hold for arbitrary $a(t)$, arbitrary $k$ and arbitrary background matter sources, we can now proceed to solve them. For solutions we shall focus on the $k<0$ cosmology that is phenomenologically preferred in the conformal case, though a similar analysis could be made for $k=0$ or $k>0$ conformal cosmologies if desired.

\section{Solving the Fluctuation Equations}
\label{S4}

In all there are ten independent fluctuation equations, four for the scalars [(\ref{3.4z}), (\ref{3.6z}), (\ref{3.9z}), (\ref{3.10z})], two for the vectors [(\ref{3.13z}), (\ref{3.19z})], and one for the tensor [(\ref{3.30y})]. All of these equations  have in common the appearance of the spatial derivative operator $\tilde{\nabla}_i\tilde{\nabla}^i$. If we for instance consider (\ref{3.30y}), we could satisfy it by $(\tilde{\nabla}_c\tilde{\nabla}^c-2k)T_{ij}=0$, by $(\tilde{\nabla}_a\tilde{\nabla}^a-6k)T_{ij}=0$, by $(\tilde{\nabla}_b\tilde{\nabla}^b-3k)T_{ij}=0$, or by having the term in brackets vanish. (Here $T_{ij}$ represents the entire $E_{ij}$-dependent term that appears in the term in brackets in (\ref{3.30y}).) Ignoring the possibility that $T_{ij}$ itself vanishes for  the moment (the only possibility that involves both  spatial and temporal derivatives), we would have to solve the generic $(\tilde{\nabla}_d\tilde{\nabla}^d+A_T)T_{ij}=0$ where $A_T$ is an appropriate separation constant. Analogously, there will be separation constants  $A_S$ and $A_V$ in the scalar and vector cases. (While for the moment $A_S$, $A_V$ and $A_T$ are just appropriate constants, we designate them as separation constants since in Sec. \ref{S6} they will serve as such for wave equations that contain a time dependence.) Having such a class of solutions in which $T_{ij}$ and its scalar and vector analogs do not vanish would not be even remotely desirable for fluctuation theory since only the spatial behavior of the fluctuations would be specified and nothing would then fix the time behavior. However,  by solving these $(\tilde{\nabla}_d\tilde{\nabla}^d+A_T)T_{ij}=0$ type equations and finding the eigenmodes of the $\tilde{\nabla}_d\tilde{\nabla}^d$ operator, it was shown in  \cite{Phelps2019} in the analog  Einstein gravity fluctuation case that the way that the various scalar, vector and tensor components would need to interplay with each other in the fluctuation equations is actually excluded, as the requisite interplay is not compatible with the boundary conditions at $\chi=\infty$ and $\chi=0$. Thus in the Einstein gravity case we have to solve the fluctuation equations by having the $\eta=0$ limit of the term in brackets in (\ref{3.30y}) and its scalar and vector analogs vanish, and then we are able to fix the time dependence of the fluctuations. Moreover, this also enabled us to show in \cite{Phelps2019} that the so-called decomposition theorem (viz. that the scalar, vector and tensor components separately solve the $\Delta_{\mu\nu}=0$ fluctuation equations) holds for standard Einstein cosmological fluctuation theory. 

We now apply this same analysis to conformal gravity. This will enable us to show that the decomposition theorem also holds in the conformal gravity case, and we discuss this point below. Moreover, even if we solve equations such (\ref{3.30y}) by having the term in brackets vanish we would still need to have to find the eigenmodes of the $\tilde{\nabla}_d\tilde{\nabla}^d$ operator as it appears in the bracketed term. Thus we proceed first to a study of the $\tilde{\nabla}_d\tilde{\nabla}^d$ operator, and we follow the technique developed for it in \cite{Phelps2019}.

\subsection{Scalar Fluctuations}
\label{S4a}

To study the eigenmodes of the $\tilde{\nabla}_d\tilde{\nabla}^d$ operator in the $k<0$ case of interest to us here,  we have found it convenient to set $k=-1/L^2$,  $r/L=\sinh\chi$, $p=\tau/L$, so that the background metric takes the form
\begin{eqnarray}
ds^2=L^2\Omega^2(p)\left[ dp^2-d\chi^2-\sinh^2\chi d\theta^2-\sinh^2\chi\sin^2\theta d\phi^2\right].
\label{4.1z}
\end{eqnarray}
In all allowable solutions we will require the fluctuations to be well behaved and not diverge anywhere. We shall thus require the fluctuations to go to zero at $\chi=\infty$ and to a finite value at $\chi=0$.  

For the scalar case we need to solve
\begin{eqnarray}
\left(\tilde{\nabla}_a\tilde{\nabla}^a+\frac{A_S}{L^2}\right)S=0
\label{4.2z}
\end{eqnarray}
for a generic scalar function $S$. On setting $S(\chi,\theta,\phi)=S_{\ell}(\chi)Y^m_{\ell}(\theta,\phi)$ (\ref{4.2z}) reduces to 
\begin{eqnarray}
 \frac{1}{L^2}\left[\frac{d^2}{d\chi^2}+2\frac{\cosh\chi }{\sinh\chi}\frac{d }{ d\chi}
-\frac{\ell(\ell+1)}{ \sinh^2\chi}+A_S\right]S_{\ell}=0.
\label{4.3z}
\end{eqnarray}
In the  $\chi\rightarrow \infty$ and $\chi\rightarrow 0$ limits  we take the solution to behave as $e^{\lambda \chi}$ (times an irrelevant polynomial in $\chi$), and as $\chi^n$, to thus obtain
\begin{eqnarray}
&&\lambda^2+2\lambda+A_S=0,\quad \lambda=-1\pm(1-A_S)^{1/2},
\nonumber\\
&&n(n-1)+2n-\ell(\ell+1)=0,\quad n=\ell,-\ell-1.
\label{4.4z}
\end{eqnarray}
Asymptotic convergence will thus depend on $A_S$, while finiteness at $\chi=0$ will depend on $\ell$. With there being two possible values for $\lambda$ there will be two families of solutions, which we will label $\hat{S}_{\ell}^{(1)}$ and $\hat{S}_{\ell}^{(2)}$ in the following.

Exact solutions to (\ref{4.3z}) exist in the literature (see e.g.  \cite{Bander1966,Mannheim1988,Phelps2019}). They  are known as associated Legendre functions and are of the form 
\begin{eqnarray}
S_{\ell}=\sinh^{\ell}\chi\left(\frac{1}{ \sinh\chi} \frac{d }{ d\chi}\right)^{\ell+1}f(\chi),
\label{4.5z}
\end{eqnarray}
where $f(\chi)$ obeys
\begin{eqnarray}
\left[\frac{d^3}{d\chi^3}+\nu^2\frac{d}{d\chi}\right]f(\chi)=0,\quad \nu^2=A_S-1,
\label{4.6z}
\end{eqnarray}
with $f(\chi)$ thus obeying 
\begin{eqnarray}
f(\nu^2>0)=\cos\nu\chi,~\sin\nu\chi,\quad f(\nu^2=-\mu^2<0)=\cosh\mu\chi,~\sinh\mu\chi,\quad f(\nu^2=0)=\chi,~\chi^2.
\label{4.7z}
\end{eqnarray}
Both $\nu$ and $\mu$ are continuous variables, with the class of all $\nu\geq 0$ and the class of all $\mu\geq 0$ both being complete. For each $f(\chi)$ (\ref{4.7z})  would lead to solutions of the form
\begin{eqnarray}
\hat{S}_0=\frac{1}{\sinh\chi}\frac{df}{d\chi},\quad \hat{S}_1=\frac{d\hat{S}_0}{d\chi},\quad \hat{S}_2=\sinh\chi\frac{d}{d\chi}\left[\frac{\hat{S}_1}{\sinh\chi}\right],\quad \hat{S}_3=\sinh^2\chi\frac{d}{d\chi}\left[\frac{\hat{S}_2}{\sinh^2\chi}\right],.....
\label{4.8z}
\end{eqnarray}
However, on evaluating these expressions it can happen that some of these solutions vanish. Thus for $A_S=0$ for instance where $f(\chi)=(\sinh\chi,\cosh\chi)$ the two solutions with $\ell=0$ are $\cosh\chi/\sinh\chi$ and $1$. However this would lead to the two solutions with $\ell=1$ being $1/\sinh^2\chi$ and $0$. To address this point we note that suppose we have obtained some non-zero solution $\hat{S}_{\ell}$. Then, a second solution of the form $\hat{f}_{\ell}(\chi)\hat{S}_{\ell}(\chi)$ may be found by inserting $\hat{f}_{\ell}(\chi)\hat{S}_{\ell}(\chi)$ into (\ref{4.3z}), to yield
\begin{eqnarray}
\hat{S}_{\ell}\frac{d^2 \hat{f}_{\ell}}{ d\chi^2}+2\hat{S}_{\ell}\frac{\cosh\chi }{ \sinh\chi}\frac{d \hat{f}_{\ell}}{ d\chi}+2\frac{d \hat{S}_{\ell}}{ d\chi}\frac{d \hat{f}_{\ell}}{ d\chi}=0,
\label{4.9z}
\end{eqnarray}
which integrates to
\begin{eqnarray}
\frac{d \hat{f}_{\ell}}{ d\chi}=\frac{1}{\sinh^2\chi\hat{S}_{\ell}^2},~~~~~\hat{f}_{\ell}\hat{S}_{\ell}=\hat{S}_{\ell}\int \frac{d\chi }{\sinh^2\chi\hat{S}_{\ell}^2}.
\label{4.10z}
\end{eqnarray}
Thus for $\ell=1$, from the non-trivial $A_S=0$ solution $\hat{S}_{1}=1/\sinh^2\chi$ we obtain a second solution of the form $\hat{f}_{\ell}\hat{S}_{\ell}=\cosh\chi/\sinh\chi-\chi/\sinh^2\chi$. However, once we have this second solution we can then return to (\ref{4.8z}) and use it to obtain the subsequent solutions associated with higher $\ell$ values, since use of the chain in (\ref{4.8z}) only requires that at any point the elements in it are solutions regardless of how they may or may not have been found.

\subsection{Vector Fluctuations}
\label{S4b}

In the vector sector the components of $V_i$ obey the transverseness condition
\begin{eqnarray}
\tilde\nabla_a V^a&=& \frac{V_{2} \cos\theta}{\sin\theta \sinh^2\chi} + \frac{2 V_{1} \cosh\chi}{\sinh\chi} + \partial_{1}V_{1} + \frac{\partial_{2}V_{2}}{\sinh^2\chi} + \frac{\partial_{3}V_{3}}{\sin^2\theta \sinh^2\chi}=0.
\label{4.11z}
\end{eqnarray}
On implementing this condition, the $(\chi,\theta,\phi) \equiv (1,2,3)$ components of $\tilde{\nabla}_a\tilde{\nabla}^aV^i$ take the form
\begin{eqnarray}
\tilde{\nabla}_a\tilde{\nabla}^aV^1&=&V_{1} \left(2 + \frac{2}{\sinh^2\chi}\right) + \frac{4 \cosh\chi \partial_{1}V_{1}}{\sinh\chi} + \partial_{1}\partial_{1}V_{1} + \frac{\cos\theta \partial_{2}V_{1}}{\sin\theta \sinh^2\chi} + \frac{\partial_{2}\partial_{2}V_{1}}{\sinh^2\chi} + \frac{\partial_{3}\partial_{3}V_{1}}{\sin^2\theta \sinh^2\chi},
 \nonumber\\ 
\tilde{\nabla}_a\tilde{\nabla}^aV^2&=& V_{2} \left(- \frac{2}{\sinh^4\chi} + \frac{1}{\sin^2\theta \sinh^4\chi} -  \frac{2}{\sinh^2\chi}\right) + \frac{4 V_{1} \cos\theta \cosh\chi}{\sin\theta \sinh^3\chi} + \frac{2 \cos\theta \partial_{1}V_{1}}{\sin\theta \sinh^2\chi} + \frac{\partial_{1}\partial_{1}V_{2}}{\sinh^2\chi} \nonumber \\ 
&& + \frac{2 \cosh\chi \partial_{2}V_{1}}{\sinh^3\chi} + \frac{3 \cos\theta \partial_{2}V_{2}}{\sin\theta \sinh^4\chi} + \frac{\partial_{2}\partial_{2}V_{2}}{\sinh^4\chi} + \frac{\partial_{3}\partial_{3}V_{2}}{\sin^2\theta \sinh^4\chi},
\nonumber\\ 
\tilde{\nabla}_a\tilde{\nabla}^aV^3&=& - \frac{2 V_{3}}{\sin^2\theta \sinh^2\chi} + \frac{\partial_{1}\partial_{1}V_{3}}{\sin^2\theta \sinh^2\chi} -  \frac{\cos\theta \partial_{2}V_{3}}{\sin^3\theta \sinh^4\chi} + \frac{\partial_{2}\partial_{2}V_{3}}{\sin^2\theta \sinh^4\chi} + \frac{2 \cosh\chi \partial_{3}V_{1}}{\sin^2\theta \sinh^3\chi} \nonumber \\ 
&& + \frac{2 \cos\theta \partial_{3}V_{2}}{\sin^3\theta \sinh^4\chi} + \frac{\partial_{3}\partial_{3}V_{3}}{\sin^4\theta \sinh^4\chi}.
\label{4.12z}
\end{eqnarray}

To explore the structure of the $k=-1/L^2$ vector sector we seek solutions to
\begin{eqnarray}
\left(\tilde{\nabla}_a\tilde{\nabla}^a+\frac{A_V}{L^2}\right)V_i=0
\label{4.13z}
\end{eqnarray}
for a generic $V_i$. Conveniently, we find that the equation for $V_1$ involves no mixing with $V_2$ or $V_3$, and can thus be solved directly. On setting $V_1(\chi,\theta,\phi)=g_{1,\ell}(\chi)Y_{\ell}^m(\theta,\phi)$, the equation for $V_1$ reduces to 
\begin{eqnarray}
\frac{1}{L^2}\left[\frac{d^2}{d\chi^2}+4\frac{\cosh\chi}{ \sinh\chi}\frac{d }{d\chi}
+2+A_V+\frac{2 }{ \sinh^2\chi}-\frac{\ell(\ell+1)}{ \sinh^2\chi}\right]g_{1,\ell}=0.
\label{4.14z}
\end{eqnarray}
The $\chi \rightarrow \infty$ and $\chi \rightarrow 0$ limits give
\begin{eqnarray}
&&\lambda^2+4\lambda+2+A_V=0,\quad\lambda=-2\pm (2-A_V)^{1/2},
\nonumber\\
&&n(n-1)+4n+2-\ell(\ell+1)=0,\quad n=\ell-1, -\ell-2.
\label{4.15z}
\end{eqnarray}
Asymptotic convergence will thus depend on $A_V$, while finiteness at $\chi=0$ will depend on $\ell$. However the conditions differ from the scalar ones, a point that will prove crucial below in establishing the conformal gravity decomposition theorem. With there being two possible values for $\lambda$ there will be two families of solutions, which we will label $\hat{V}_{\ell}^{(1)}$ and $\hat{V}_{\ell}^{(2)}$ in the following.

To solve (\ref{4.14z}) we set  $g_{1,\ell}=\alpha_{\ell}/\sinh\chi$, and  find that (\ref{4.14z}) takes the form
\begin{eqnarray}
\frac{1}{L^2}\left[\frac{d^2 }{d\chi^2}+2\frac{\cosh\chi}{ \sinh\chi}\frac{d }{d\chi}
-\frac{\ell(\ell+1) }{\sinh^2\chi}+A_V-1\right]\alpha_{\ell}=0.
\label{4.16z}
\end{eqnarray}
We recognize (\ref{4.16z}) as being in the form given in  (\ref{4.3z}), which we discussed above, with $\nu^2=A_V-2$.

\subsection{Tensor Fluctuations}
\label{S4c}

For $k=-1/L^2$ the transverse-traceless tensor sector modes need to satisfy 
\begin{eqnarray}
 \tilde{\gamma}^{ab}T_{ab}&=& T_{11} + \frac{T_{22}}{\sinh^2\chi} + \frac{T_{33}}{\sin^2\theta \sinh^2\chi} =0,
\nonumber\\
\tilde\nabla_a T^{a 1}&=& - \frac{\cosh\chi T_{22}}{\sinh^3\chi} -  \frac{\cosh\chi T_{33}}{\sin^2\theta \sinh^3\chi} + \frac{\cos\theta T_{12}}{\sin\theta \sinh^2\chi} + \frac{2 \cosh\chi T_{11}}{\sinh\chi} + \partial_{1}T_{11} + \frac{\partial_{2}T_{12}}{\sinh^2\chi} \nonumber \\ 
&& + \frac{\partial_{3}T_{13}}{\sin^2\theta \sinh^2\chi}=0, \nonumber\\
\tilde\nabla_a T^{a 2}&=& - \frac{\cos\theta T_{33}}{\sin^3\theta \sinh^4\chi} + \frac{\cos\theta T_{22}}{\sin\theta \sinh^4\chi} + \frac{2 \cosh\chi T_{12}}{\sinh^3\chi} + \frac{\partial_{1}T_{12}}{\sinh^2\chi} + \frac{\partial_{2}T_{22}}{\sinh^4\chi} + \frac{\partial_{3}T_{23}}{\sin^2\theta \sinh^4\chi}=0,
\nonumber\\
\tilde\nabla_a T^{a 3}&=& \frac{\cos\theta T_{23}}{\sin^3\theta \sinh^4\chi} + \frac{2 \cosh\chi T_{13}}{\sin^2\theta \sinh^3\chi} + \frac{\partial_{1}T_{13}}{\sin^2\theta \sinh^2\chi} + \frac{\partial_{2}T_{23}}{\sin^2\theta \sinh^4\chi} + \frac{\partial_{3}T_{33}}{\sin^4\theta \sinh^4\chi}=0.
\label{4.17z}
\end{eqnarray}
Under these conditions the components of $\tilde{\nabla}_a\tilde{\nabla}^aT^{ij}$ evaluate to
\begin{eqnarray}
\tilde{\nabla}_a\tilde{\nabla}^aT^{11}&=& T_{11} \left(6 + \frac{6}{\sinh^2\chi}\right) + \frac{6 \cosh\chi \partial_{1}T_{11}}{\sinh\chi} + \partial_{1}\partial_{1}T_{11} + \frac{\cos\theta \partial_{2}T_{11}}{\sin\theta \sinh^2\chi} + \frac{\partial_{2}\partial_{2}T_{11}}{\sinh^2\chi} + \frac{\partial_{3}\partial_{3}T_{11}}{\sin^2\theta \sinh^2\chi},
 \nonumber\\ 
\tilde{\nabla}_a\tilde{\nabla}^aT^{22}&=& \frac{4 T_{22}}{\sinh^6\chi} -  \frac{4 T_{22}}{\sin^2\theta \sinh^6\chi} + \frac{4 T_{11}}{\sinh^4\chi} -  \frac{2 T_{22}}{\sinh^4\chi} -  \frac{2 T_{11}}{\sin^2\theta \sinh^4\chi} + \frac{2 T_{11}}{\sinh^2\chi} -  \frac{2 \cosh\chi \partial_{1}T_{22}}{\sinh^5\chi} \nonumber \\ 
&& + \frac{\partial_{1}\partial_{1}T_{22}}{\sinh^4\chi} + \frac{4 \cosh\chi \partial_{2}T_{12}}{\sinh^5\chi} + \frac{\cos\theta \partial_{2}T_{22}}{\sin\theta \sinh^6\chi} + \frac{\partial_{2}\partial_{2}T_{22}}{\sinh^6\chi} -  \frac{4 \cos\theta \partial_{3}T_{23}}{\sin^3\theta \sinh^6\chi} + \frac{\partial_{3}\partial_{3}T_{22}}{\sin^2\theta \sinh^6\chi},
 \nonumber\\ 
\tilde{\nabla}_a\tilde{\nabla}^aT^{33}&=& \frac{2T_{33}} {\sin^4\theta\sinh^6\chi}\left(1-{\sinh^2\chi}\right) + T_{11} \left(\frac{2}{\sin^4\theta \sinh^4\chi} + \frac{2}{\sin^2\theta \sinh^2\chi}\right) -  \frac{4 \cos\theta \cosh\chi T_{12}}{\sin^3\theta \sinh^5\chi} 
\nonumber \\ 
&& -  \frac{4 \cos\theta \partial_{1}T_{12}}{\sin^3\theta \sinh^4\chi} -  \frac{2 \cosh\chi \partial_{1}T_{33}}{\sin^4\theta \sinh^5\chi} + \frac{\partial_{1}\partial_{1}T_{33}}{\sin^4\theta \sinh^4\chi} + \frac{4 \cos\theta \partial_{2}T_{11}}{\sin^3\theta \sinh^4\chi} + \frac{\cos\theta \partial_{2}T_{33}}{\sin^5\theta \sinh^6\chi} \nonumber \\ 
&& + \frac{\partial_{2}\partial_{2}T_{33}}{\sin^4\theta \sinh^6\chi} + \frac{4 \cosh\chi \partial_{3}T_{13}}{\sin^4\theta \sinh^5\chi} + \frac{\partial_{3}\partial_{3}T_{33}}{\sin^6\theta \sinh^6\chi},
\nonumber\\ 
\tilde{\nabla}_a\tilde{\nabla}^aT^{12}&=& T_{12} \left(- \frac{1}{\sin^2\theta \sinh^4\chi} -  \frac{2}{\sinh^2\chi}\right) + \frac{2 \cosh\chi \partial_{1}T_{12}}{\sinh^3\chi} + \frac{\partial_{1}\partial_{1}T_{12}}{\sinh^2\chi} + \frac{2 \cosh\chi \partial_{2}T_{11}}{\sinh^3\chi} \nonumber \\ 
&& + \frac{\cos\theta \partial_{2}T_{12}}{\sin\theta \sinh^4\chi} + \frac{\partial_{2}\partial_{2}T_{12}}{\sinh^4\chi} -  \frac{2 \cos\theta \partial_{3}T_{13}}{\sin^3\theta \sinh^4\chi} + \frac{\partial_{3}\partial_{3}T_{12}}{\sin^2\theta \sinh^4\chi},
 \nonumber\\ 
\tilde{\nabla}_a\tilde{\nabla}^aT^{13}&=& - \frac{2 T_{13}}{\sin^2\theta \sinh^2\chi} + \frac{2 \cosh\chi \partial_{1}T_{13}}{\sin^2\theta \sinh^3\chi} + \frac{\partial_{1}\partial_{1}T_{13}}{\sin^2\theta \sinh^2\chi} -  \frac{\cos\theta \partial_{2}T_{13}}{\sin^3\theta \sinh^4\chi} + \frac{\partial_{2}\partial_{2}T_{13}}{\sin^2\theta \sinh^4\chi} \nonumber \\ 
&& + \frac{2 \cosh\chi \partial_{3}T_{11}}{\sin^2\theta \sinh^3\chi} + \frac{2 \cos\theta \partial_{3}T_{12}}{\sin^3\theta \sinh^4\chi} + \frac{\partial_{3}\partial_{3}T_{13}}{\sin^4\theta \sinh^4\chi},
 \nonumber\\ 
\tilde{\nabla}_a\tilde{\nabla}^aT^{23}&=& T_{23} \left(\frac{2(1-\sinh^2\chi)}{\sin^2\theta\sinh^6\chi} -  \frac{1}{\sin^4\theta \sinh^6\chi}\right) + \frac{2 \cos\theta \partial_{1}T_{13}}{\sin^3\theta \sinh^4\chi} -  \frac{2 \cosh\chi \partial_{1}T_{23}}{\sin^2\theta \sinh^5\chi} + \frac{\partial_{1}\partial_{1}T_{23}}{\sin^2\theta \sinh^4\chi}\nonumber \\ 
&& + \frac{2 \cosh\chi \partial_{2}T_{13}}{\sin^2\theta \sinh^5\chi} + \frac{\cos\theta \partial_{2}T_{23}}{\sin^3\theta \sinh^6\chi} + \frac{\partial_{2}\partial_{2}T_{23}}{\sin^2\theta \sinh^6\chi} + \frac{2 \cosh\chi \partial_{3}T_{12}}{\sin^2\theta \sinh^5\chi} + \frac{2 \cos\theta \partial_{3}T_{22}}{\sin^3\theta \sinh^6\chi} 
\nonumber \\ 
&& + \frac{\partial_{3}\partial_{3}T_{23}}{\sin^4\theta \sinh^6\chi}.
\label{4.18z}
\end{eqnarray}

Following our analysis of the vector sector, in the $k=-1/L^2$ tensor sector we seek solutions to
\begin{eqnarray}
\left(\tilde{\nabla}_a\tilde{\nabla}^a+\frac{A_T}{L^2}\right)T_{ij}=0
\label{4.19z}
\end{eqnarray}
for a generic tensor $T_{ij}$. Conveniently, we find that the equation for $T_{11}$ involves no mixing with any other components of $T_{ij}$, and can thus be solved directly. On setting $T_{11}(\chi,\theta,\phi)=h_{11,\ell}(\chi)Y_{\ell}^m(\theta,\phi)$, the equation for $T_{11}$ reduces to 
\begin{eqnarray}
\frac{1}{L^2}\left[\frac{d^2}{d\chi^2}+6\frac{\cosh\chi}{ \sinh\chi}\frac{d }{d\chi}
+6+\frac{6 }{ \sinh^2\chi}-\frac{\ell(\ell+1)}{ \sinh^2\chi}+A_T\right]h_{11,\ell}=0.
\label{4.20z}
\end{eqnarray}

To determine the $\chi \rightarrow \infty$ and $\chi \rightarrow 0$ limits, we take the solutions to behave as $e^{\lambda\chi}$ (times an irrelevant polynomial in $\chi$) and $\chi^n$ in these two limits. For (\ref{4.20z}) the limits give
\begin{eqnarray}
&&\lambda^2+6\lambda+6+A_T=0,\quad \lambda=-3\pm(3-A_T)^{1/2},
\nonumber\\
&&n(n-1)+6n+6-\ell(\ell+1)=0,\quad n=\ell-2, -\ell-3.
\label{4.21z}
\end{eqnarray}
Asymptotic convergence will thus depend on $A_T$, while finiteness at $\chi=0$ will depend on $\ell$. However the conditions differ from both the scalar and vector ones, a point that will prove crucial below in establishing the conformal gravity decomposition theorem. With there being two possible values for $\lambda$ there will be two families of solutions, which we will label $\hat{T}_{\ell}^{(1)}$ and $\hat{T}_{\ell}^{(2)}$ in the following.

To solve (\ref{4.20z}) we set $h_{11,\ell}=\gamma_{\ell}/\sinh^2\chi$ to obtain:
\begin{eqnarray}
\frac{1}{L^2} \left[\frac{d^2}{d\chi^2}+2\frac{\cosh\chi}{\sinh\chi}\frac{d}{d\chi}
-\frac{\ell(\ell+1) }{ \sinh^2\chi}-2+A_T\right]\gamma_{\ell}=0.
\label{4.22z}
\end{eqnarray}
We recognize (\ref{4.22z}) as being (\ref{4.3z}) where $\nu^2=A_T-3$.

\subsection{Master Formalism for Scalar, Vector and Tensor Modes and Their Normalization}
\label{S4d}

In \cite{Bander1966} a scalar field master equation  of the form
\begin{eqnarray}
\left[\frac{d^2}{d\chi^2}+(F-1)\frac{\cosh\chi }{\sinh\chi}\frac{d }{ d\chi}
-\frac{\beta(\beta+F-2)}{\sinh^2\chi}+\nu^2+\left(\frac{F-1}{2}\right)^2\right]Z_{\nu,\beta}(\chi)=0
\label{4.23y}
\end{eqnarray}
was presented that holds for the radial modes of a scalar field propagating in $F$ spatial dimensions with angular momentum   $\ell=\beta +(F-3)/2$. However, we can also use this master equation to describe the propagation of scalar, vector and tensor radial modes in three spatial dimensions.  Specifically, we see that comparison with (\ref{4.3z}) shows that for the scalar  we have $F=3$, $\nu^2+1=A_S$, $\beta=\ell$. 
Comparison with (\ref{4.14z}) shows that for the vector we have $F=5$, $\nu^2+2=A_V$, $\beta=\ell-1$.
Comparison with (\ref{4.20z}) shows that for the tensor we have $F=7$, $\nu^2+3=A_T$, $\beta=\ell-2$. The radial equations for scalar modes in five and seven spatial dimensions thus respectively correspond to the radial equations for vectors and tensors in three spatial dimensions.

For the case in which the $f(\nu)$ of (\ref{4.7z}) is given by $f(\nu)=\cos\nu\chi$, this being the case that will prove to be relevant in Secs. \ref{S6}, \ref{S7} and \ref{S8}, the modes are normalized according to \cite{Bander1966}
\begin{eqnarray}
Z_{\nu,\beta}(\chi)=A(\nu,\beta,F)\sinh^{\beta}\chi\left(\frac{1}{\sinh\chi}\frac{d}{d\chi}\right)^{(F-1)/2+\beta}\cos\nu\chi,
\label{4.24y}
\end{eqnarray}
where
\begin{eqnarray}
A(\nu,\beta,F)=\frac{2^{1/2}}{[\pi \nu^2(\nu^2+1^2)(\nu^2+2^2)......(\nu^2+((F-3)/2+\beta)^2)]^{1/2}}.
\label{4.25y}
\end{eqnarray}
With the integration measure for $d\chi^2+\sinh^2\chi d\theta^2+\sinh^2\chi\sin^2\theta d\phi^2$ being $\sinh^2\chi \sin\theta$ as needed for the normalization of $Z_{\nu,\beta}(\chi)Y^m_{\ell}(\theta,\phi)$, for $ Z_{\nu,\beta}(\chi)$ itself the integration measure is $\sinh^2\chi$. And with this normalization the modes obey the Dirac delta function orthonormality  condition \cite{Bander1966}
\begin{eqnarray}
\int_0^{\infty}d\chi\sinh^2\chi \sinh^{(F-3)/2}\chi Z_{\nu_1,\beta_1}(\chi) \sinh^{(F-3)/2}\chi Z^*_{\nu_2,\beta_2}(\chi)=\delta_{\beta_1,\beta_2}\delta(\nu_1-\nu_2).
\label{4.26y}
\end{eqnarray}
The extra $\sinh^{(F-3)/2}\chi$ factors that have been introduced here take the values $1,~\sinh\chi,~\sinh^2\chi$ for $F=3,5,7$. While not relevant for the scalar case, for the vector and the tensor modes these are precisely the factors needed to go from (\ref{4.14z}) to (\ref{4.16z}) and to go from (\ref{4.20z}) to (\ref{4.22z}) \cite{footnote1}.

To understand the emergence of the $\delta(\nu_1-\nu_2)$ term we directly evaluate the scalar mode case with $F=3$, $\ell_1=\ell_2=0$, $\beta_1=\beta_2=0$, viz. the scalar field $\hat{S}_0$ given in (\ref{4.8z}). And with both $\nu_1$ and $\nu_2$ positive we obtain
\begin{eqnarray}
A(\nu_1,0,3)A(\nu_2,0,3)\int_0^{\infty}d\chi\sinh^2\chi  \frac{\nu_1\sin\nu_1\chi\nu_2\sin\nu_2\chi}{\sinh^2\chi}
&&=\frac{2}{\pi}\frac{1}{2}\frac{1}{(2i)^2}\int_{-\infty}^{\infty}d\chi(e^{i\nu_1\chi}-e^{-i\nu_1\chi})(e^{i\nu_2\chi}-e^{-i\nu_2\chi})
\nonumber\\
&&=\delta(\nu_1-\nu_2)-\delta(\nu_1+\nu_2)=\delta(\nu_1-\nu_2).
\label{4.27y}
\end{eqnarray}

For the vector mode case with $F=5$, $\ell_1=\ell_2=1$, $\beta_1=\beta_2=0$, and  with both $\nu_1$ and $\nu_2$ again positive,  we obtain
\begin{align}
 &A(\nu_1,0,5)A(\nu_2,0,5)\int_0^{\infty}d\chi\sinh^2\chi  
 \frac{d}{d\chi}\left[-\frac{\nu_1\sin\nu_1\chi}{\sinh\chi}\right]  \frac{d}{d\chi}\left[-\frac{\nu_2\sin\nu_2\chi}{\sinh\chi}\right]
\nonumber\\
 &=A(\nu_1,0,5)A(\nu_2,0,5)\nu_1\nu_2\int_0^{\infty}d\chi
 \left[-\nu_1\cos\nu_1\chi+\frac{\cosh\chi}{\sinh\chi}\sin\nu_1\chi\right] 
 \left[-\nu_2\cos\nu_2\chi+\frac{\cosh\chi}{\sinh\chi}\sin\nu_2\chi\right]
 \nonumber\\
 &=A(\nu_1,0,5)A(\nu_2,0,5)\frac{\nu_1\nu_2}{2}\int_{-\infty}^{\infty}d\chi
 \bigg{[}\nu_1\nu_2\cos\nu_1\chi\cos\nu_2\chi+\sin\nu_1\chi\sin\nu_2\chi
 \nonumber\\
&+\frac{\sin\nu_1\chi\sin\nu_2\chi}{\sinh^2\chi}
-\frac{\cosh\chi}{\sinh\chi}\nu_2\sin\nu_1\chi\cos\nu_2\chi
-\frac{\cosh\chi}{\sinh\chi}\nu_1\sin\nu_2\chi\cos\nu_1\chi\bigg{]} 
\nonumber\\
&=A(\nu_1,0,5)A(\nu_2,0,5)\frac{\nu_1\nu_2}{2}\int_{-\infty}^{\infty}d\chi
 \bigg{[}\nu_1\nu_2\cos\nu_1\chi\cos\nu_2\chi+\sin\nu_1\chi\sin\nu_2\chi
-\frac{d}{d\chi}\left(\frac{\sin\nu_1\chi\sin\nu_2\chi\cosh\chi}{\sinh\chi}\right)\bigg{]}.
\label{4.28y}
\end{align}
With the surface term oscillating away (when integrated with a good test function), it follows that 
\begin{eqnarray}
 &&\frac{2}{\pi\nu_1(\nu_1^2+1^2)^{1/2}\nu_2(\nu_2^2+1^2)^{1/2}}\int_0^{\infty}d\chi\sinh^2\chi  
 \frac{d}{d\chi}\left[-\frac{\nu_1\sin\nu_1\chi}{\sinh\chi}\right]  \frac{d}{d\chi}\left[-\frac{\nu_2\sin\nu_2\chi}{\sinh\chi}\right]
=\delta(\nu_1-\nu_2),
\label{4.29y}
\end{eqnarray}
just as required. Because of the relation between (\ref{4.14z}) and (\ref{4.16z}), this calculation is identical to that of the normalization of the $\ell=1$ scalar mode $A(\nu,1,3)\hat{S}_1$ introduced in (\ref{4.8z}).

A similar analysis holds for tensor modes with $F=7$, $\ell_1=\ell_2=2$, $\beta_1=\beta_2=0$ and positive $\nu_1$ and $\nu_2$. If as in \cite{Bander1966} we define 
\begin{align}
q_p(\nu,\chi)=\sinh^p\chi\left(\frac{1}{\sinh\chi}\frac{d}{d\chi}\right)^pf(\nu,\chi),
\label{4.30y}
\end{align}
where as before $f(\nu,\chi)=\cos\nu\chi$,  then we obtain
\begin{align}
&q_3(\nu,\chi)=\left(\frac{d}{d\chi}-\frac{2\cosh\chi}{\sinh\chi}\right)q_2(\nu,\chi),\quad \left(\frac{d}{d\chi}+\frac{2\cosh\chi}{\sinh\chi}\right)q_3(\nu,\chi)=-(\nu^2+2^2)q_2(\nu,\chi),
\nonumber\\
&q_2(\nu,\chi)=\left(\frac{d}{d\chi}-\frac{\cosh\chi}{\sinh\chi}\right)q_1(\nu,\chi),\quad 
\left(\frac{d}{d\chi}+\frac{\cosh\chi}{\sinh\chi}\right)q_2(\nu,\chi)=-(\nu^2+1^2)q_1(\nu,\chi),\quad q_1(\nu,\chi)=\frac{d f(\nu,\chi)}{d\chi}.
\label{4.31y}
\end{align}
Following some integrations by parts we obtain
\begin{align}
&\int_0^{\infty}d\chi q_3(\nu_1,\chi)q_3(\nu_2,\chi)=\int_0^{\infty}d\chi \left(\frac{d}{d\chi}-\frac{2\cosh\chi}{\sinh\chi}\right)q_2(\nu_1,\chi)q_3(\nu_2,\chi)
\nonumber\\
&=-\int_0^{\infty}d\chi q_2(\nu_1,\chi)\left(\frac{d}{d\chi}+\frac{2\cosh\chi}{\sinh\chi}\right)q_3(\nu_2,\chi)
=(\nu_2^2+2^2)\int_0^{\infty}d\chi q_2(\nu_1,\chi)q_2(\nu_2,\chi)
\nonumber\\
&=(\nu_2^2+2^2)\int_0^{\infty}d\chi \left(\frac{d}{d\chi}-\frac{\cosh\chi}{\sinh\chi}\right)q_1(\nu_1,\chi)q_2(\nu_2,\chi)
=-(\nu_2^2+2^2)\int_0^{\infty}d\chi q_1(\nu_1,\chi)\left(\frac{d}{d\chi}+\frac{\cosh\chi}{\sinh\chi}\right)q_2(\nu_2,\chi)
\nonumber\\
&=(\nu_2^2+2^2)(\nu_2^2+1^2)\int_0^{\infty}d\chi q_1(\nu_1,\chi)q_1(\nu_2,\chi)=\nu_1\nu_2(\nu_2^2+2^2)(\nu_2^2+1^2)\int_0^{\infty}d\chi \sin\nu_1\chi\sin\nu_2\chi
\nonumber\\
&=\frac{\pi}{2}\nu_1\nu_2(\nu_2^2+2^2)(\nu_2^2+1^2)\delta(\nu_1-\nu_2).
\label{4.32y}
\end{align}
From (\ref{4.32y}) it follows that 
\begin{align}
 &\frac{2}{\pi\nu_1(\nu_1^2+1^2)^{1/2}(\nu_1^2+2^2)^{1/2}\nu_2(\nu_2^2+1^2)^{1/2}(\nu_2+2^2)^{1/2}}
 \nonumber\\
 &\times \int_0^{\infty}d\chi\sinh^4\chi  \frac{d}{d\chi}\left[\frac{1}{\sinh\chi}
 \frac{d}{d\chi}\left(-\frac{\nu_1\sin\nu_1\chi}{\sinh\chi}\right)\right]  
 \frac{d}{d\chi}\left[\frac{1}{\sinh\chi}
 \frac{d}{d\chi}\left(-\frac{\nu_2\sin\nu_2\chi}{\sinh\chi}\right)\right] 
=\delta(\nu_1-\nu_2),
\label{4.33y}
\end{align}
just as required.
Because of the relation between (\ref{4.20z}) and (\ref{4.22z}), this calculation is identical to that of the normalization of the $\ell =2$ scalar mode $A(\nu,2,3)\hat{S}_2$ introduced in (\ref{4.8z}) \cite{footnote1a}.

Comparing with (\ref{4.4z}), (\ref{4.15z}) and  (\ref{4.21z}) we see that the scalar, vector and tensor modes behave asymptotically as $e^{\lambda \chi}$ where respectively $\lambda=-1\pm (1-A_S)^{1/2}=-1\pm i\nu$, $\lambda=-2\pm (2-A_V)^{1/2}=-2\pm i\nu$, $\lambda=-3\pm (3-A_T)^{1/2}=-3\pm i\nu$. For real $\nu$ these modes are all suppressed at $\chi=\infty$, with the scalar, vector and tensor modes respectively converging as $e^{-\chi}$, $e^{-2\chi}$ and $e^{-3\chi}$. However, that does not make them normalizable since the $\sinh^2\chi$, $\sinh^4\chi$ and $\sinh^6\chi$ factors in the respective integration measures for bilinear products of the modes diverge as $e^{2\chi}$, $e^{4\chi}$, $e^{6\chi}$. Thus, just like plane waves (the modes appropriate to $k=0$), the scalar, vector and tensor modes have to be Dirac delta function normalized \cite{footnote2}. Thus in general we have to distinguish between asymptotic boundedness and normalizability, with our boundedness criterion having to be that modes have to fall off at least as fast as needed to match the growth in  the relevant integration measure \cite{footnote3}.

Having now set up a second-order derivative formalism given in (\ref{4.3z}), (\ref{4.14z}) and (\ref{4.20z}) and shown its compatibility with the master equation approach  in the respective scalar, vector and tensor sectors, we need to use the formalism to solve the higher-derivative equations, four for the scalars [(\ref{3.4z}), (\ref{3.6z}), (\ref{3.9z}), (\ref{3.10z})], two for the vectors [(\ref{3.13z}), (\ref{3.19z})] and one for the tensors [(\ref{3.30y})]. We leave the details to the Appendix, and in Secs. \ref{S4e} and \ref{S5} we show how this will enable us to obtain a conformal gravity decomposition theorem.

\subsection{Impossibility of Reconciling the Scalar, Vector and Tensor Solutions}
\label{S4e}

While equations such as (\ref{3.30y}) contain the spatial derivative operator $(\tilde{\nabla}_c\tilde{\nabla}^c-2k)(\tilde{\nabla}_a\tilde{\nabla}^a-6k)(\tilde{\nabla}_b\tilde{\nabla}^b-3k)$, this derivative operator does not appear in the second-order derivative fluctuation equation $\eta \delta W_{ij}-\Delta_{ij}=0$ itself. In the Appendix we show that there are $\chi$-dependent solutions to $(\tilde{\nabla}_c\tilde{\nabla}^c-2k)(\tilde{\nabla}_a\tilde{\nabla}^a-6k)(\tilde{\nabla}_b\tilde{\nabla}^b-3k)T_{ij}=0$ that can meet boundary conditions at $\chi=\infty$ and $\chi=0$ without requiring $T_{ij}$ itself to vanish. (Here $T_{ij}$ represents the entire $E_{ij}$-dependent term that appears in the term in brackets in (\ref{3.30y}).) A similar situation also holds in the vector sector. In the solutions in which $T_{ij}$ does not itself vanish, the dependence on $\chi$ is fixed but not the dependence on $\tau$. (To fix the dependence on $\tau$ we would need $T_{ij}$ and its scalar and vector analogs to vanish, just as we discuss in Sec. \ref{S6}.) Now we note that since (\ref{4.3z}), (\ref{4.14z}) and (\ref{4.20z}) are different equations, the various scalar, vector and tensor modes would have differing behaviors in $\chi$. In the event that we do realize  (\ref{3.30y}) and its analogs by not having $T_{ij}$ and its analogs vanish, the only way for the modes to then be able to satisfy $\eta \delta W_{\mu\nu}-\Delta_{\mu\nu}=0$ is by mutual cancellation of their respective spatial dependencies. We now show that this cannot be the case, with  $\eta \delta W_{\mu\nu}-\Delta_{\mu\nu}=0$ thus having to split into separate scalar, vector and tensor sectors, to thus give a conformal gravity decomposition theorem.

To this end we note that since some scalar mode terms appear with two $\tilde{\nabla}_a$ derivatives in the $\Delta_{ij}$ sector, the vector sector terms appear with one $\tilde{\nabla}_a$ derivative and some of the tensor sector terms appear with none, we need to compare derivatives of scalars with vectors and derivatives of vectors with tensors. To see how it would be possible to obtain such a needed common $\chi$ behavior we differentiate the scalar field (\ref{4.3z}) with respect to $\chi$, and obtain
\begin{eqnarray}
 \left[\frac{d^2}{d\chi^2}+4\frac{\cosh\chi}{\sinh\chi}\frac{d }{ d\chi}
+\frac{2}{\sinh^2\chi}-\frac{\ell(\ell+1)}{\sinh^2\chi}+4+A_S\right]\frac{d S_{\ell}}{d \chi}
+2A_S\frac{\cosh\chi}{\sinh\chi}S_{\ell}=0.
\label{4.34y}
\end{eqnarray}
Comparing with the vector (\ref{4.14z}) we see that up to an overall normalization we can identify $d S_{\ell}/d\chi$ with the vector $g_{1,\ell}$ for modes that obey $A_S=0$ and $A_V=2$, so that these particular scalar and vector modes can interface. 

Similarly, if we differentiate the vector field (\ref{4.14z}) with respect to $\chi$ we  obtain
\begin{eqnarray}
\left[\frac{d^2}{d\chi^2}+6\frac{\cosh\chi}{ \sinh\chi}\frac{d }{d\chi}
+10+A_V+\frac{6 }{ \sinh^2\chi}-\frac{\ell(\ell+1)}{ \sinh^2\chi}\right]\frac{d g_{1,\ell}}{d \chi}
+2(2+A_V)\frac{\cosh\chi}{\sinh\chi}g_{1,\ell}=0.
\label{4.35y}
\end{eqnarray}
Comparing with the tensor (\ref{4.20z}) we see that up to an overall normalization we can identify $d g_{1,\ell}/d\chi$ with the tensor $h_{11,\ell}$ for modes that obey $A_V=-2$ and $A_T=2$, so that these particular vector and tensor modes can interface. Thus while we can interface $A_S=0$ and $A_V=2$, we cannot interface $A_V=2$ with any of the tensor modes. Rather, we must interface the $A_V=-2$ vector modes with the  $A_T=2$ tensor modes \cite{footnote4}. 

Now for scalar modes with $A_S=0$ solutions we have $\nu=i$, and  the relevant $f(\nu^2)$ given in (\ref{4.7z}) are $\cosh \chi$ and $\sinh \chi$. Similarly, for the vector modes with $A_V=2$ we have $\nu=0$ and $f(\nu^2)=\chi, \chi^2$.
Consequently, the first few $\hat{S}^{(i)}_{\ell}$, $i=1,2$ solutions to $(\tilde{\nabla}_a\tilde{\nabla}^a+A_S)S=0$ with $A_S=0$ and the first few $\hat{V}^{(i)}_{\ell}$, $i=1,2$ solutions to $(\tilde{\nabla}_a\tilde{\nabla}^a+A_V)V_i=0$ with $A_V=2$ are of the form \cite{Phelps2019}
\begin{align}
&\hat{S}^{(1)}_{0}(A_S=0)=\frac{\cosh\chi}{\sinh\chi},\quad \hat{S}^{(2)}_{0}(A_S=0)=1,
\nonumber\\
&\hat{S}^{(1)}_{1}(A_S=0)=\frac{1}{\sinh^2\chi},\quad \hat{S}^{(2)}_{1}(A_S=0)=\frac{\cosh\chi}{\sinh\chi}-\frac{\chi}{\sinh^2\chi},
\nonumber\\
&\hat{S}^{(1)}_{2}(A_S=0)=\frac{\cosh\chi}{\sinh^3\chi},\quad \hat{S}^{(2)}_{2}(A_S=0)=1+\frac{3}{\sinh^2\chi}-\frac{3\chi\cosh\chi}{\sinh^3\chi},
\nonumber\\
&\hat{S}^{(1)}_{3}(A_S=0)=\frac{4}{\sinh^2\chi}+\frac{5}{\sinh^4\chi},\quad \hat{S}^{(2)}_{3}(A_S=0)=
\frac{2\cosh\chi}{\sinh\chi}+\frac{15\cosh\chi}{\sinh^3\chi}-\frac{12\chi}{\sinh^2\chi}-\frac{15\chi}{\sinh^4\chi}.
\label{4.36y}
\end{align}
\begin{eqnarray}
\hat{V}^{(1)}_0(A_V=2)&=&\frac{1}{ \sinh^2\chi},\quad \hat{V}^{(2)}_0(A_V=2)=\frac{\chi }{ \sinh^2\chi},
\nonumber\\
\hat{V}^{(1)}_1(A_V=2)&=&\frac{\cosh \chi }{ \sinh^3\chi},\quad \hat{V}^{(2)}_1(A_V=2)=\frac{1}{ \sinh^2\chi}-\frac{\chi\cosh\chi}{ \sinh^3\chi},
\nonumber\\
\hat{V}^{(1)}_2(A_V=2)&=&\frac{2}{ \sinh^2\chi}+\frac{3}{\sinh^4\chi},\quad \hat{V}^{(2)}_2(A_V=2)=\frac{3\cosh\chi}{\sinh^3\chi}-\frac{2\chi}{\sinh^2\chi}-\frac{3\chi }{\sinh^4\chi},
\nonumber\\
\hat{V}^{(1)}_3(A_V=2)&=&\frac{2\cosh\chi}{\sinh^3\chi}+\frac{5\cosh\chi}{\sinh^5\chi},\quad \hat{V}^{(2)}_3(A_V=2)=\frac{11}{\sinh^2\chi}+\frac{15}{\sinh^4\chi}-\frac{6\chi\cosh\chi}{\sinh^3\chi}-\frac{15\chi\cosh\chi }{\sinh^5\chi},~~~
\label{4.37y}
\end{eqnarray}
viz. two solutions for each $\ell$ value, with $\ell$ being the lower index.
From this pattern we see that the $\hat{V}^{(2)}_{\ell}(A_V=2)$ solutions with $\ell \geq1$ are bounded at  $\chi=\infty$ and well behaved at $\chi=0$. However, the $\hat{S}^{(i)}_{\ell}$ solutions that are bounded at $\chi=\infty$ are badly-behaved at $\chi=0$, while the solutions that are  well behaved at $\chi=0$ are unbounded at $\chi=\infty$. Thus all of these $A_S=0$ solutions are excluded by a requirement that solutions be  bounded at $\chi=\infty$ and be well behaved at $\chi=0$. Hence we cannot interface the scalar $A_S=0$ solutions with the vector $A_V=2$ solutions or make (\ref{4.4z}) and (\ref{4.15z}) be compatible, and so an interface between $A_S=0$ and $A_V=2$ is excluded. 

For the other possible interface, viz. that  between the $\hat{V}^{(2)}_{\ell}(A_V=-2)$ ($\nu^2=-4$, $f(\nu)=\cosh 2\chi,~\sinh 2\chi$) vector modes and the $\hat{T}^{(2)}_{\ell}(A_T=2)$ ($\nu^2=-1$, $f(\nu)=\cosh \chi,~\sinh \chi$) tensor modes, the first few relevant mode solutions are \cite{Phelps2019} 
\begin{eqnarray}
&&\hat{V}^{(1)}_0(A_V=-2)=\frac{\cosh\chi}{\sinh\chi},\quad \hat{V}^{(2)}_0(A_V=-2)=2+\frac{1}{\sinh^2\chi},
\nonumber\\
&&\hat{V}^{(1)}_1(A_V=-2)=1,\quad \hat{V}^{(2)}_1(A_V=-2)=2\frac{\cosh\chi}{\sinh\chi}-\frac{\cosh\chi}{\sinh^3\chi},
\nonumber\\
&&\hat{V}^{(1)}_2(A_V=-2)=2\frac{\cosh\chi}{\sinh\chi}-\frac{3\cosh\chi}{\sinh^3\chi}+\frac{3\chi}{\sinh^4\chi},\quad \hat{V}^{(2)}_2(A_V=-2)=\frac{1}{\sinh^4\chi},
\nonumber\\
&&\hat{V}^{(1)}_3(A_V=-2)=2-\frac{5}{\sinh^2\chi}-\frac{15}{\sinh^4\chi}+\frac{15\chi\cosh\chi}{\sinh^5\chi},\quad \hat{V}^{(2)}_3(A_V=-2)=\frac{\cosh\chi}{\sinh^5\chi}.
\label{4.38y}
\end{eqnarray}
\begin{align}
&\hat{T}^{(1)}_{0}(A_T=2)=\frac{\cosh\chi}{\sinh^3\chi},\quad \hat{T}^{(2)}_{0}(A_T=2)=\frac{1}{\sinh^2\chi},
\nonumber\\
&\hat{T}^{(1)}_{1}(A_T=2)=\frac{1}{\sinh^4\chi},\quad \hat{T}^{(2)}_{1}(A_T=2)=\frac{\cosh\chi}{\sinh^3\chi}-\frac{\chi}{\sinh^4\chi},
\nonumber\\
&\hat{T}^{(1)}_{2}(A_T=2)=\frac{\cosh\chi}{\sinh^5\chi},\quad \hat{T}^{(2)}_{2}(A_T=2)=\frac{1}{\sinh^2\chi}+\frac{3}{\sinh^4\chi}-\frac{3\chi\cosh\chi}{\sinh^5\chi},
\nonumber\\
&\hat{T}^{(1)}_{3}(A_T=2)=\frac{4}{\sinh^4\chi}+\frac{5}{\sinh^6\chi},\quad \hat{T}^{(2)}_{3}(A_T=2)=
\frac{2\cosh\chi}{\sinh^3\chi}+\frac{15\cosh\chi}{\sinh^5\chi}-\frac{12\chi}{\sinh^4\chi}-\frac{15\chi}{\sinh^6\chi}.
\label{4.39y}
\end{align}
From this pattern we see that all of the $\hat{T}^{(2)}_{\ell}(A_T=2)$ solutions are bounded at $\chi=\infty$ and all $\hat{T}^{(2)}_{\ell}(A_T=2)$ solutions with $\ell\geq 2$ are well behaved at $\chi=0$. However, none of the $\hat{V}^{(1)}_{\ell}(A_V=-2)$ vanish at $\chi=\infty$,  and while the $\hat{V}^{(2)}_{\ell}(A_V=-2)$ with $\ell \geq 2$ are bounded at $\chi=\infty$ they diverge at $\chi=0$. Hence we cannot interface the $A_V=-2$ vector solutions with the  $A_T=2$ tensor solutions or make (\ref{4.15z}) and (\ref{4.21z}) be compatible, and so an interface between $A_V=-2$ and $A_T=2$ is excluded. Consequently, we can only satisfy equations  such as (\ref{3.30y}) by having the $E_{ij}$-dependent term in brackets vanish. The vanishing of this particular term and its scalar and vector analogs will then fix the dependence of the fluctuations on the conformal time $\tau$.

\section{The Conformal Gravity Decomposition Theorem}
\label{S5}

For a decomposition theorem for $\eta\delta W_{\mu\nu}=\Delta_{\mu\nu}$ to hold,  the ten $\eta\delta W_{\mu\nu}=\Delta_{\mu\nu}$ conditions must break up into separate scalar, vector and tensor sectors of the form 
\begin{eqnarray}
&& - \frac{2\eta}{3\Omega^2} (\tilde\nabla_a\tilde\nabla^a + 3k)\tilde\nabla_b\tilde\nabla^b \alpha
 =6 \dot{\Omega}^2 \Omega^{-2}(\alpha-\dot\gamma) + \delta \hat{\rho} \Omega^2 + 2 \dot{\Omega} \Omega^{-1} \tilde{\nabla}_{a}\tilde{\nabla}^{a}\gamma,
 \label{5.1z}
 \end{eqnarray}
 \begin{eqnarray}
&&-\frac{2\eta}{3\Omega^2}  \tilde\nabla_i (\tilde\nabla_a\tilde\nabla^a + 3k)\dot\alpha
=-2 \dot{\Omega} \Omega^{-1} \tilde{\nabla}_{i}(\alpha - \dot\gamma) + 2 k \tilde{\nabla}_{i}\gamma 
+(-4 \dot{\Omega}^2 \Omega^{-3}  + 2 \overset{..}{\Omega} \Omega^{-2}  - 2 k \Omega^{-1}) \tilde{\nabla}_{i}\hat{V},
\label{5.2z}
 \end{eqnarray}
 \begin{align}
&\frac{\eta}{2\Omega^2}(\tilde\nabla_b \tilde\nabla^b-\partial_{\tau}^2-2k)(\tilde\nabla_c \tilde\nabla^c+2k)(B_i-\dot{E}_i)
 =k(B_i-\dot E_i)+ \frac{1}{2} \tilde{\nabla}_{a}\tilde{\nabla}^{a}(B_{i} - \dot{E}_{i})
+ (-4 \dot{\Omega}^2 \Omega^{-3} + 2 \overset{..}{\Omega} \Omega^{-2} - 2 k \Omega^{-1})V_{i},
 \label{5.3z}
 \end{align}
 \begin{align}
 &-\frac{\eta}{3 \Omega^2} \left[ \tilde{\gamma}_{ij} \tilde\nabla_a\tilde\nabla^a (\tilde\nabla_b \tilde\nabla^b +2k-\partial_{\tau}^2)\alpha - \tilde\nabla_i\tilde\nabla_j(\tilde\nabla_a\tilde\nabla^a - 3\partial_{\tau}^2)\alpha \right]
\nonumber\\
&= \tilde{\gamma}_{ij}\big[ 2 \dot{\Omega}^2 \Omega^{-2}(\alpha-\dot\gamma)
-2  \dot{\Omega} \Omega^{-1}(\dot\alpha -\ddot\gamma)-4\ddot\Omega\Omega^{-1}(\alpha-\dot\gamma)+ \Omega^2 \delta \hat{p}-\tilde\nabla_a\tilde\nabla^a( \alpha + 2\dot\Omega \Omega^{-1}\gamma) \big] 
+\tilde\nabla_i\tilde\nabla_j( \alpha + 2\dot\Omega \Omega^{-1}\gamma),
\label{5.4z}
 \end{align}
 \begin{align}
&& \frac{\eta}{2 \Omega^2} \left[ \tilde\nabla_i (\tilde\nabla_a\tilde\nabla^a -2k-\partial_{\tau}^2) (\dot{B}_j-\ddot{E}_j) 
+  \tilde\nabla_j ( \tilde\nabla_a\tilde\nabla^a -2k-\partial_{\tau}^2) (\dot{B}_i-\ddot{E}_i)\right]
\nonumber\\
&&=\dot{\Omega} \Omega^{-1} \tilde{\nabla}_{i}(B_{j}-\dot E_j)+\frac{1}{2} \tilde{\nabla}_{i}(\dot{B}_{j}-\ddot{E}_j)
+\dot{\Omega} \Omega^{-1} \tilde{\nabla}_{j}(B_{i}-\dot E_i)+\frac{1}{2} \tilde{\nabla}_{j}(\dot{B}_{i}-\ddot{E}_i),
\label{5.5z}
 \end{align}
 \begin{eqnarray}
&& \frac{\eta}{\Omega^2}\left[ (\tilde\nabla_b \tilde\nabla^b-\partial_{\tau}^2-2k)^2+4k\partial_{\tau}^2 \right] E_{ij}
=- \overset{..}{E}_{ij} - 2 k E_{ij} - 2 \dot{E}_{ij} \dot{\Omega} \Omega^{-1} + \tilde{\nabla}_{a}\tilde{\nabla}^{a}E_{ij}.
\label{5.6z}
\end{eqnarray}
However, even if such a decomposition were to occur in this form, initially this is not enough information as there are only nine pieces of information. The vector sector has four equations and four degrees of freedom (the two-component $B_i-\dot{E}_i$ and the two-component $V_i$), and the tensor sector has two degrees of freedom (the two-component $E_{ij}$). However in the scalar sector there are only three equations [(\ref{5.1z}), (\ref{5.2z}) and (\ref{5.4z})], with there thus only being a total of nine pieces of information. However, since $\tilde{\gamma}_{ij}$ and $ \tilde{\nabla}_i\tilde{\nabla}_j$ transform differently under three-dimensional rotations (\ref{5.4z}) actually breaks up into two sectors according to 
 \begin{align}
 &-\frac{\eta}{3 \Omega^2}  \tilde\nabla_a\tilde\nabla^a (\tilde\nabla_b \tilde\nabla^b +2k-\partial_{\tau}^2)\alpha = 2 \dot{\Omega}^2 \Omega^{-2}(\alpha-\dot\gamma)
-2  \dot{\Omega} \Omega^{-1}(\dot\alpha -\ddot\gamma)-4\ddot\Omega\Omega^{-1}(\alpha-\dot\gamma)+ \Omega^2 \delta \hat{p}-\tilde\nabla_a\tilde\nabla^a( \alpha + 2\dot\Omega \Omega^{-1}\gamma), 
\label{5.7z}
 \end{align}
and
 \begin{align}
 & \frac{\eta}{3 \Omega^2} (\tilde\nabla_a\tilde\nabla^a - 3\partial_{\tau}^2)\alpha =  \alpha + 2\dot\Omega \Omega^{-1}\gamma,
\label{5.8z}
 \end{align}
 and now we do have ten pieces of information.

In addition we note that the non-trivial solution to (\ref{5.2z}) is given by 
 \begin{eqnarray}
&&-\frac{2\eta}{3\Omega^2}  (\tilde\nabla_a\tilde\nabla^a + 3k)\dot\alpha
=-2 \dot{\Omega} \Omega^{-1} (\alpha - \dot\gamma) + 2 k\gamma 
+(-4 \dot{\Omega}^2 \Omega^{-3}  + 2 \overset{..}{\Omega} \Omega^{-2}  - 2 k \Omega^{-1}) \hat{V},
\label{5.9z}
 \end{eqnarray}
while  (\ref{5.5z}) yields
\begin{align}
&& \frac{\eta}{2 \Omega^2} (\tilde\nabla_a\tilde\nabla^a -2k-\partial_{\tau}^2) (\dot{B}_i-\ddot{E}_i)
=\dot{\Omega} \Omega^{-1}(B_{i}-\dot E_i)+\frac{1}{2}(\dot{B}_{i}-\ddot{E}_i).
\label{5.10z}
\end{align}
If the decomposition theorem is to be valid then the scalar (\ref{5.1z}), (\ref{5.7z}), (\ref{5.8z}), (\ref{5.9z}), the vector (\ref{5.3z}), (\ref{5.10z}), and the tensor (\ref{5.6z}) all need to hold, with  (\ref{5.1z}) being automatic since $\eta \delta W_{00}-\Delta_{00}$ only contains scalars to begin with. 

Since the $\chi\rightarrow\infty$, $\chi =0$ boundary conditions for the separated fluctuation equations given in Sec. \ref{S3} exclude the vanishing of expressions such as $(\tilde{\nabla}_c\tilde{\nabla}^c-2k)(\tilde{\nabla}_a\tilde{\nabla}^a-6k)(\tilde{\nabla}_b\tilde{\nabla}^b-3k)E_{ij}$, we must instead set the bracketed term in (\ref{3.30y}) and its analogs to zero. When this is done we find that  because of the boundary conditions (\ref{5.8z}), (\ref{5.9z}),  (\ref{5.3z}), (\ref{5.10z}) and (\ref{5.6z}) respectively follow from (\ref{3.9z}), (\ref{3.6z}), (\ref{3.13z}), (\ref{3.19z}) and (\ref{3.30y}). However, (\ref{5.7z}) does not follow this way. To derive (\ref{5.7z}) we note that with the boundary conditions  every sector of $\eta\delta W_{ij}-\Delta_{ij}$ other than  the $\tilde{\gamma}_{ij}$ sector then does obey the decomposition theorem. However, since  the relation $\eta\delta W_{ij}=\Delta_{ij}$ does hold it follows that the $\tilde{\gamma}_{ij}$ sector must obey the decomposition theorem too. 

We thus extend the decomposition theorem to conformal gravity, and note that the ten-component decomposition theorem equations  (\ref{5.1z}), (\ref{5.7z}), (\ref{5.8z}), (\ref{5.9z}), (\ref{5.3z}), (\ref{5.10z}) and (\ref{5.6z}) are both gauge invariant and exact without approximation. With $\Omega(\tau)$ having to obey (\ref{2.7z}), once we specify a form for $p_m/\rho_m$ and thus a form for $\delta p/\delta \rho$, which according to (\ref{2.22y}) is equal to  $\delta p_m/\delta \rho_m$, we can in principle then solve the theory completely in any cosmological epoch.

\section{Solution in the Recombination Era}
\label{S6}

As noted above, at recombination we can set $\Omega(\tau_R)=2S_0(k/2\Lambda)^{1/2}\exp[(-k)^{1/2}\tau_R]$, $\dot{\Omega}(\tau_R)/\Omega(\tau_R)=(-k)^{1/2}$, as expressed in conformal time. Thus at recombination   (\ref{5.1z}), (\ref{5.3z}), (\ref{5.6z}), (\ref{5.7z}), (\ref{5.8z}), (\ref{5.9z}) and (\ref{5.10z}) reduce to 
\begin{eqnarray}
&& - \frac{2\eta}{3\Omega^2} (\tilde\nabla_a\tilde\nabla^a + 3k)\tilde\nabla_b\tilde\nabla^b \alpha
 =-6k (\alpha-\dot\gamma) + 2 (-k)^{1/2} \tilde{\nabla}_{a}\tilde{\nabla}^{a}\gamma,
 \label{6.1z}
 \end{eqnarray}
 \begin{align}
&(\tilde\nabla_a \tilde\nabla^a+2k)\left[\frac{\eta}{2\Omega^2}(\tilde\nabla_b \tilde\nabla^b-\partial_{\tau}^2-2k)(B_i-\dot{E}_i)
 -\frac{1}{2}(B_{i} - \dot{E}_{i})\right]=0,
 \label{6.2z}
 \end{align}
 \begin{eqnarray}
&& \frac{\eta}{\Omega^2}\left[ (\tilde\nabla_b \tilde\nabla^b-\partial_{\tau}^2-2k)^2+4k\partial_{\tau}^2 \right] E_{ij}
=- \overset{..}{E}_{ij} - 2 k E_{ij} - 2 \dot{E}_{ij} (-k)^{1/2} + \tilde{\nabla}_{a}\tilde{\nabla}^{a}E_{ij},
\label{6.3z}
\end{eqnarray}
 \begin{align}
 &-\frac{\eta}{3 \Omega^2}  \tilde\nabla_a\tilde\nabla^a (\tilde\nabla_b \tilde\nabla^b +2k-\partial_{\tau}^2)\alpha = 2k(\alpha-\dot\gamma)
-2  (-k)^{1/2}(\dot\alpha -\ddot\gamma)-\tilde\nabla_a\tilde\nabla^a( \alpha + 2(-k)^{1/2}\gamma),
\label{6.4z}
 \end{align}
 \begin{align}
 & \frac{\eta}{3 \Omega^2} (\tilde\nabla_a\tilde\nabla^a - 3\partial_{\tau}^2)\alpha =  \alpha + 2(-k)^{1/2}\gamma,
\label{6.5z}
 \end{align}
 \begin{eqnarray}
&&-\frac{2\eta}{3\Omega^2}  (\tilde\nabla_a\tilde\nabla^a + 3k)\dot\alpha
=-2 (-k)^{1/2}(\alpha - \dot\gamma) + 2 k\gamma,
\label{6.6z}
 \end{eqnarray}
\begin{align}
&& \frac{\eta}{2 \Omega^2} (\tilde\nabla_a\tilde\nabla^a -2k-\partial_{\tau}^2) (\dot{B}_i-\ddot{E}_i)
=(-k)^{1/2}(B_{i}-\dot E_i)+\frac{1}{2}(\dot{B}_{i}-\ddot{E}_i).
\label{6.7z}
\end{align}
Just as anticipated  above, at recombination the  matter fields and $\Lambda$ contributions automatically drop out 
leaving us with just the gravitational contributions. (The contributions of $\delta \hat{\rho}$ and $\delta \hat{p}$ are suppressed by factors of order $\Omega^2 \sim \exp[2(-k)^{1/2}\tau_R]$ as $\tau_R\rightarrow -\infty$, while combinations such as $-4 \dot{\Omega}^2 \Omega^{-3}  + 2 \overset{..}{\Omega} \Omega^{-2}  - 2 k \Omega^{-1}$ vanish identically at $\tau=\tau_R$.) 

\subsection{The Scalar Sector}
\label{S6a}

For the scalar sector we only have two independent degrees of freedom at recombination, $\alpha$ and $\gamma$, but we have four equations, (\ref{6.1z}), (\ref{6.4z}), (\ref{6.5z}) and (\ref{6.6z}). There must thus be two relations between them. With $\dot{\Omega}/\Omega=(-k)^{1/2}$ they are
\begin{align}
&\frac{d}{d\tau}\left(3(-k)^{1/2}(\ref{6.6z})+(\ref{6.1z})\right)=\tilde{\nabla}_b\tilde{\nabla}^b(\ref{6.6z})+(-k)^{1/2}(\ref{6.1z})
-2(-k)^{1/2}\left[3(-k)^{1/2}(\ref{6.6z})+(\ref{6.1z})-(\tilde{\nabla}_b\tilde{\nabla}^b+3k)(\ref{6.5z})\right],
\nonumber\\
&\left(\frac{d}{d\tau} +2(-k)^{1/2}\right)(\ref{6.6z})=(\tilde{\nabla}_b\tilde{\nabla}^b+2k)(\ref{6.5z})+(\ref{6.4z}).
\label{6.8x}
\end{align}
For the two remaining  relations we note first that
\begin{eqnarray}
&&3(-k)^{1/2}(\ref{6.6z})+(\ref{6.1z}) -(\tilde{\nabla}_b\tilde{\nabla}^b+3k)(\ref{6.5z})
=(\tilde{\nabla}_b\tilde{\nabla}^b+3k)\left[\frac{\eta}{\Omega^2}(\ddot{\alpha}-2(-k)^{1/2}\dot{\alpha}-\tilde{\nabla}_b\tilde{\nabla}^b\alpha)+\alpha\right]=0,
\label{6.9x}
\end{eqnarray}
and can thus set \cite{footnoteE}
\begin{eqnarray}
\eta[\ddot{\alpha}-2(-k)^{1/2}\dot{\alpha}-\tilde{\nabla}_b\tilde{\nabla}^b\alpha]=-\Omega^2\alpha.
\label{6.10x}
\end{eqnarray}
From (\ref{6.10x}) we can fix $\alpha$, and then from (\ref{6.5z}) we can determine $\gamma$ according to 
 \begin{align}
 \gamma=\frac{1}{2(-k)^{1/2}} \left[\frac{\eta}{3 \Omega^2} (\tilde\nabla_a\tilde\nabla^a - 3\partial_{\tau}^2)\alpha -  \alpha\right].
 \label{6.11x}
 \end{align}

With $k=-1$ (i.e., $L^2=1$) we now look at  separable solutions to (\ref{6.10x}) of the dimensionless form $(\tilde{\nabla}_i\tilde{\nabla}^i+A_S)\alpha=0$ just as in (\ref{4.2z}), and obtain
\begin{eqnarray}
\frac{\eta}{\Omega^2}(\ddot{\alpha}-2\dot{\alpha}+A_S\alpha)+\alpha=0
.
\label{6.12x}
\end{eqnarray}
As introduced, at this stage the $A_S$ separation constant is arbitrary.  We will fix its value below.
To solve (\ref{6.12x}) it is simplest to convert to comoving time by setting $d/d\tau=a(t)d/dt$, where $a(t)=\Omega(\tau)$.  And with $\Omega(\tau)=e^{\tau}$, $a(t)=t$ we obtain
\begin{eqnarray}
\left[\eta\left(\partial_t^2 -\frac{1}{t}\partial_t +\frac{A_S}{t^2}\right)+1\right]\alpha=0,\quad 
\left[\eta\left(\partial_t^2 +\frac{1}{t}\partial_t +\frac{A_S-1}{t^2}\right)+1\right]\left(\frac{\alpha}{t}\right)=0.
\label{6.13x}
\end{eqnarray}
The solution to the second equation in (\ref{6.13x}) is a  Bessel function, and on excluding the irregular Bessel function since it is badly behaved at $t=0$, at recombination the solution to (\ref{6.13x}) is given in comoving time and conformal time by
\begin{align}
&\alpha=tJ_{\mu}(t/\eta^{1/2}),\quad \alpha=e^{\tau}J_{\mu}(e^{\tau}/\eta^{1/2}),\quad \mu=\pm(1-A_S)^{1/2}.
\label{6.14x}
\end{align}
With these solutions behaving as $t(t/\eta^{1/2})^{\mu}$ near $t=0$, we see that we get oscillating solutions if $A_S>1$. Also we note that if $\eta$ had been negative (i.e., if the gravitational coupling constant $\alpha_g$ had been positive) the $A_S>1$ solutions would have been modified Bessel functions with a totally different behavior as a function of $t$.

Since we introduced an overall factor of $\Omega^2(\tau)$ in the definition of the fluctuations in (\ref{2.11z}), the full scalar sector fluctuations are given by $\Omega^2\alpha$ and $\Omega^2\gamma$. Thus at recombination we see that $t^2\alpha$ behaves as $t^{\pm(1-A_S)^{1/2} +3}$ in comoving time, while according to (\ref{6.11x}), the leading behavior of  $t^2\gamma$ is also $t^{\pm(1-A_S)^{1/2} +3}$.  With the dependence on the spatial coordinate $\chi$ as $\chi\rightarrow \infty$ being given in (\ref{4.4z}) as $e^{\lambda \chi}$ where  $\lambda =-1\pm(1-A_S)^{1/2}$, we see exactly the same $\pm(1-A_S)^{1/2}$ dependence  as in the behavior in $t$. Moreover, if $A_S$ is real and obeys $A_S>1$, then in the scalar sector the solutions would behave as $e^{-\chi}$ times an oscillating function, and thus be well behaved and delta function normalizable as $\chi \rightarrow \infty$. Since according to (\ref{4.4z}) scalar sector solutions with $\ell \geq 0$ are well behaved at $\chi=0$, for real $A_S>1$ all $\ell \geq 0$ scalar sector solutions are well behaved at both $\chi=0$ and $\chi=\infty$.

\subsection{The Vector Sector}
\label{S6b}

As discussed in the Appendix, for the $B_i-\dot{E}_i$ sector we can factor out the $\tilde\nabla_a \tilde\nabla^a+2k$ factor in (\ref{6.2z}), and thus obtain
 \begin{align}
&\eta(\tilde\nabla_b \tilde\nabla^b-\partial_{\tau}^2-2k)(B_i-\dot{E}_i)
 =\Omega^2(B_{i} - \dot{E}_{i}).
 \label{6.15z}
 \end{align}
On taking the $\tau$ derivative we obtain
 \begin{align}
\eta(\tilde\nabla_b \tilde\nabla^b-\partial_{\tau}^2-2k)(\dot{B}_i-\ddot{E}_i)
 &=\Omega^2(\dot{B}_{i} - \ddot{E}_{i})+2\Omega\dot{\Omega}(B_i-\dot{E}_i)
 \nonumber\\
 & =\Omega^2(\dot{B}_{i} - \ddot{E}_{i})+2\Omega^2(-k)^{1/2}(B_i-\dot{E}_i).
 \label{6.16z}
 \end{align}
 Comparing with (\ref{6.7z}) we obtain
 \begin{align}
 &\dot{B}_{i} - \ddot{E}_{i}+2(-k)^{1/2}(B_i-\dot{E}_i)
 =2(-k)^{1/2}(B_{i}-\dot E_i)+\dot{B}_{i}-\ddot{E}_i.
  \label{6.17z}
 \end{align}
We thus establish that (\ref{6.2z}) and (\ref{6.7z}) are consistent, with (\ref{6.15z}) being a first integral of (\ref{6.7z}).

On setting $k=-1$ we look at separable solutions  to (\ref{6.15z}) of the dimensionless form $(\tilde\nabla_b \tilde\nabla^b+A_V)(B_i-\dot{E}_i)=0$ just as in (\ref{4.13z}). As with the scalar sector, at this stage the $A_V$ separation constant is arbitrary.  We will fix its value below. With this separation constant (\ref{6.15z}) takes the form
 \begin{align}
&\left[\eta\left(A_V+\frac{\partial^2}{\partial \tau ^2}-2\right)+\Omega^2\right](B_i-\partial_{\tau} E_i)=0.
 \label{6.18z}
 \end{align}
To solve (\ref{6.18z}) we rewrite it in comoving coordinates, and at recombination obtain
\begin{align}
&\left[\eta\left(\frac{A_V-2}{t^2} +\frac{\partial^2}{\partial t^2}+\frac{1}{t}\frac{\partial}{\partial t}\right)+1\right]\left(B_i-t\partial_t  E_i\right)=0,
\label{6.19z}
\end{align}
with solution in comoving time and conformal time of the form
\begin{align}
&B_i-t\partial_t E_i=\epsilon_iJ_{\rho}(t/\eta^{1/2}),\quad B_i-\partial_\tau E_i=\epsilon_iJ_{\rho}(e^{\tau}/\eta^{1/2}),\quad \rho=\pm(2-A_V)^{1/2},
\label{6.20z}
\end{align}
where $\epsilon_i$ is a transverse polarization vector. 

Since we introduced an overall factor of $\Omega^2(\tau)$ in the definition of the fluctuations in (\ref{2.11z}), the full vector  sector fluctuation is given by $t^2(B_i-t\partial_t E_i)$. Thus at recombination the solutions behave in comoving time as $t^2(B_i-t\partial_t E_i)\sim t^{\pm (2-A_V)^{1/2}+2}$. With the dependence on the spatial coordinate $\chi$ as $\chi\rightarrow \infty$ being given in (\ref{4.15z}) as $e^{\lambda \chi}$ where  $\lambda =-2\pm(2-A_V)^{1/2}$, we see exactly the same $\pm(2-A_V)^{1/2}$ dependence  as in the behavior in $t$. Moreover, if $A_V$ is real and obeys $A_V>2$, then in the vector sector the solutions would behave as $e^{-2\chi}$ times an oscillating function, and thus be well behaved and delta function normalizable as $\chi \rightarrow \infty$. Since according to (\ref{4.15z}) vector sector solutions with $\ell \geq 1$ are well behaved at $\chi=0$, for real $A_V>2$ all  $\ell \geq 1$ vector sector solutions are well behaved at both $\chi=0$ and $\chi=\infty$.

\subsection{The Tensor Sector}
\label{S6c}

Because we are able to set $\dot{\Omega}\Omega^{-1}=(-k)^{1/2}$ in (\ref{5.6z}), the resulting (\ref{6.3z}) turns out to be factorizable, and it takes the form:
\begin{align}
&\left[\frac{\eta}{\Omega^2}(\tilde{\nabla}_a\tilde{\nabla^a}-2k-\partial_{\tau}^2+2(-k)^{1/2}\partial_{\tau})-1\right]\left[\tilde{\nabla}_b\tilde{\nabla}^b-2k-\partial_{\tau}^2-2(-k)^{1/2}\partial_{\tau}\right]E_{ij}=0.
\label{6.21z}
\end{align}
With $k=-1$ we look for separable solutions of the dimensionless form $(\tilde{\nabla}_a\tilde{\nabla^a}+A_T)E_{ij}=0$ just as in (\ref{4.19z}). As with $A_S$ and $A_V$, at this stage the $A_T$ separation constant is arbitrary.  We will fix its value below, but first we look for solutions to 
\begin{align}
&\left[\frac{\eta}{\Omega^2}(\partial_{\tau}^2 -2\partial_{\tau}-2+A_T)+1\right]\left[\partial_{\tau}^2 +2\partial_{\tau}-2+A_T\right]E_{ij}=0.
\label{6.22z}
\end{align}
Equation (\ref{6.22z}) can be rewritten in comoving time as
\begin{align}
&\left[\frac{\eta}{a^2}(a^2\partial_{t}^2 +(a\partial_ta-2a)\partial_t-2+A_T)+1\right]\left[ a^2\partial_{t}^2 +(a\partial_ta+2a)\partial_t-2+A_T\right]E_{ij}=0.
\label{6.23z}
\end{align}
With $a(t)=t$ we can rewrite (\ref{6.23z}) as
\begin{align}
&\left[\frac{\eta}{t^2}(t^2\partial_{t}^2 -t\partial_t-2+A_T)+1\right]\left[ t^2\partial_{t}^2 +3t\partial_t-2+A_T\right]E_{ij}
\nonumber\\
&=\left[\frac{\eta}{t^2}(t^2\partial_{t}^2 -t\partial_t-2+A_T)+1\right]\left[t\left[ \partial_{t}^2 +\frac{1}{t}\partial_t+\frac{A_T-3}{t^2}\right](tE_{ij})\right]
\nonumber\\
&=t \left[ \eta\left(\partial_{t}^2 +\frac{1}{t}\partial_t+\frac{A_T-3}{t^2}\right)+1\right] \left[ \partial_{t}^2 +\frac{1}{t}\partial_t+\frac{A_T-3}{t^2}\right](tE_{ij})=0.
\label{6.24y}
\end{align}

There are two classes of solutions, and with $f=tE_{ij}$ they symbolically  obey
\begin{eqnarray}
\left[ \partial_{t}^2 +\frac{1}{t}\partial_t+\frac{A_T-3}{t^2}\right]f=Df=0,
\label{6.25y}
\end{eqnarray}
\begin{eqnarray}
\left[ \partial_{t}^2 +\frac{1}{t}\partial_t+\frac{A_T-3}{t^2}+\frac{1}{\eta}\right]g=\left[D+\frac{1}{\eta}\right]g=0,\quad Df=g,
\label{6.26y}
\end{eqnarray}
with (\ref{6.25y}) serving to define the derivative operator $D$.
For the first class of solutions we can set
\begin{align}
f_1=t^{\pm (3-A_T)^{1/2}}.
\label{6.27y}
\end{align}
For the second class of solutions we have
\begin{align}
g=J_{\sigma}(t/\eta^{1/2}),\quad  \sigma=\pm (3-A_T)^{1/2}.
\label{6.28y}
\end{align}
However, for this second class of solutions we also have
\begin{align}
D\left(g+\frac{f_2}{\eta}\right)=0,
\label{6.29y}
\end{align}
and thus we can set $g+f_2/\eta=t^{\sigma}$. Thus finally with $f=tE_{ij}$ the general solution to (\ref{6.23z}) in comoving time and conformal time is given by
\begin{align}
&E_{ij}=\epsilon_{ij}\left[at^{\sigma -1}+bt^{-1}J_{\sigma}(t/\eta^{1/2})\right],\quad E_{ij}=\epsilon_{ij}\left[ae^{(\sigma-1)\tau}+be^{-\tau}J_{\sigma}(e^{\tau}/\eta^{1/2})\right],\quad  \sigma=\pm(3-A_T)^{1/2},
\label{6.30y}
\end{align}
where $\epsilon_{ij}$ is a transverse-traceless polarization tensor and $a$ and $b$ are time-independent coefficients.

As we see, at recombination both comoving time solutions behave as $t^{\sigma -1}$. Since we introduced an overall factor of $\Omega^2(\tau)$ in the definition of the fluctuations in (\ref{2.11z}), the full tensor sector fluctuation is given by $\Omega^2E_{ij}$. Thus at recombination the solutions behave in comoving time as $t^2E_{ij}\sim t^{\pm (3-A_T)^{1/2}+1}$. With the dependence on the spatial coordinate $\chi$ as $\chi\rightarrow \infty$ being given in (\ref{4.21z}) as $e^{\lambda \chi}$ where  $\lambda =-3\pm(3-A_T)^{1/2}$, we see exactly the same $\pm(3-A_T)^{1/2}$ dependence  as in the behavior in $t$.  Finally we note that if $A_T$ is real and obeys $A_T>3$, then in the tensor sector the solutions would behave as $e^{-3\chi}$ times an oscillating function, and thus be well behaved and delta function normalizable as $\chi \rightarrow \infty$. Since according to (\ref{4.21z}) tensor sector solutions with $\ell \geq 2$ are well behaved at $\chi=0$, for real $A_T>3$ all $\ell \geq 2$ tensor sector solutions are well behaved at both $\chi=0$ and $\chi=\infty$. 

The time behaviors that we have found for the scalars, vectors and tensors in (\ref{6.14x}), (\ref{6.20z}) and (\ref{6.30y}) are of the respective forms $\alpha=t^{\pm (1-A_S)^{1/2}+1}$, $B_i-t\partial_t E_i=t^{\pm (2-A_V)^{1/2}}$, $E_{ij}=t^{\pm(3-A_T)^{1/2}-1}$, and according to the master equation discussed in Sec. \ref{S4d} are thus of the form $\alpha=t^{\pm i\nu+1}$, $B_i-t\partial_t E_i=t^{\pm i\nu}$, $E_{ij}=t^{\pm i\nu-1}$. We thus see a drop in powers of $t$ as we go from scalar to vector to tensor. To understand this we note that because of the Bianchi identity for the background we have $\nabla_{\mu}(\eta W^{\mu\nu}-\Delta_{(0)}^{\mu\nu})=0$. Thus on perturbing we have 
\begin{eqnarray}
\partial_{\mu}(\eta \delta W^{\mu\nu}-\Delta^{\mu\nu})&+&\Gamma^{\mu}_{\mu\sigma}(\eta \delta W^{\sigma\nu}-\Delta^{\sigma\nu})+ \Gamma^{\nu}_{\mu\sigma}(\eta \delta W^{\mu\sigma}-\Delta^{\mu\sigma})
\nonumber\\
&+&\delta\Gamma^{\mu}_{\mu\sigma}(\eta W^{\sigma\nu}-\Delta_{(0)}^{\sigma\nu})+\delta \Gamma^{\nu}_{\mu\sigma}(\eta W^{\mu\sigma}-\Delta_{(0)}^{\mu\sigma})=0.
\label{6.31y}
\end{eqnarray}
But in the background $\eta W^{\sigma\nu}-\Delta_{(0)}^{\sigma\nu}=0$. Thus the perturbed equations of motion obey 
\begin{eqnarray}
\nabla_{\mu}(\eta \delta W^{\mu\nu}-\Delta^{\mu\nu})=0.
\label{6.32y}
\end{eqnarray}
We thus have 
\begin{eqnarray}
\nabla_0(\eta \delta W^{00}-\Delta^{00})+\nabla_i(\eta \delta W^{i0}-\Delta^{i0})=0, \quad \nabla_0(\eta \delta W^{0j}-\Delta^{0j})+\nabla_i(\eta \delta W^{ij}-\Delta^{ij})=0.
\label{6.33y}
\end{eqnarray}
Consequently, as we go from scalar to vector we lose one power of $t$, and as we go from vector to tensor we lose another power of $t$.

Having now obtained the structure of the solutions we still need to fix appropriate values for $A_S$, $A_V$ and $A_T$. To do this we turn to an alternate way of solving the problem based on the traceless part $K_{\mu\nu}=h_{\mu\nu}-\tfrac{1}{4}g_{\mu\nu}g^{\rho\sigma}h_{\rho \sigma}$ of the fluctuations. This will naturally lead us to oscillating solutions, just as we would like.

\section{Fixing the Separation Constants}
\label{S7}

\subsection{The $K_{\mu\nu}$ Basis}
\label{S7a}

To fix the separation constants we compare the above  scalar, vector, tensor results with an alternate approach to the problem, one which involves none of the scalar, vector, tensor separation constants at all. Specifically, in the approach presented in \cite{Mannheim2012a, Amarasinghe2019} $\delta W_{\mu\nu}$ was developed in terms of the traceless fluctuation $K_{\mu\nu}=h_{\mu\nu}-\tfrac{1}{4}g_{\mu\nu}g^{\alpha\beta}h_{\alpha\beta}$ where $g_{\mu\nu}+h_{\mu\nu}$ is the full fluctuation. The utility of using $K_{\mu\nu}$ is that for any background geometry that is conformal to flat Minkowski, viz. of the form
\begin{eqnarray}
ds^2=\Omega^2(x)[d\tau^2-dx^2-dy^2-dz^2],
\label{7.1z}
\end{eqnarray}
where $\Omega(x)$ is an arbitrary function of the coordinates, $\delta W_{\mu\nu}$ is given without approximation and without any choice of gauge as  \cite{Amarasinghe2019} 
\begin{eqnarray}
\delta W_{\mu\nu}&=&\frac{1}{2}\Omega^{-2}\bigg{(}\partial_{\sigma}\partial^{\sigma}\partial_{\tau}\partial^{\tau}[\Omega^{-2}K_{\mu\nu}]
-\partial_{\sigma}\partial^{\sigma}\partial_{\mu}\partial^{\alpha}[\Omega^{-2}K_{\alpha\nu}]
-\partial_{\sigma}\partial^{\sigma}\partial_{\nu}\partial^{\alpha}[\Omega^{-2}K_{\alpha\mu}]
\nonumber\\
&+&\frac{2}{3}\partial_{\mu}\partial_{\nu}\partial^{\alpha}\partial^{\beta}[\Omega^{-2}K_{\alpha\beta}]+\frac{1}{3}\eta_{\mu\nu}\partial_{\sigma}\partial^{\sigma}\partial^{\alpha}\partial^{\beta}[\Omega^{-2}K_{\alpha\beta}]\bigg{)}.
\label{7.2z}
\end{eqnarray} 
With $\delta W_{\mu\nu}$ being traceless it is written in terms of the nine traceless components of $h_{\mu\nu}$, viz. $K_{\mu\nu}$. The great utility of (\ref{7.2z}) is that it only involves the Minkowski fourth-order derivative operator, to thus involve none of the separation constants of the wave equation associated with the conformal time Robertson-Walker metric $ds^2=\Omega^2(\tau)[d\tau^2-dr^2/(1-kr^2)-r^2d\theta^2-r^2\sin^2\theta d\phi^2]$. It will of course involve the separation constants of the $\partial_{\sigma}\partial^{\sigma}\partial_{\tau}\partial^{\tau}$ operator, but they are just the standard momentum variables. 

To take advantage of (\ref{7.2z}) we must rewrite the conformal time Robertson-Walker metric with negative $k$ in the form given in (\ref{7.1z}). Following e.g.  \cite{Amarasinghe2019}, on conveniently setting $k=-1/L^2$ and introducing ${\rm sinh} \chi=r/L$,  the conformal time metric then takes the form
\begin{eqnarray}
ds^2=L^2a^2(p)\left[dp^2-d\chi^2 -{\rm sinh}^2\chi d\theta^2-{\rm sinh}^2\chi \sin^2\theta d\phi^2\right],
\label{7.3z}
\end{eqnarray}
where $p=\tau/L$. Next we introduce
\begin{eqnarray}
p^{\prime}+r^{\prime}=\tanh[(p+\chi)/2],\qquad p^{\prime}-r^{\prime}=\tanh[(p-\chi)/2],\qquad p^{\prime}=\frac{\sinh p}{\cosh p+\cosh \chi},\qquad r^{\prime}=\frac{\sinh \chi}{\cosh p+\cosh \chi},
\label{7.4z}
\end{eqnarray}
so that
\begin{eqnarray}
dp^{\prime 2}-dr^{\prime 2}&=&\frac{1}{4}[dp^2-d\chi^2]{\rm sech}^2[(p+\chi)/2]{\rm sech}^2[(p-\chi)/2],
\nonumber\\
\frac{1}{4}(\cosh p+\cosh \chi)^2&=&{\rm \cosh}^2[(p+\chi)/2]{\rm \cosh}^2[(p-\chi)/2]=\frac{1}{[1-(p^{\prime}+r^{\prime})^2][1-(p^{\prime}-r^{\prime})^2]}.
\label{7.5z}
\end{eqnarray}
With these transformations the line element takes the conformal to flat form
\begin{eqnarray}
ds^2=\frac{4L^2a^2(p)}{[1-(p^{\prime}+r^{\prime})^2][1-(p^{\prime}-r^{\prime})^2]}\left[dp^{\prime 2}-dr^{\prime 2} -r^{\prime 2}d\theta^2-r^{\prime 2} \sin^2\theta d\phi^2\right].
\label{7.6z}
\end{eqnarray}
The spatial sector can then be written in Cartesian form
\begin{eqnarray}
ds^2=L^2a^2(p)(\cosh p+\cosh \chi)^2\left[dp^{\prime 2}-dx^{\prime 2} -dy^{\prime 2} -dz^{\prime 2}\right],
\label{7.7z}
\end{eqnarray}
where $r^{\prime}=(x^{\prime 2}+ y^{\prime 2}+z^{\prime 2})^{1/2}$.  The metric is now written in the conformal to flat form given in (\ref{7.1z}), and we can identify its $\Omega^2(x)$ factor as $\Omega^2(x)=L^2a^2(p)(\cosh p+\cosh \chi)^2$. We note that in transforming from (\ref{7.3z})  to (\ref{7.7z}) we have only made coordinate transformations and not made any conformal transformation. 

To determine the implications of (\ref{7.2z}) in a conformal to flat geometry such as that given in (\ref{7.7z}), we note  that in the gauge $\partial^{\mu}[\Omega^{-2}K_{\mu\nu}]=0$, the perturbation tensor  $\delta W_{\mu\nu}$ takes the form
\begin{eqnarray}
\delta W_{\mu\nu}=\frac{1}{2}\Omega^{-2}\partial_{\sigma}\partial^{\sigma}\partial_{\tau}\partial^{\tau}[\Omega^{-2}K_{\mu\nu}].
\label{7.8z}
\end{eqnarray} 
The great utility of (\ref{7.8z}) is that not only does it only involve the flat Minkowski wave operator, it is even diagonal in the $(\mu,\nu)$ indices.  If we ignore $\Delta_{\mu\nu}$ (the case considered in \cite{Amarasinghe2019}) we need to solve $\eta \delta W_{\mu\nu}=0$, and the solution then is of the form
\begin{eqnarray}
K_{\mu\nu}(x^{\prime})=\Omega^{2}(x^{\prime})\left[A_{\mu\nu}e^{ik\cdot x^{\prime}}+B_{\mu\nu}(n\cdot x^{\prime})e^{ik\cdot x^{\prime}}\right],\quad k_{\mu}k^{\mu}=0,
\label{7.9z}
\end{eqnarray}
together with the  complex conjugate solution. Here $n_{\mu}$ is a  spacetime-independent reference vector, $A_{\mu\nu}$ and $B_{\mu\nu}$ are traceless polarization tensors, with the $A_{\mu\nu}$ solution being a standard massless plane wave and with the massless $B_{\mu\nu}$ solution growing as $(n\cdot x^{\prime})$.  In the $\partial^{\mu}[\Omega^{-2}K_{\mu\nu}]=0$ gauge the solutions obey  $ik^{\mu}B_{\mu\nu}=0$, $ik^{\mu}A_{\mu\nu}+n^{\mu}B_{\mu\nu}=0$ \cite{Amarasinghe2019}. From (\ref{7.9z}) we see the natural emergence of massless plane wave solutions to the theory, with separation constants that are just the momentum variables associated with plane wave fluctuations.

If we now do include $\Delta_{\mu\nu}$ we can integrate $\eta \delta W_{\mu\nu}=\Delta_{\mu\nu}$ with the retarded Green's function associated with the $\partial_{\sigma}\partial^{\sigma}\partial_{\tau}\partial^{\tau}$ operator. This Green's function is of the form $\theta(t-r)/8\pi$ \cite{Mannheim2007}  as it obeys 
\begin{equation}
(\partial_t^2-\nabla^2)^2\left(\frac{\theta(t-r)}{8\pi}\right)=(\partial_t^2-\nabla^2)\left(\frac{\delta(t-r)}{4\pi r}\right)
=\delta^4(x). 
\label{7.10z}
\end{equation}
Consequently, the solution to $\eta \delta W_{\mu\nu}=\Delta_{\mu\nu}$ in the metric given by (\ref{7.7z}) is given by
\begin{eqnarray}
K_{\mu\nu}(x^{\prime})=\frac{2\Omega^2(x^{\prime})}{8\pi \eta}\int d^4x^{\prime\prime}\theta[p^{\prime}-p^{\prime\prime}-|\textbf{x}^{\prime}-\textbf{x}^{\prime\prime}|]\Omega^2(x^{\prime\prime})\Delta_{\mu\nu}(x^{\prime\prime}).
\label{7.11z}
\end{eqnarray}
For sources that are localized in space and oscillating in time the solution given in (\ref{7.11z}) will approach (\ref{7.9z}) far from the sources. Once we have (\ref{7.11z}) we can transform back from (\ref{7.7z}) to (\ref{7.3z}) by general coordinate transformations. And with a plane wave $\exp[ik(p^{\prime}-r^{\prime})]$ transforming into  $\exp[ik\tanh[(p-\chi)/2]]$, the oscillating nature of the solution persists, and  thus we can set all three of $(1-A_S)^{1/2}$, $(2-A_V)^{1/2}$ and $(3-A_T)^{1/2}$ equal to $i\nu$ where $\nu$ is a continuous positive parameter. 

To reinforce these remarks we note that since we are in a conformal invariant theory, as well as make coordinate transformations we can also make conformal transformations. With the gauge condition that we are using being conformal invariant \cite{footnoteG}, by a conformal transformation we can transform the metric in (\ref{7.7z}) into the  completely flat metric $ds^2=dp^{\prime 2}-dx^{\prime 2} -dy^{\prime 2} -dz^{\prime 2}$. Under this conformal transformation (\ref{7.2z}) will transform into 
\begin{eqnarray}
\delta W_{\mu\nu}&=&\frac{1}{2}\bigg{(}\partial_{\sigma}\partial^{\sigma}\partial_{\tau}\partial^{\tau}K_{\mu\nu}
-\partial_{\sigma}\partial^{\sigma}\partial_{\mu}\partial^{\alpha}K_{\alpha\nu}
-\partial_{\sigma}\partial^{\sigma}\partial_{\nu}\partial^{\alpha}K_{\alpha\mu}
\nonumber\\
&+&\frac{2}{3}\partial_{\mu}\partial_{\nu}\partial^{\alpha}\partial^{\beta}K_{\alpha\beta}+\frac{1}{3}\eta_{\mu\nu}\partial_{\sigma}\partial^{\sigma}\partial^{\alpha}\partial^{\beta}K_{\alpha\beta}\bigg{)}.
\label{7.12z}
\end{eqnarray} 
At the same time our conformal invariant gauge condition will transform into $\partial_{\mu}K^{\mu\nu}=0$, so that (\ref{7.12z}) will reduce to 
\begin{eqnarray}
\delta W_{\mu\nu}&=&\frac{1}{2}\partial_{\sigma}\partial^{\sigma}\partial_{\tau}\partial^{\tau}K_{\mu\nu}.
\label{7.13z}
\end{eqnarray} 
Thus again plane waves solutions emerge. Finally, to ensure that oscillating solutions in the $K_{\mu\nu}$ sector do propagate through to the scalar, vector, tensor basis we now relate the two sets of bases, something that is actually of interest in its own right as it is a strictly kinematic procedure that does not involve the imposition of any gravitational equation of motion at all.

\subsection{Matching the Scalar, Vector, Tensor  Basis with the $K_{\mu\nu}$ Basis}
\label{S7b}

To achieve the required matching of the two sets of bases we recall from above that for fluctuations around a $k\neq 0$ conformal time Robertson-Walker metric of the form 
\begin{align}
ds^2&=-(g_{\mu\nu}+h_{\mu\nu})dx^{\mu}dx^{\nu}=\Omega^2(\tau)\left[d\tau^2-\frac{dr^2}{1-kr^2}-r^2d\theta^2-r^2\sin^2\theta d\phi^2\right]
\nonumber\\
&+\Omega^2(\tau)\left[2\phi d\tau^2 -2(\tilde{\nabla}_i B +B_i)d\tau dx^i - [-2\psi\tilde{\gamma}_{ij} +2\tilde{\nabla}_i\tilde{\nabla}_j E + \tilde{\nabla}_i E_j + \tilde{\nabla}_j E_i + 2E_{ij}]dx^i dx^j\right],
\label{7.14x}
\end{align}
the perturbed $\delta W_{\mu\nu}$ is given by 
\begin{eqnarray}
\delta W_{00}&=& - \frac{2}{3\Omega^2} (\tilde\nabla_a\tilde\nabla^a + 3k)\tilde\nabla_b\tilde\nabla^b \alpha,
 \nonumber\\ 
\delta W_{0i}&=& -\frac{2}{3\Omega^2}  \tilde\nabla_i (\tilde\nabla_a\tilde\nabla^a + 3k)\dot\alpha
+\frac{1}{2\Omega^2}\left[ (\tilde\nabla_b \tilde\nabla^b-\partial_{\tau}^2-2k)(\tilde\nabla_c \tilde\nabla^c+2k)\right](B_i-\dot{E}_i),
  \nonumber\\ 
\delta W_{ij}&=& -\frac{1}{3 \Omega^2} \left[ \tilde{\gamma}_{ij} \tilde\nabla_a\tilde\nabla^a (\tilde\nabla_b \tilde\nabla^b +2k-\partial_{\tau}^2)\alpha - \tilde\nabla_i\tilde\nabla_j(\tilde\nabla_a\tilde\nabla^a - 3\partial_{\tau}^2)\alpha \right]
\nonumber\\
&& +\frac{1}{2 \Omega^2} \left[ \tilde\nabla_i (\tilde\nabla_a\tilde\nabla^a -2k-\partial_{\tau}^2) (\dot{B}_j-\ddot{E}_j) 
+  \tilde\nabla_j ( \tilde\nabla_a\tilde\nabla^a -2k-\partial_{\tau}^2) (\dot{B}_i-\ddot{E}_i)\right]
\nonumber\\
&&+ \frac{1}{\Omega^2}\left[ (\tilde\nabla_b \tilde\nabla^b-\partial_{\tau}^2-2k)^2+4k\partial_{\tau}^2 \right] E_{ij}.
\label{7.15x}
\end{eqnarray}
It must thus be possible to match the scalar $\alpha$, the vector $B_i-\dot{E}_i$, and the tensor $E_{ij}$ with the $K_{\mu\nu}$. And since the remaining scalar $\gamma$ of the scalar, vector, tensor expansion of $h_{\mu\nu}$  does not appear in $\delta W_{\mu\nu}$, in any matching $\gamma$ (a tenth degree of freedom)  must involve the trace of $h_{\mu\nu}$. As we now show, not only is it possible to have such a matching, the matching is purely kinematical and requires no reference to any gravitational tensor such as $\delta W_{\mu\nu}$ at all.  

To achieve the matching it is convenient to set $h_{\mu\nu}=\Omega^2(\tau)f_{\mu\nu}$, so that from (\ref{7.14x}) we obtain
\begin{align}
&f_{\tau\tau}=-2\phi,\quad f_{\tau i}=\tilde{\nabla}_i B +B_i,\quad f_{ij}=-2\psi\tilde{\gamma}_{ij} +2\tilde{\nabla}_i\tilde{\nabla}_j E + \tilde{\nabla}_i E_j + \tilde{\nabla}_j E_i + 2E_{ij}.
\label{7.16x}
\end{align}
Then with $\alpha=\phi+\psi+\partial_{\tau}B-\partial_{\tau}^2E$, $\gamma = - \Omega[\partial_{\tau}\Omega]^{-1}\psi + B - \partial_{\tau} E$, solely by taking appropriate  derivatives  of (\ref{7.16x}) we obtain \cite{Phelps2019} 
\begin{align}
&(3k+\tilde{\nabla}^b\tilde{\nabla}_b)\tilde{\nabla}^a\tilde{\nabla}_a\alpha=-\frac{1}{2}(3k+\tilde{\nabla}^b\tilde{\nabla}_b)\tilde{\nabla}^i\tilde{\nabla}_if_{\tau \tau}
\nonumber\\
&+\frac{1}{4}\tilde{\nabla}^a\tilde{\nabla}_a\left(-2kf-\tilde{\nabla}^b\tilde{\nabla}_bf+\tilde{\nabla}^m\tilde{\nabla}^nf_{mn}\right)
+\partial_{\tau}(3k+\tilde{\nabla}^b\tilde{\nabla}_b)\tilde{\nabla}^if_{\tau i}-\frac{1}{4}\partial^2_{\tau}\left(3\tilde{\nabla}^m\tilde{\nabla}^nf_{mn}-\tilde{\nabla}^a\tilde{\nabla}_af\right),
\label{7.17x}
\end{align}
\begin{align}
&(3k+\tilde{\nabla}^b\tilde{\nabla}_b)\tilde{\nabla}^a\tilde{\nabla}_a\gamma
\nonumber\\
&=-\frac{1}{4}\Omega[\partial_{\tau}\Omega]^{-1}\tilde{\nabla}^a\tilde{\nabla}_a\left(-2kf-\tilde{\nabla}^b\tilde{\nabla}_bf+\tilde{\nabla}^m\tilde{\nabla}^nf_{mn}\right)
+(3k+\tilde{\nabla}^b\tilde{\nabla}_b)\tilde{\nabla}^if_{\tau i}-\frac{1}{4}\partial_{\tau}\left(3\tilde{\nabla}^m\tilde{\nabla}^nf_{mn}-\tilde{\nabla}^a\tilde{\nabla}_af\right),
\label{7.18x}
\end{align}
where $f=\tilde{\gamma}^{ij}f_{ij}$.

With $h_{\mu\nu}=\Omega^2f_{\mu\nu}$  we introduce $K_{\mu\nu}=\Omega^2k_{\mu\nu}$ so that $k_{\mu\nu}=f_{\mu\nu}-\tfrac{1}{4}g_{\mu\nu}(-f_{\tau\tau}+f)$,  and obtain 
\begin{eqnarray}
f_{\tau\tau}=\frac{4}{3}k_{\tau\tau}-\frac{1}{3}f, \quad f_{\tau i}=k_{\tau i},\quad f_{ij}=k_{ij}+\frac{1}{3}\tilde{\gamma}_{ij}[f-k_{\tau\tau}].
\label{7.19x}
\end{eqnarray}
On now inserting (\ref{7.19x}) into (\ref{7.17x}) and (\ref{7.18x}) we obtain
\begin{align}
(3k+\tilde{\nabla}^b\tilde{\nabla}_b)\tilde{\nabla}^a\tilde{\nabla}_a\alpha&=-\frac{1}{4}(8k+3\tilde{\nabla}^b\tilde{\nabla}_b)\tilde{\nabla}_c\tilde{\nabla}^ck_{\tau \tau}+\frac{1}{4}\tilde{\nabla}_d\tilde{\nabla}^d\tilde{\nabla}^e\tilde{\nabla}^fk_{ef}
\nonumber\\
&+\partial_{\tau}(3k+\tilde{\nabla}^b\tilde{\nabla}_b)\tilde{\nabla}^ik_{\tau i}
-\frac{1}{4}\partial^2_{\tau}\left(3\tilde{\nabla}^m\tilde{\nabla}^nk_{mn}-\tilde{\nabla}_a\tilde{\nabla}^ak_{\tau\tau}\right),
\label{7.20x}
\end{align}
\begin{align}
&(3k+\tilde{\nabla}^b\tilde{\nabla}_b)\tilde{\nabla}^a\tilde{\nabla}_a\gamma
\nonumber\\
&=-\frac{1}{4}\Omega[\partial_{\tau}\Omega]^{-1}\tilde{\nabla}^a\tilde{\nabla}_a\left(-2kf-\frac{2}{3}\tilde{\nabla}^b\tilde{\nabla}_bf-\frac{1}{3}\tilde{\nabla}^b\tilde{\nabla}_bk_{\tau\tau}+\tilde{\nabla}^m\tilde{\nabla}^nk_{mn}\right)
+(3k+\tilde{\nabla}^b\tilde{\nabla}_b)\tilde{\nabla}^ik_{\tau i}
\nonumber\\
&-\frac{1}{4}\partial_{\tau}\left(3\tilde{\nabla}^m\tilde{\nabla}^nk_{mn}-\tilde{\nabla}^a\tilde{\nabla}_ak_{\tau\tau}\right).
\label{7.21x}
\end{align}
With $\alpha$ but not $\gamma$ appearing in $\delta W_{\mu\nu}$, and with the trace of $h_{\mu\nu}$ not appearing in $\delta W_{\mu\nu}$, it follows that $f$ must not appear in $\alpha$ but must appear in $\gamma$, just as we see.

For the vector sector we have \cite{Phelps2019} 
\begin{align}
(\tilde{\nabla}^a\tilde{\nabla}_a-2k)(\tilde{\nabla}^i\tilde{\nabla}_i +2k)(B_j-\partial_{\tau}E_j)&=(\tilde{\nabla}^i\tilde{\nabla}_i +2k)(\tilde{\nabla}^a\tilde{\nabla}_af_{\tau j}-2kf_{\tau j}
-\tilde{\nabla}_j\tilde{\nabla}^af_{\tau a})
\nonumber
\\
&-\partial_\tau\tilde{\nabla}^a\tilde{\nabla}_a\tilde{\nabla}^if_{ij}
+\partial_\tau\tilde{\nabla}_j\tilde{\nabla}^a\tilde{\nabla}^bf_{ab}
+2k\partial_\tau\tilde{\nabla}^if_{ij},
\label{7.22x}
\end{align}
and thus with use of the first relation in (\ref{3.1z}) obtain
\begin{align}
(\tilde{\nabla}^a\tilde{\nabla}_a-2k)(\tilde{\nabla}^i\tilde{\nabla}_i +2k)(B_j-\partial_{\tau}E_j)&=(\tilde{\nabla}^i\tilde{\nabla}_i +2k)(\tilde{\nabla}^a\tilde{\nabla}_ak_{\tau j}-2kk_{\tau j}
-\tilde{\nabla}_j\tilde{\nabla}^ak_{\tau a})
\nonumber\\
&-\partial_\tau\tilde{\nabla}^a\tilde{\nabla}_a\tilde{\nabla}^ik_{ij}
+\partial_\tau\tilde{\nabla}_j\tilde{\nabla}^a\tilde{\nabla}^bk_{ab}
+2k\partial_\tau\tilde{\nabla}^ik_{ij}.
\label{7.23x}
\end{align}
As we see, again $f$ drops out just as it should, since like $\alpha$, $B_i-\partial_{\tau}E_i$ also appears in $\delta W_{\mu\nu}$.

For the tensor sector we have \cite{Phelps2019}
\begin{align}
&2(\tilde{\nabla}^a\tilde{\nabla}_a-2k)(\tilde{\nabla}^b\tilde{\nabla}_b-3k)E_{ij}
=(\tilde{\nabla}^a\tilde{\nabla}_a-2k)(\tilde{\nabla}^b\tilde{\nabla}_b-3k)f_{ij}
\nonumber\\
&+\frac{1}{2}\tilde{\nabla}_i\tilde{\nabla}_j\left[\tilde{\nabla}^a\tilde{\nabla}^bf_{ab}+(\tilde{\nabla}^a\tilde{\nabla}_a+4k)f\right]-(\tilde{\nabla}^a\tilde{\nabla}_a-3k)(\tilde{\nabla}_i\tilde{\nabla}^bf_{jb}+\tilde{\nabla}_j\tilde{\nabla}^bf_{ib})
\nonumber\\
&+\frac{1}{2}\tilde{\gamma}_{ij}\left[(\tilde{\nabla}^a\tilde{\nabla}_a-4k)\tilde{\nabla}^b\tilde{\nabla}^cf_{bc}
-(\tilde{\nabla}_a\tilde{\nabla}^a\tilde{\nabla}_b\tilde{\nabla}^b-2k\tilde{\nabla}^a\tilde{\nabla}^a+4k^2)f\right].
\label{7.24x}
\end{align}
On substituting (\ref{7.19x}) and using the first relation in (\ref{3.2z}) with $A_j=\tilde{\nabla}_jf$ we obtain 
\begin{align}
&2(\tilde{\nabla}^a\tilde{\nabla}_a-2k)(\tilde{\nabla}^b\tilde{\nabla}_b-3k)E_{ij}
=(\tilde{\nabla}^a\tilde{\nabla}_a-2k)(\tilde{\nabla}^b\tilde{\nabla}_b-3k)k_{ij}
\nonumber\\
&+\frac{1}{2}\tilde{\nabla}_i\tilde{\nabla}_j\tilde{\nabla}^a\tilde{\nabla}^bk_{ab}
-(\tilde{\nabla}^a\tilde{\nabla}_a-3k)\left(\tilde{\nabla}_i\tilde{\nabla}^bk_{jb}+\tilde{\nabla}_j\tilde{\nabla}^bk_{ib}
\right)+\frac{1}{2}\tilde{\gamma}_{ij}(\tilde{\nabla}^a\tilde{\nabla}_a-4k)\tilde{\nabla}^b\tilde{\nabla}^ck_{bc}
\nonumber\\
&+\frac{1}{2}\tilde{\nabla}_i\tilde{\nabla}_j\left(\tilde{\nabla}_b\tilde{\nabla}^b+4k\right)k_{\tau\tau}
-\frac{1}{2}\tilde{\gamma}_{ij}\left(\tilde{\nabla}_a\tilde{\nabla}^a\tilde{\nabla}_b\tilde{\nabla}^b-2k\tilde{\nabla}_b\tilde{\nabla}^b+4k^2\right)k_{\tau\tau}.
\label{7.25x}
\end{align}
As we see again $f$ drops out just as it should, since like $\alpha$ and  $B_i-\partial_{\tau}E_i$, $E_{ij}$ also appears in $\delta W_{\mu\nu}$.

One can directly check the validity of (\ref{7.20x}), (\ref{7.21x}), (\ref{7.23x}) and (\ref{7.25x})  by substituting (\ref{7.19x}) and (\ref{7.16x}) into their right-hand sides.  Now in arriving at (\ref{7.20x}), (\ref{7.21x}), (\ref{7.23x}) and (\ref{7.25x})  we made no gauge choice. Then with $\alpha$, $\gamma$, $B_i-\partial_{\tau}E_i$, and $E_{ij}$ all being gauge invariant, it follows that the right-hand sides of (\ref{7.20x}), (\ref{7.21x}), (\ref{7.23x}) and (\ref{7.25x}) have to be gauge invariant too, with their invariance under $h_{\mu\nu}\rightarrow h_{\mu\nu}-\nabla_{\mu}\epsilon_{\nu}-\nabla_{\nu}\epsilon_{\mu}$ being explicitly established  in \cite{Phelps2019}. (Alternatively, one could start by showing that the right-hand sides of the purely kinematic (\ref{7.20x}), (\ref{7.21x}), (\ref{7.23x}) and (\ref{7.25x}) are invariant under $h_{\mu\nu}\rightarrow h_{\mu\nu}-\nabla_{\mu}\epsilon_{\nu}-\nabla_{\nu}\epsilon_{\mu}$ and then infer that $\alpha$, $\gamma$, $B_i-\partial_{\tau}E_i$, and $E_{ij}$ are indeed gauge invariant.) As anticipated in Sec. \ref{S1az}, the relations in (\ref{7.20x}), (\ref{7.21x}), (\ref{7.23x}) and (\ref{7.25x}) generalize the study that  we made in Sec. \ref{S1az} on the decomposition of a vector into its transverse and longitudinal components. And whether we can go from differential equations to integral relations for $\alpha$, $\gamma$, $B_i-\partial_{\tau}E_i$, and $E_{ij}$ will also depend on boundary conditions. Thus just as in the simple example given in Sec. \ref{S1az}, establishing the decomposition theorem for  $\alpha$, $\gamma$, $B_i-\partial_{\tau}E_i$, and $E_{ij}$ depends on the same boundary conditions that are needed to establish their very existence in the first place. 

Finally, since the relations given in (\ref{7.20x}), (\ref{7.21x}), (\ref{7.23x}) and (\ref{7.25x}) are gauge invariant, we can evaluate them in any gauge we like. Choosing the  $\partial^{\mu}[\Omega^{-2}K_{\mu\nu}]=0$ gauge in which  $\delta W_{\mu\nu}$ takes the form given in (\ref{7.8z}), from (\ref{7.20x}), (\ref{7.21x}), (\ref{7.23x}) and (\ref{7.25x}) we see that oscillating solutions in the $K_{\mu\nu}$ sector do indeed propagate to the scalar, vector and tensor sectors. With the $K_{\mu\nu}$ fluctuations being plane waves, it follows that $A_S$, $A_V$ and $A_T$ must be oscillating continuum modes, to thus fix $A_S \geq 1$ in the scalar (\ref{6.14x}),  $A_V \geq 2$ in the vector (\ref{6.20z}), and $A_T\geq 3$ in the tensor (\ref{6.30y}), i.e., continua that start at $A_S=1$, $A_V = 2$, and $A_T= 3$.

\section{The Full Solution at Recombination}
\label{S8}

Given that we did find plane wave solutions by analyzing the $K_{\mu\nu}$ basis, we thus look for scalar, vector and tensor mode solutions that oscillate in both time and space. For the spatial behavior of the scalar modes we see from (\ref{4.4z}) that we get oscillatory behavior in $\chi$ for all continuous values of $A_S$ that obey $A_S>1$. For the vector modes we see from (\ref{4.15z}) that we get oscillatory behavior in $\chi$ for all $A_V$ that obey $A_V>2$. For the tensor modes we see from (\ref{4.21z}) that we get oscillatory behavior in $\chi$ for all $A_T$ that obey $A_T>3$. Noting next that  Bessel functions with pure imaginary index have a leading small $t$ behavior of the form 
\begin{eqnarray}
J_{i\nu}(t)\rightarrow t^{i\nu}=\cos(\nu \log t)+i\sin(\nu \log t),
\label{8.1z}
\end{eqnarray}
from  (\ref{6.14x}), (\ref{6.20z}) and (\ref{6.30y})  we see that we get oscillatory behavior in time in the scalar, vector and tensor sectors under precisely these same $A_S>1$, $A_V>2$, $A_T>3$ conditions. These then are the required ranges for the scalar, vector and tensor separation constants.

As far as the spatial behavior is concerned, these solutions belong to the $f(\nu)=(\cos\nu\chi, \sin\nu\chi)$ sector of  options for the function $f(\nu)$ as listed in (\ref{4.7z}), with the class of all $\nu\geq 0$ solutions being complete (just like the spherical waves that they would become if were to set $k=0$ in (\ref{1.18y})). For the scalar modes the solutions to (\ref{4.3z}) have $\nu^2=A_S-1$. For the vector modes the solutions to (\ref{4.14z}) have $\nu^2=A_V-2$. For the tensor  modes the solutions to (\ref{4.20z}) have $\nu^2=A_T-3$. Thus all solutions in the scalar, vector and tensor mode sectors are indeed oscillatory in space, just as we want.

Since according to (\ref{4.4z}) spatial sector scalar mode solutions behave as $e^{\lambda \chi}$ at large $\chi$ where 
$\lambda =-1 \pm(1-A_S)^{1/2}$, both solutions with any given $A_S>1$ (i.e., either sign of the square root) will be suppressed at large $\chi$ (and thus at large $r$), and thus automatically satisfy the asymptotic boundary and normalization conditions that we require of all fluctuations. Since we also see from (\ref{4.4z}) that one of the two solutions will be well behaved at $\chi=0$, for any $A_S>1$ there will always be one solution that meets the boundary conditions at both $\chi=\infty$ and $\chi=0$ (viz. $r=\infty$ and $r=0$). Now according to  (\ref{4.8z}) $\hat{S}_0(\chi)$ is given by $(df/d\chi)/\sinh\chi$. This solution will be well behaved at $\chi=0$ if we choose the $f(\nu)=\cos\nu \chi$ family. With this choice of $f(\nu)$ the first few solutions given in (\ref{4.8z}) are of the form 
\begin{align}
&\hat{S}_0(\chi)=-\frac{\nu\sin\nu\chi}{\sinh\chi},~~ \hat{S}_1(\chi)=\frac{ \nu\sin\nu\chi\cosh\chi}{\sinh^2 \chi}-\frac{\nu^2 \cos\nu\chi}{\sinh \chi},~~\hat{S}_2(\chi)=\frac{ 3\nu^2\cos\nu\chi\cosh\chi}{\sinh^2 \chi}-\frac{\nu(2-\nu^2) \sin\nu\chi}{\sinh \chi}-\frac{ 3\nu\sin\nu\chi}{\sinh^3 \chi},
\label{8.2z}
\end{align}
to be normalized as described above using the master equation given in Sec. \ref{S4d}.
All of these solutions are well behaved at both $\chi=0$ and $\chi=\infty$, with all integer $\ell\geq 0$ being allowed. We note that all of these solutions are even in $\nu$. Thus while choosing either of the two roots $\nu =\pm (A_S-1)^{1/2}$ affects the $t^{i\nu}$ behavior in  time it does not affect the behavior in space. 

According to (\ref{4.16z}) the first few allowed spatial sector vector mode solutions that satisfy the asymptotic boundary and normalization conditions are of the form 
\begin{align}
&\hat{V}_1(\chi)=\frac{\nu\sin\nu\chi\cosh\chi}{\sinh^3 \chi}-\frac{\nu^2 \cos\nu\chi}{\sinh^2 \chi}=\frac{\hat{S}_1(\chi)}{\sinh\chi},~~
\hat{V}_2(\chi)=\frac{ 3\nu^2\cos\nu\chi\cosh\chi}{\sinh^3 \chi}-\frac{\nu(2-\nu^2) \sin\nu\chi}{\sinh^2 \chi}-\frac{ 3\nu\sin\nu\chi}{\sinh^4 \chi}=\frac{\hat{S}_2(\chi)}{\sinh\chi},
\label{8.3z}
\end{align}
with all integer $\ell\geq 1$ being allowed. All of these solutions are also even in  $\nu$.

Similarly, according to (\ref{4.22z}) the first allowed spatial sector tensor mode solution that satisfies the asymptotic boundary and normalization conditions is of the form 
\begin{eqnarray}
\hat{T}_2(\chi)=\frac{ 3\nu^2\cos\nu\chi\cosh\chi}{\sinh^4 \chi}-\frac{\nu(2-\nu^2) \sin\nu\chi}{\sinh^3 \chi}-\frac{ 3\nu\sin\nu\chi}{\sinh^5 \chi}=\frac{\hat{V}_2(\chi)}{\sinh\chi}=\frac{\hat{S}_2(\chi)}{\sinh^2\chi},
\label{8.4z}
\end{eqnarray}
with all integer $\ell\geq 2$ being allowed. This  solution is also even in  $\nu$.

From (\ref{6.14x}), (\ref{6.20z}) and (\ref{6.30y}) we see that the behavior in time is of the form $tJ_{i\nu}(t/\eta^{1/2})$, $J_{i\nu}(t/\eta^{1/2})$ and $t^{-1}J_{i\nu}(t/\eta^{1/2})$ in the respective cases. 
Finally, on multiplying by  $Y^m_{\ell}(\theta,\phi)$, introducing the polarization vectors and tensors given in (\ref{6.20z}) and (\ref{6.30y}) and the $A(\nu,\beta,F)$ normalization factors with $\beta=\ell-(F-3)/2$ that are given in (\ref{4.25y}), the full structure for the allowed modes is given by
\begin{align}
\alpha&= A(\nu,\ell,3)\hat{S}_{\ell}(\chi)Y^m_{\ell}(\theta,\phi)tJ_{i\nu}(t/\eta^{1/2}),
\quad
\gamma=-\frac{\eta}{6t^2}\left(\nu^2+1+3\partial^2_t\right)\alpha-\frac{\alpha}{2},
\nonumber\\
B_i-t\partial_t E_i&=A(\nu,\ell -1,5)\epsilon_i\hat{V}_{\ell}(\chi)Y^m_{\ell}(\theta,\phi)J_{i\nu}(t/\eta^{1/2}),
\quad
E_{ij}=A(\nu,\ell -2,7)\epsilon_{ij}\hat{T}_{\ell}(\chi)Y^m_{\ell}(\theta,\phi)t^{-1}J_{i\nu}(t/\eta^{1/2}),
\label{8.5z}
\end{align}
for all real and positive $\nu$. In the conformal theory these scalar, vector and tensor mode solutions are exact to one part in $10^4$ at recombination \cite{footnoteK}.

\begin{acknowledgments}
The author acknowledges useful conversations with A. Amarasinghe, T. Liu, D. Norman, Dr. M. Phelps and Dr. K. Shankar.
\end{acknowledgments}

\appendix
\numberwithin{equation}{section}
\setcounter{equation}{0}

\section{Some Typical Solutions to the Scalar, Vector and Tensor Equations}
\label{SA}

In Sec. \ref{S3} we obtained higher-derivative, decoupled fluctuation equations for the scalar, vector and tensor fluctuations. For the scalars we obtained equations such as (\ref{3.9z}), viz. 
\begin{align}
&-\frac{2\eta}{3 \Omega^2}\tilde{\nabla}_i\tilde{\nabla}^i(\tilde{\nabla}_j\tilde{\nabla}^j+3k)( 3\partial_{\tau}^2 -\tilde\nabla_a\tilde\nabla^a)\alpha-2\tilde{\nabla}_i\tilde{\nabla}^i(\tilde{\nabla}_j\tilde{\nabla}^j+3k)(\alpha + 2\dot\Omega \Omega^{-1}\gamma)=0.
\label{A.1}
\end{align}
For the vectors we obtained equations such as (\ref{3.19z}), viz.
\begin{eqnarray}
&&\frac{\eta}{2\Omega^2}(\tilde{\nabla}_a\tilde{\nabla}^a-2k)(\tilde{\nabla}_b\tilde{\nabla}^b+k) (\tilde{\nabla}_{c}\tilde{\nabla}^{c}+2k)(\tilde\nabla_a\tilde\nabla^a -2k-\partial_{\tau}^2) (\dot{B}_i-\ddot{E}_i)
\nonumber\\
&&-(\tilde{\nabla}_a\tilde{\nabla}^a-2k)(\tilde{\nabla}_b\tilde{\nabla}^b+k)(\tilde{\nabla}_{c}\tilde{\nabla}^{c}+2k)\left[\frac{1}{2}(\dot{B}_i-\ddot{E}_i)+\dot{\Omega}\Omega^{-1}(B_i-\dot{E}_i)\right]=0.
\label{A.2}
\end{eqnarray}
For the tensors we obtained equations such as (\ref{3.30y}), viz.
\begin{eqnarray}
&&(\tilde{\nabla}_c\tilde{\nabla}^c-2k)(\tilde{\nabla}_a\tilde{\nabla}^a-6k)(\tilde{\nabla}_b\tilde{\nabla}^b-3k)
\nonumber\\
&&\times
\bigg{[}\frac{\eta}{\Omega^2}\left[ (\tilde\nabla_b \tilde\nabla^b-\partial_{\tau}^2-2k)^2+4k\partial_{\tau}^2 \right] E_{ij}
+ \overset{..}{E}_{ij} +2 k E_{ij} +2  \dot{\Omega} \Omega^{-1}\dot{E}_{ij} - \tilde{\nabla}_{d}\tilde{\nabla}^{d}E_{ij}\bigg{]}=0.
\label{A.3}
\end{eqnarray}
Then in Sec. \ref{S4} we presented a general procedure for solving the associated wave equations for the scalar, vector and tensor fluctuations. We now show how to use this procedure to solve (\ref{A.1}), (\ref{A.2}) and (\ref{A.3}).

\section{The Scalar Sector}
\label{SB}

We can write (\ref{A.1}) in the generic form
\begin{align}
&\tilde{\nabla}_i\tilde{\nabla}^i(\tilde{\nabla}_j\tilde{\nabla}^j+3k)S(\tau,\chi,\theta,\phi)=0,
\label{B.1}
\end{align}
where
\begin{align}
S(\tau,\chi,\theta,\phi)=-\frac{2\eta}{3 \Omega^2} (3\partial_{\tau}^2 -\tilde\nabla_a\tilde\nabla^a)\alpha-2(\alpha + 2\dot\Omega \Omega^{-1}\gamma).
\label{B.2}
\end{align}
Since $\tilde{\nabla}_i\tilde{\nabla}^i$ and $\tilde{\nabla}_j\tilde{\nabla}^j+3k$ commute with each other there are three possibilities: that $\tilde{\nabla}_i\tilde{\nabla}^iS(\tau,\chi,\theta,\phi)=0$, that $(\tilde{\nabla}_j\tilde{\nabla}^j+3k)S(\tau,\chi,\theta,\phi)=0$ or that $S(\tau,\chi,\theta,\phi)=0$. Of these three options only the last one fixes the $\tau$ dependence of $S(\tau,\chi,\theta,\phi)$. We thus need to find a way to exclude the first two options. This will be done with boundary conditions on $\chi$.

On introducing a separation constant $A_S$ and extracting out the angular $Y^m_{\ell}(\theta, \phi)$ behavior, we found that the radial $S_{\ell}(\chi)$ obeyed the second-order differential equation given in (\ref{4.3z}). With $k<0$ we thus need to solve this equation for $A_S=0$ and $A_S=-3$. Explicit solutions in these two cases have been given in \cite{Phelps2019}. For $A_S=0$ the first few solutions are of the form
\begin{align}
&\hat{S}^{(1)}_{0}(A_S=0)=\frac{\cosh\chi}{\sinh\chi},\quad \hat{S}^{(2)}_{0}(A_S=0)=1,
\nonumber\\
&\hat{S}^{(1)}_{1}(A_S=0)=\frac{1}{\sinh^2\chi},\quad \hat{S}^{(2)}_{1}(A_S=0)=\frac{\cosh\chi}{\sinh\chi}-\frac{\chi}{\sinh^2\chi},
\nonumber\\
&\hat{S}^{(1)}_{2}(A_S=0)=\frac{\cosh\chi}{\sinh^3\chi},\quad \hat{S}^{(2)}_{2}(A_S=0)=1+\frac{3}{\sinh^2\chi}-\frac{3\chi\cosh\chi}{\sinh^3\chi},
\nonumber\\
&\hat{S}^{(1)}_{3}(A_S=0)=\frac{4}{\sinh^2\chi}+\frac{5}{\sinh^4\chi},\quad \hat{S}^{(2)}_{3}(A_S=0)=
\frac{2\cosh\chi}{\sinh\chi}+\frac{15\cosh\chi}{\sinh^3\chi}-\frac{12\chi}{\sinh^2\chi}-\frac{15\chi}{\sinh^4\chi},
\label{B.3}
\end{align}
where as before there are two solutions for each $\ell$ value, with $\ell$ being the lower index.
From this pattern we see that the solutions that are bounded at $\chi=\infty$ are badly-behaved at $\chi=0$, while the solutions that are  well-behaved at $\chi=0$ are unbounded at $\chi=\infty$. Thus all of these $A_S=0$ solutions are excluded by a requirement that solutions be  bounded at $\chi=\infty$ and be well-behaved at $\chi=0$.

For $A_S=-3$ the first few solutions are of the form
\begin{eqnarray}
&&\hat{S}^{(1)}_0(A_S=-3)=\cosh\chi,\quad \hat{S}^{(2)}_0(A_S=-3)=2\sinh\chi+\frac{1}{\sinh\chi},
\nonumber\\
&&\hat{S}^{(1)}_1(A_S=-3)=\sinh\chi,\quad \hat{S}^{(2)}_1(A_S=-3)=2\cosh\chi-\frac{\cosh\chi}{\sinh^2\chi},
\nonumber\\
&&\hat{S}^{(1)}_2(A_S=-3)=2\cosh\chi-\frac{3\cosh\chi}{\sinh^2\chi}+\frac{3\chi}{\sinh^3\chi},\quad \hat{S}^{(2)}_2(A_S=-3)=\frac{1}{\sinh^3\chi},
\nonumber\\
&&\hat{S}^{(1)}_3(A_S=-3)=2\sinh\chi-\frac{5}{\sinh\chi}-\frac{15}{\sinh^3\chi}+\frac{15\chi\cosh\chi}{\sinh^4\chi},\quad \hat{S}^{(2)}_3(A_S=-3)=\frac{\cosh\chi}{\sinh^4\chi}.
\label{B.4}
\end{eqnarray}
From this pattern we again see that the solutions that are bounded at $\chi=\infty$ are badly-behaved at $\chi=0$, while the solutions that are  well-behaved at $\chi=0$ are unbounded at $\chi=\infty$. Thus all of these $A_S=-3$ solutions are also excluded by a requirement that solutions be  bounded at $\chi=\infty$ and be well-behaved at $\chi=0$. 

Thus for the scalar sector the only option left is that $S(\tau,\chi,\theta,\phi)$ as given in (\ref{B.2}) vanishes. In the recombination era (\ref{B.2}) reduces to (\ref{6.5z}), and in Sec. \ref{S6} we solved (\ref{6.5z}) and the three other scalar sector equations (\ref{6.1z}), (\ref{6.6z}) and (\ref{6.4z}), equations that can be derived from (\ref{3.4z}), (\ref{3.6z}), and a linear combination of (\ref{3.9z}) and (\ref{3.10z}) by a treatment analogous to the one we have just given for (\ref{3.9z}). Hence for the scalar sector we see that boundary conditions enable us to exclude the spatial derivative conditions $\tilde{\nabla}_i\tilde{\nabla}^iS(\tau,\chi,\theta,\phi)=0$ and $(\tilde{\nabla}_j\tilde{\nabla}^j+3k)S(\tau,\chi,\theta,\phi)=0$, and we only need to consider $S(\tau,\chi,\theta,\phi)=0$, which thereby enables us to fix the $\tau$ behavior of the scalar sector.

\section{The Vector Sector}
\label{SC}

For the vector sector we can write (\ref{A.2}) in the generic form
\begin{align}
&(\tilde{\nabla}_a\tilde{\nabla}^a-2k)(\tilde{\nabla}_b\tilde{\nabla}^b+k) (\tilde{\nabla}_{c}\tilde{\nabla}^{c}+2k)
V_i(\tau,\chi,\theta,\phi)=0,
\label{C.1}
\end{align}
where
\begin{align}
V_i(\tau,\chi,\theta,\phi)=
&\frac{\eta}{2\Omega^2}(\tilde\nabla_a\tilde\nabla^a -2k-\partial_{\tau}^2) (\dot{B}_i-\ddot{E}_i)
-\frac{1}{2}(\dot{B}_i-\ddot{E}_i)-\dot{\Omega}\Omega^{-1}(B_i-\dot{E}_i).
\label{C.2}
\end{align}
Since $\tilde{\nabla}_i\tilde{\nabla}^i-2k$, $\tilde{\nabla}_i\tilde{\nabla}^i+k$ and $\tilde{\nabla}_j\tilde{\nabla}^j+2k$ all commute with each other there are four possibilities: that $(\tilde{\nabla}_i\tilde{\nabla}^i-2k)V_i(\tau,\chi,\theta,\phi)=0$, that $(\tilde{\nabla}_i\tilde{\nabla}^i+k)V_i(\tau,\chi,\theta,\phi)=0$, that $(\tilde{\nabla}_i\tilde{\nabla}^i+2k)V_i(\tau,\chi,\theta,\phi)=0$, or that $V_i(\tau,\chi,\theta,\phi)=0$. Of these four options only the last one fixes the $\tau$ dependence of $V_i(\tau,\chi,\theta,\phi)$. We thus need to find a way to exclude the first three options. 

On introducing a separation constant $A_V$ and extracting out the angular $Y^m_{\ell}(\theta, \phi)$ behavior, we found that the radial $g_{1,\ell}(\chi)$ obeyed the second-order differential equation given in (\ref{4.14z}). With $k<0$ we thus need to solve this equation for $A_V=2$, $A_V=-1$  and $A_V=-2$. Explicit solutions for $A_V=2$ have been given above in (\ref{4.37y}). For the $A_V=2$ solutions  we see that the $\hat{V}^{(2)}_{\ell}(A_V=2)$ solutions with $\ell \geq1$ are bounded at  $\chi=\infty$ and well behaved at $\chi=0$. Thus if implement (\ref{C.1}) by $(\tilde{\nabla}_a\tilde{\nabla}^a+2)V_i=0$,  we are not forced to $V_i=0$.

Solutions for $A_V=-1$, $A_V=-2$  have been given in \cite{Phelps2019}. For $A_V=-1$ the first few solutions are of the form
\begin{eqnarray}
&&\hat{V}^{(1)}_0(A_V=-1)=\frac{e^{\chi\surd{3}}}{\sinh^2\chi},\quad \hat{V}^{(2)}_0(A_V=-1)=\frac{e^{-\chi\surd{3}}}{\sinh^2\chi},
\nonumber\\
&&\hat{V}^{(1)}_1(A_V=-1)=\frac{e^{\chi\surd{3}}}{\sinh^3\chi}\left[\surd{3}\sinh\chi-\cosh\chi\right],\quad \hat{V}^{(2)}_1(A_V=-1)=\frac{e^{-\chi\surd{3}}}{\sinh^3\chi}\left[-\surd{3}\sinh\chi-\cosh\chi\right],
\nonumber\\
&&\hat{V}^{(1)}_2(A_V=-1)=\frac{e^{\chi\surd{3}}}{\sinh^4\chi}\left[3-3\surd{3}\cosh\chi\sinh\chi+5\sinh^2\chi
\right],
\nonumber\\
 &&\hat{V}^{(2)}_2(A_V=-1)=\frac{e^{-\chi\surd{3}}}{\sinh^4\chi}\left[3+3\surd{3}\cosh\chi\sinh\chi+5\sinh^2\chi\right],
\nonumber\\
&&\hat{V}^{(1)}_3(A_V=-1)=\frac{e^{\chi\surd{3}}}{\sinh^5\chi}\left[15\surd{3}\sinh\chi+14\surd{3}\sinh^3\chi
-15\cosh\chi-24\cosh\chi\sinh^2\chi\right],
\nonumber\\
&&\hat{V}^{(2)}_3(A_V=-1)=\frac{e^{-\chi\surd{3}}}{\sinh^5\chi}\left[-15\surd{3}\sinh\chi-14\surd{3}\sinh^3\chi
-15\cosh\chi-24\cosh\chi\sinh^2\chi\right].
\label{C.3}
\end{eqnarray}
All of these solutions are bounded at $\chi=\infty$ and all $\hat{V}^{(1)}_{\ell}(A_V=-1)-\hat{V}^{(2)}_{\ell}(A_V=-1)$ with $\ell\geq 1$ are well-behaved at $\chi=0$.  Thus if implement (\ref{C.1}) by $(\tilde{\nabla}_a\tilde{\nabla}^a-1)V_i=0$,  we are not forced to $V_i=0$. 

For $A_V=-2$ the first few solutions are of the form
\begin{eqnarray}
&&\hat{V}^{(1)}_0(A_V=-2)=\frac{\cosh\chi}{\sinh\chi},\quad \hat{V}^{(2)}_0(A_V=-2)=2+\frac{1}{\sinh^2\chi},
\nonumber\\
&&\hat{V}^{(1)}_1(A_V=-2)=1,\quad \hat{V}^{(2)}_1(A_V=-2)=2\frac{\cosh\chi}{\sinh\chi}-\frac{\cosh\chi}{\sinh^3\chi},
\nonumber\\
&&\hat{V}^{(1)}_2(A_V=-2)=2\frac{\cosh\chi}{\sinh\chi}-\frac{3\cosh\chi}{\sinh^3\chi}+\frac{3\chi}{\sinh^4\chi},\quad \hat{V}^{(2)}_2(A_V=-2)=\frac{1}{\sinh^4\chi},
\nonumber\\
&&\hat{V}^{(1)}_3(A_V=-2)=2-\frac{5}{\sinh^2\chi}-\frac{15}{\sinh^4\chi}+\frac{15\chi\cosh\chi}{\sinh^5\chi},\quad \hat{V}^{(2)}_3(A_V=-2)=\frac{\cosh\chi}{\sinh^5\chi}.
\label{C.4}
\end{eqnarray}
Of these solutions the only ones that are  bounded at $\chi=\infty$ are $\hat{V}^{(2)}_2(A_V=-2)$ and $\hat{V}^{(2)}_3(A_V=-2)$. However, they are not well-behaved at $\chi=0$. Since they thus can  be excluded by boundary conditions at $\chi=\infty$ and $\chi=0$, if we set $(\tilde{\nabla}_a\tilde{\nabla}^a-2)V_i=0$,  the only allowed solution will be $V_i=0$. As we see, boundary conditions are not capable of excluding the $A_V=2$ and $A_V=-1$ cases. We address this concern in Sec. \ref{S4d}. Because of its overall $\tilde{\nabla}_a\tilde{\nabla}^a-2k$ factor, we note that considerations similar to our treatment of (\ref{3.19z}) also apply to (\ref{3.13z}), the other vector sector equation.

\section{The Tensor Sector}
\label{SD}

For the tensor sector we can write (\ref{A.3}) in the generic form
\begin{eqnarray}
&&(\tilde{\nabla}_c\tilde{\nabla}^c-2k)(\tilde{\nabla}_a\tilde{\nabla}^a-6k)(\tilde{\nabla}_b\tilde{\nabla}^b-3k)T_{ij}(\tau,\chi,\theta,\phi)=0,
\label{D.1}
\end{eqnarray}
where
\begin{eqnarray}
&&T_{ij}(\tau,\chi,\theta,\phi)=\frac{\eta}{\Omega^2}\left[ (\tilde\nabla_b \tilde\nabla^b-\partial_{\tau}^2-2k)^2+4k\partial_{\tau}^2 \right] E_{ij}
+ \overset{..}{E}_{ij} +2 k E_{ij} +2  \dot{\Omega} \Omega^{-1}\dot{E}_{ij} - \tilde{\nabla}_{d}\tilde{\nabla}^{d}E_{ij}.
\label{D.2}
\end{eqnarray}
Since $\tilde{\nabla}_i\tilde{\nabla}^i-2k$, $\tilde{\nabla}_i\tilde{\nabla}^i-6k$ and $\tilde{\nabla}_j\tilde{\nabla}^j-3k$ all commute with each other there are four possibilities: that $(\tilde{\nabla}_i\tilde{\nabla}^i-2k)T_{ij}(\tau,\chi,\theta,\phi)=0$, that $(\tilde{\nabla}_i\tilde{\nabla}^i-6k)T_{ij}(\tau,\chi,\theta,\phi)=0$, that $(\tilde{\nabla}_i\tilde{\nabla}^i-3k)T_{ij}(\tau,\chi,\theta,\phi)=0$, or that $T_{ij}(\tau,\chi,\theta,\phi)=0$. Of these four options only the last one fixes the $\tau$ dependence of $T_{ij}(\tau,\chi,\theta,\phi)$. We thus need to find a way to exclude the first three options. 

On introducing a separation constant $A_T$ and extracting out the angular $Y^m_{\ell}(\theta, \phi)$ behavior, we found that the radial $h_{11,\ell}(\chi)$ obeyed the second-order differential equation given in (\ref{4.20z}). With $k<0$ we thus need to solve this equation for $A_T=2$, $A_T=6$  and $A_T=3$. Explicit solutions for $A_T=2$ have been given above in (\ref{4.39y}). For the $A_T=2$ solutions  we see that the $\hat{T}^{(2)}_{\ell}(A_T=2)$ solutions with $\ell \geq 2$ are bounded at  $\chi=\infty$ and well behaved at $\chi=0$. Thus if implement (\ref{D.1}) by $(\tilde{\nabla}_a\tilde{\nabla}^a+2)T_{ij}=0$,  we are not forced to $T_{ij}=0$.

Solutions for $A_T=6$, $A_T=3$  have been given in \cite{Phelps2019}. For $A_T=3$ the first few solutions are of the form
\begin{eqnarray}
\hat{T}^{(1)}_0(A_T=3)&=&\frac{1}{ \sinh^3\chi},\quad \hat{T}^{(2)}_0(A_T=3)=\frac{\chi }{\sinh^3\chi},
\nonumber\\
\hat{T}^{(1)}_1(A_T=3)&=&\frac{\cosh \chi }{ \sinh^4\chi},\quad \hat{T}^{(2)}_1(A_T=3)=\frac{1}{ \sinh^3\chi}-\frac{\chi\cosh\chi}{\sinh^4\chi},
\nonumber\\
\hat{T}^{(1)}_2(A_T=3)&=&\frac{2}{ \sinh^3\chi}+\frac{3}{\sinh^5\chi},\quad \hat{T}^{(2)}_2(A_T=3)=\frac{3\cosh\chi}{\sinh^4\chi}-\frac{2\chi}{\sinh^3\chi}-\frac{3\chi }{\sinh^5\chi},
\nonumber\\
\hat{T}^{(1)}_3(A_T=3)&=&\frac{2\cosh\chi}{\sinh^4\chi}+\frac{5\cosh\chi}{\sinh^6\chi},\quad \hat{T}^{(2)}_3(A_T=3)=\frac{11}{\sinh^3\chi}+\frac{15}{\sinh^5\chi}-\frac{6\chi\cosh\chi}{\sinh^4\chi}-\frac{15\chi\cosh\chi }{\sinh^6\chi}.~~~
\label{D.3}
\end{eqnarray}
All of these solutions are bounded at $\chi=\infty$ and all $\hat{T}^{(2)}_{\ell}(A_T=3)$ with $\ell\geq 2$ are well-behaved at $\chi=0$. Thus if implement (\ref{D.1}) by $(\tilde{\nabla}_a\tilde{\nabla}^a+3)T_{ij}=0$,  we are not forced to $T_{ij}=0$.

A similar outcome occurs for $A_T=6$, and even though we do not evaluate the $A_T=6$ solutions explicitly, according to (\ref{4.21z}) all solutions to $(\tilde{\nabla}_a\tilde{\nabla}^a+6)T_{11}=0$ with $A_T=6$ are bounded at $\chi=\infty$ (behaving as $e^{-3\chi}\cos(\surd{3}\chi)$ and $e^{-3\chi}\sin(\surd{3}\chi)$), with one set of these solutions being well-behaved at $\chi=0$ for all $\ell \geq 2$. Thus if implement (\ref{D.1}) by $(\tilde{\nabla}_a\tilde{\nabla}^a+6)T_{ij}=0$,  we are not forced to $T_{ij}=0$.

We thus see that while boundary conditions at $\chi=\infty$ and at $\chi=0$ will force us to set $S(\tau,\chi,\theta,\phi)=0$ in the scalar sector, they do not force us to set $V_i(\tau,\chi,\theta,\phi)=0$ or $T_{ij}(\tau,\chi,\theta,\phi)=0$ in the vector and tensor sectors. However, solutions in which $V_i(\tau,\chi,\theta,\phi)$ and $T_{ij}(\tau,\chi,\theta,\phi)$ are non-vanishing will each have their own specific dependence on $\chi$. In Sec. \ref{S4e} we show that these various $\chi$ dependencies do not line up with each other in the original coupled second-order $\eta \delta W_{\mu\nu}-\Delta_{\mu\nu}=0$ fluctuation equations themselves, and in the end that is what forces us to   $V_i(\tau,\chi,\theta,\phi)=0$, $T_{ij}(\tau,\chi,\theta,\phi)=0$.  Thus solving the higher-derivative (\ref{3.9z}), (\ref{3.19z}) and (\ref{3.30y}) (and analogously (\ref{3.4z}), (\ref{3.6z}), (\ref{3.10z}) and (\ref{3.13z})) do not lead us to any discrete allowed values for $A_S$, $A_V$ or $A_T$. Rather, the analysis of Sec. \ref{S7} shows that they each possess a continuum of values with $A_S\geq 1$, $A_V\geq 2$ and $A_T\geq 3$. In Sec. \ref{S6} we solved the $S(\tau,\chi,\theta,\phi)=0$, $V_i(\tau,\chi,\theta,\phi)=0$ and $T_{ij}(\tau,\chi,\theta,\phi)=0$ equations in the recombination era. In those solutions the factors $\mu$, $\rho$ and $\sigma$ that appear in the scalar (\ref{6.14x}), the vector (\ref{6.20z}) and the tensor (\ref{6.28y}) are all pure imaginary.


\begin{thebibliography}{99}
\bibitem{footnoteY} That quantum mechanics is even relevant on large macroscopic scales is already manifest in the CMB itself as it is described by an intrinsically quantum-mechanical  black-body radiation spectrum. Also, the stability of both white dwarf stars and neutron stars is provided by the Fermi-Dirac statistics of the spin one-half particles in the star, with the Chandrasekhar mass $(\hbar c/G)^{3/2}/m_p^2$ of white dwarf stars intrinsically depending on $\hbar$.
\bibitem{footnoteA} With the antisymmetric Maxwell tensor $F_{\mu\nu}$ being invariant under a local electromagnetic gauge transformation and with the Maxwell action being of the form $I_{\rm M}=\int d^4x(-g)^{1/2}F_{\mu\nu}F^{\mu\nu}$, we see that $I_{\rm W}$ is the  gravitational analog of $I_{\rm M}$. Also we note that under a gauge transformation $g_{\mu\nu}(x)$ transforms into itself, while under a conformal transformation $A_{\mu}(x)$ transforms into itself. In consequence, like $I_{\rm W}$ the Maxwell action $I_{\rm M}$ is also locally conformal invariant.
\bibitem{footnoteA1} In order to determine the force at any given point on a spherical surface in a spherical distribution of matter sources in Newtonian gravity, one only needs to consider matter interior to the spherical surface. (The force falls as like $1/r^2$ while the solid angle grows like $r^2$, so the exterior cancels.) Thus in Newtonian gravity any given set of dark matter particles would only affect the rotation curve of the galaxy in which that particular set of dark matter particles  is itself located. However, if the potential is changed, the solid angle does not change, and so now there is no exterior cancellation. Thus for the linear potentials of conformal gravity matter outside of any given galaxy does affect the  rotation curve within that galaxy.
\bibitem{footnoteA2} Interestingly, in analog to (\ref{1.39y}) we show in Sec. \ref{S7a}  that the retarded Green's function associated with the flat four-space massless fourth-order $\partial_{\sigma}\partial^{\sigma}\partial_{\tau}\partial^{\tau}$ derivative operator of relevance to conformal gravity is of the form $\theta(t-r)/8\pi $. It thus takes support in the entire $t^2-r^2\geq 0$ region with $t>0$, i.e., both on and within the forward light cone. In contrast,  the retarded Green's function associated with the flat four-space massless second-order  $\partial_{\sigma}\partial^{\sigma}$ operator is $\delta(t-r)/4\pi r$, which only takes support on the forward light cone.
\bibitem{footnoteB} For the full comoving time $a(t)$ given in (\ref{1.32y}) the integral $\int dt/a(t)$ is an analytic continuation of an elliptic integral, which is why the conformal time (\ref{2.8z}) is somewhat intractable.
\bibitem{footnoteC} From this point we set $c=1$, so $t$ or $\tau$ means $ct$ or $c\tau$.
\bibitem{footnoteD} Since one can make gauge transformations with gauge functions that diverge asymptotically, a function that is bounded in any specifically chosen gauge could become unbounded in a different one. However, for the gauge-invariant functions that appear in the fluctuation equations this cannot occur, and if they are  bounded in one gauge they are bounded in every gauge.
\bibitem{footnote2} There is however a slight difference. Even though the $k=-1$ modes fall off as $e^{-\chi}$, plane waves do not fall off at spatial infinity. Nonetheless, one can build localized wave packets out of plane waves, and just as one can do so for plane waves one could also do so for the mode solutions given in (\ref{4.24y}).
\bibitem{footnote3} Since maximally  symmetric three-spaces with metric $d\chi^2+\sinh^2\chi d\theta^2+\sinh^2\chi \sin^2\theta d\phi^2$ are conformal to flat, the delta function normalization that is needed for plane waves propagating in a flat space translates into the delta function normalization for the $Z_{\nu,\beta}(\chi)$ that is given  in (\ref{4.26y}), with the continuous parameter $\nu$ replacing the continuous momentum  variable $k$.
\bibitem{footnote4} In consequence, there is no choice of $A_S$ and $A_T$ for which we can bypass the vector equation and have the second derivative of the scalar wave equation (\ref{4.3z}) become the tensor equation (\ref{4.20z}).
\bibitem{footnoteE} In \cite{Phelps2019} and in the Appendix it is shown that for $k=-1$ there are no solutions to $(\tilde{\nabla}_a\tilde{\nabla}^a+3k)S=0$ where $S$ is a scalar that are well behaved at both $\chi =\infty$ and $\chi=0$. Thus we can factor out the overall $\tilde{\nabla}_a\tilde{\nabla}^a+3k$ factor on the right-hand side of  (\ref{6.9x}).
\bibitem{footnoteG} In \cite{Mannheim2012a,Amarasinghe2019} we introduced a gauge condition, $\nabla_{\nu}K^{\mu\nu}-K^{\mu\nu}\Gamma^{\alpha}_{\alpha \nu}=0$,  that is preserved under a local conformal transformation $g_{\mu\nu}(x)\rightarrow \Omega^2(x)g_{\mu\nu}(x)$. With  $\nabla_{\nu}K^{\mu\nu}-K^{\mu\nu}\Gamma^{\alpha}_{\alpha \nu}=
\partial_{\nu}K^{\mu\nu}+\Gamma^{\mu}_{\nu \sigma}K^{\sigma\nu}$, for the background metric given in (\ref{7.1z})  we have
$\partial_{\nu}K^{\mu\nu}+2\Omega^{-1}K^{\mu\nu}\partial_{\nu}\Omega=0$. Then with $K^{\mu\nu}=\eta^{\mu\alpha}\eta^{\nu\beta}\Omega^{-4}K_{\alpha\beta}$ we have $\partial^{\beta}K_{\sigma\beta}-2\Omega^{-1}K_{\sigma\beta}\partial^{\beta}\Omega=\Omega^{2}\partial^{\beta}[\Omega^{-2}K_{\sigma\beta}]=0$. Thus the condition $\partial^{\beta}[\Omega^{-2}K_{\sigma\beta}]=0$ that we are using here is conformal invariant.
\bibitem{footnoteK} It would be of interest to see whether the one part in $10^4$ contribution identified in (\ref{1.38y}) might be related to the one part in $10^5$  temperature  fluctuation in the CMB.
\end{thebibliography}
\end{document}